\documentclass[aps, twocolumn, superscriptaddress, nofootinbib, prx, 10pt]{revtex4-2}

\usepackage{graphicx,enumerate,verbatim,bbold}
\usepackage{amsmath,amssymb,amsthm,float,mathrsfs}
\usepackage{dsfont}
\usepackage[T1]{fontenc}
\usepackage[utf8]{inputenc}
\usepackage{lmodern}
\usepackage{fancyhdr}
\usepackage{xcolor}
\usepackage{braket}
\definecolor{columbiablue}{RGB}{0,114,206}
\usepackage[colorlinks,allcolors=columbiablue]{hyperref}
\usepackage[capitalize]{cleveref}
\usepackage[normalem]{ulem}

\newcommand{\mm}{\mathbf{m}}

\renewcommand{\vec}{\mathbf}

\newcommand{\rrop}{\hat{\rr}}

\newcommand{\Rop}{\hat{R}}
\newcommand{\sop}{\hat{s}}
\newcommand{\sdop}{\hat{s}^\dag}
\newcommand{\rhop}{\hat{\rho}}
\newcommand{\Li}{\text{Li}}

\newcommand{\avg}[1]{\ensuremath{\langle #1 \rangle}}
\newcommand{\ketbra}[2]{\left|{#1}\middle\rangle\middle\langle{#2}\right|}

\newcommand{\dagg}{^{\dag}}

\newcommand{\bdop}{\hat b^{\dagger}}
\newcommand{\bop}{\hat b}

\newcommand{\Hop}{\hat{H}}

\newcommand{\spl}{\hat \sigma^{+}}
\newcommand{\smi}{\hat \sigma^{-}}

\newcommand{\im}{\text{i}}

\newcommand{\w}{\omega}

\newcommand{\id}{\mathbb{1}}
\newcommand{\W}{\Omega}

\newcommand{\eg}{\textit{e.g.}}
\newcommand{\ie}{i.e., }

\newcommand{\hc}{{\rm H.c.}}

\newcommand{\pare}[1]{\left( {#1} \right)}
\newcommand{\spare}[1]{\left[ {#1} \right]}

\newcommand{\be}{\begin{equation}}
\newcommand{\ee}{\end{equation}}
\newcommand{\bea}{\begin{eqnarray}}
\newcommand{\eea}{\end{eqnarray}}

\newcommand{\norm}[1]{\lVert {#1} \rVert}

\newcommand{\inv}[1]{\frac{1}{{#1}}}

\newcommand{\tr}{\textrm{tr}}

 \newcommand{\sgn}[1]{\mathop{\mathrm{sgn}}(#1)}

\renewcommand{\Im}{\text{Im}}
\renewcommand{\Re}{\text{Re}}

\newcommand{\pa}[1]{\partial_{#1}}

\newcommand{\jj}{\mathbf{j}}
\newcommand{\kk}{\mathbf{k}}
\newcommand{\nn}{\mathbf{n}}
\newcommand{\pp}{\mathbf{p}}
\newcommand{\qq}{\mathbf{q}}
\newcommand{\rr}{\mathbf{r}}

\newcommand{\uu}{\mathbf{u}}

\newcommand{\EE}{\mathbf{E}}

\newcommand{\PP}{\mathbf{P}}

\def\dpp{\boldsymbol{\wp}}

\definecolor{daniel}{RGB}{120,0,220}

\begin{document}

\title{Polaron-Polaritons in Subwavelength Arrays of Trapped Atoms}

\author{Kristian Knakkergaard Nielsen}
\thanks{These authors contributed equally to this work.}
\affiliation{Niels Bohr Institute, University of Copenhagen, Jagtvej 128, DK-2200 Copenhagen, Denmark}
\affiliation{Max-Planck-Institut für Quantenoptik, Hans-Kopfermann-Strasse 1, 85748 Garching, Germany}
\author{Lukas Wangler}
\thanks{These authors contributed equally to this work.}
\affiliation{ICFO-Institut de Ciencies Fotoniques, The Barcelona Institute of Science and Technology, 08860 Castelldefels (Barcelona), Spain}
\affiliation{Max-Planck-Institut für Quantenoptik, Hans-Kopfermann-Strasse 1, 85748 Garching, Germany}
\author{David Castells-Graells}
\affiliation{PlanQC GmbH, Lichtenbergstr. 8, 85748 Garching b. München, Germany}
\affiliation{Max-Planck-Institut für Quantenoptik, Hans-Kopfermann-Strasse 1, 85748 Garching, Germany}
\affiliation{Munich Center for Quantum Science and Technology, Schellingstrasse 4, D-80799 München, Germany}
\author{J. Ignacio Cirac}
\affiliation{Max-Planck-Institut für Quantenoptik, Hans-Kopfermann-Strasse 1, 85748 Garching, Germany}
\affiliation{Munich Center for Quantum Science and Technology, Schellingstrasse 4, D-80799 München, Germany}
\author{Ana Asenjo-Garcia}
\affiliation{Department of Physics, Columbia University, New York, NY 10027, USA}
\author{Daniel Malz}
\affiliation{Department of Mathematical Sciences, University of Copenhagen, 2100 Copenhagen, Denmark}
\author{Cosimo C. Rusconi}
\affiliation{Instituto de F\'isica Fundamental - Consejo Superior de Investigaciones Cient\'ifica (CSIC), Madrid, Espa\~na}
\affiliation{Department of Physics, Columbia University, New York, NY 10027, USA}
\affiliation{Max-Planck-Institut für Quantenoptik, Hans-Kopfermann-Strasse 1, 85748 Garching, Germany}

\date{\today}

\begin{abstract}
    Subwavelength arrays of atoms trapped in optical lattices or tweezers are inherently susceptible to deformations: Optomechanical forces displace atoms within their trapping potential and produce lattice distortions, which in turn modify the optical response of the array. We show that this optomechanical coupling hybridizes collective atomic excitations (polaritons) with phonons, forming polaron–polaritons---the fundamental quasiparticles governing light-matter interactions in arrays of trapped atoms. Using analytical polaron theory and numerical simulations, we find that: (1) phonons can strongly enhance the decay of subradiant states, but also enable their efficient excitation; (2) transport of dark excitations remains remarkably robust even at low trap frequencies, except when a polariton can resonantly scatter phonons; and (3) motion reduces the reflectivity of a two-dimensional atomic mirror; by identifying design principles that mitigate this degradation, we recover reflectivity above 99\% under realistic conditions. Our findings lay the foundation for analyzing motional effects in key applications and suggest new ways to harness them in state-of-the-art experiments.
\end{abstract}

\maketitle 

\section{Introduction}

Realizing controlled interactions between atoms and photons is a central goal in quantum optics, as this capability underpins a wide range of quantum technologies \cite{Chang2018}.
Achieving such interactions with high efficiency is challenging because atoms can absorb photons from a target optical mode and scatter them into inaccessible channels, leading to significant loss of quantum information. Two well-established strategies to improve the branching ratio into the target mode exploit collective enhancement in disordered atomic ensembles~\cite{Hammerer2010} and Purcell enhancement in optical cavities~\cite{Chang2018}.

Subwavelength atomic arrays---ordered arrangements of atoms with lattice spacings smaller than the characteristic dipole-radiation wavelength---have recently emerged as a new paradigm in which destructive interference is harnessed to suppress emission into unwanted optical modes. This enables precisely controlled photon scattering into selected radiation modes---a concept known as \emph{selective radiance}~\cite{AsenjoGarcia2017PRX}.
For example, two-dimensional (2D) atomic arrays with subwavelength lattice spacing exhibit highly directional light–matter coupling and can act as nearly perfect mirrors for incident light~\cite{BettlesPRL2016,Shahmoon2017}. These properties may enable single-photon quantum memories~\cite{Manzoni2018,Guimond2019,Solomons2024}, photon–photon gates~\cite{Moreno-Cardoner2021}, and sources of non-classical light~\cite{Porras2008,Bekenstein2020,Zhang2022} with high efficiency and high fidelity. Notably, protocols exploiting selective radiance have been predicted to yield polynomial~\cite{Manzoni2018,Moreno-Cardoner2021}, and in some cases even exponential~\cite{AsenjoGarcia2017PRX}, improvements in the scaling of errors compared with protocols using disordered ensembles~\cite{Gorshkov2007,Thompson2017}. 

Additionally, subwavelength atomic arrays have attracted interest for hosting subradiant excitations---collective excitations with strongly suppressed radiative decay~\cite{AsenjoGarcia2017PRX}. These excitations provide a subspace protected from decoherence that can be harnessed for long-lived quantum information storage~\cite{Guimond2019,Ballantine2021,RubiesBigorda2022PRR} or for realizing topological~\cite{syzranov2016,Perczel2017,Perczel2017PRA} and many-body phenomena~\cite{Henriet2019,Fayard2021,GreinerSuperradiance,Wang2025}. Selective radiance and subradiance in atomic arrays also enable the realization of all-atomic nanophotonic structures~\cite{moreno2019,Masson2020PRR,Brechtelsbauer2021,Patti2021,CastellsGraells2021,BuckleyBonanno2022,RubiesBigorda2022PRR,Andreoli2023}, avoiding the technical challenges associated with placing emitters near dielectric solids and providing access to engineered interactions with unconventional photonic baths. 

However, the theoretical understanding of subwavelength atomic arrays has largely been developed within idealized models in which the atoms are perfectly pinned at fixed positions. In experimental implementations~\cite{Glicenstein2020,Rui2020,Srakaew2023, Holman2024, Hutson2024, Hofer2024, Lu2025, GreinerSuperradiance, Holman2026}, atoms are trapped in optical lattices or tweezers, which inevitably allow for atomic motion [Fig.~\ref{fig:fig1}(a)]. 
Such motion modifies the interatomic separations and thus the photon-mediated interactions, while photons themselves exert forces on the atoms.
This induces a coupling between the collective atom-light excitations (polaritons) and motional excitations (phonons). 
This coupling opens additional dissipation channels, leading to photon scattering into uncontrolled directions. 
Indeed, recent pioneering experiments on the reflectance of 2D subwavelength arrays~\cite{Rui2020, Srakaew2023} have shown that atomic motion can strongly degrade the reflectivity compared to idealized pinned-atom models. 
A systematic understanding of the impact of atomic motion is therefore essential both for establishing realistic performance bounds and for optimizing experimental platforms to realize the full potential of subwavelength atomic arrays.

In this work, we develop a microscopic framework that captures the intertwined dynamics of internal and motional degrees of freedom in subwavelength arrays of trapped atoms. We show that the interplay between these degrees of freedom naturally gives rise to polaron-polaritons---hybrid excitations composed of polaritons dressed by lattice vibrations---and argue that these quasiparticles constitute the fundamental collective excitations in arrays of trapped atoms.

The polaron-polariton viewpoint reveals that motion reshapes the collective optical properties primarily through phonon-assisted scattering processes. These processes become particularly pronounced near resonances with motional sidebands, which typically occur when the atomic trap frequency $\nu$ is comparable to or smaller than the single-atom natural linewidth $\gamma_0$. This regime is realized in most of the current experimental implementations of subwavelength atomic arrays~\cite{Glicenstein2020,Rui2020,Srakaew2023,GreinerSuperradiance, Lu2025,Holman2026}. Due to resonances, the optical properties can differ drastically from the idealized scenario of perfectly pinned atoms.

The crucial role of resonant phonon-assisted scattering has been overlooked by previous works. Earlier studies have investigated the effects of atomic motion predominantly in the limiting regimes of fast ($\nu\gg\gamma_0$) or slow ($\nu\ll\gamma_0$) motion~\cite{Porras2008,Olmos2013,Bettles2015,Damanet2016,Shahmoon2017,Guimond2019,AsenjoGarcia2019,Needham2019,Shahmoon2019,Shahmoon2020,shahmoon_arxiv2020a,shahmoon_arxiv2020b,Rui2020,Bettles2020,Masson2020PRR,Rusconi2021,GutierrezJauregui2022,CastellsGraells2024}. In these limits, motion has often been approximated as averaged or static disorder~\cite{Porras2008,Olmos2013,Bettles2015,Shahmoon2017,AsenjoGarcia2019,Needham2019,Rui2020,Bettles2020,Masson2020PRR,Rusconi2021,GutierrezJauregui2022,CastellsGraells2024}. Specifically, in the fast-motion regime, dipole–dipole interactions are averaged over the spatial wavefunction~\cite{Porras2008} (fast-motion approximation), while in the slow-motion regime atomic positions are sampled from frozen disorder configurations~\cite{Porras2008} (frozen-motion approximation). Other approaches have restricted attention to regimes where phonon scattering is non-resonant~\cite{Shahmoon2019,Shahmoon2020,shahmoon_arxiv2020a,shahmoon_arxiv2020b,olmos2025}, did not analyze motional backaction on the internal dynamics~\cite{Robicheaux2019, Suresh2021, Suresh2022, RubiesBigorda2024}, focused on few-atom systems~\cite{Berman1997, Robicheaux2025} or studied the problem numerically for a ring of up to thirty atoms~\cite{eltohfa2025}. However, the use of regime-specific approximations and numerical viewpoints has made it difficult to identify general principles governing the impact of atomic motion.

In contrast, our polaron-polariton theory provides a physically transparent framework that applies to atoms confined in the Lamb–Dicke regime over a wide range of trap frequencies $\nu/\gamma_0$, bridging the limits of fast ($\nu\gg\gamma_0$) and slow ($\nu\ll\gamma_0$) motion. Crucially, resonant phonon-assisted scattering acts as the central organizing principle governing how atomic motion reshapes the collective optical properties.
Our theory has direct relevance for ongoing experiments operating across different motional regimes~\cite{Glicenstein2020, Rui2020, Srakaew2023, Hutson2024, Hofer2024, GreinerSuperradiance, Holman2026}, and clarifies when simpler treatments of fast or frozen motion remain valid.

\subsection{Summary of results}

\begin{figure}
    \includegraphics[width=\columnwidth]{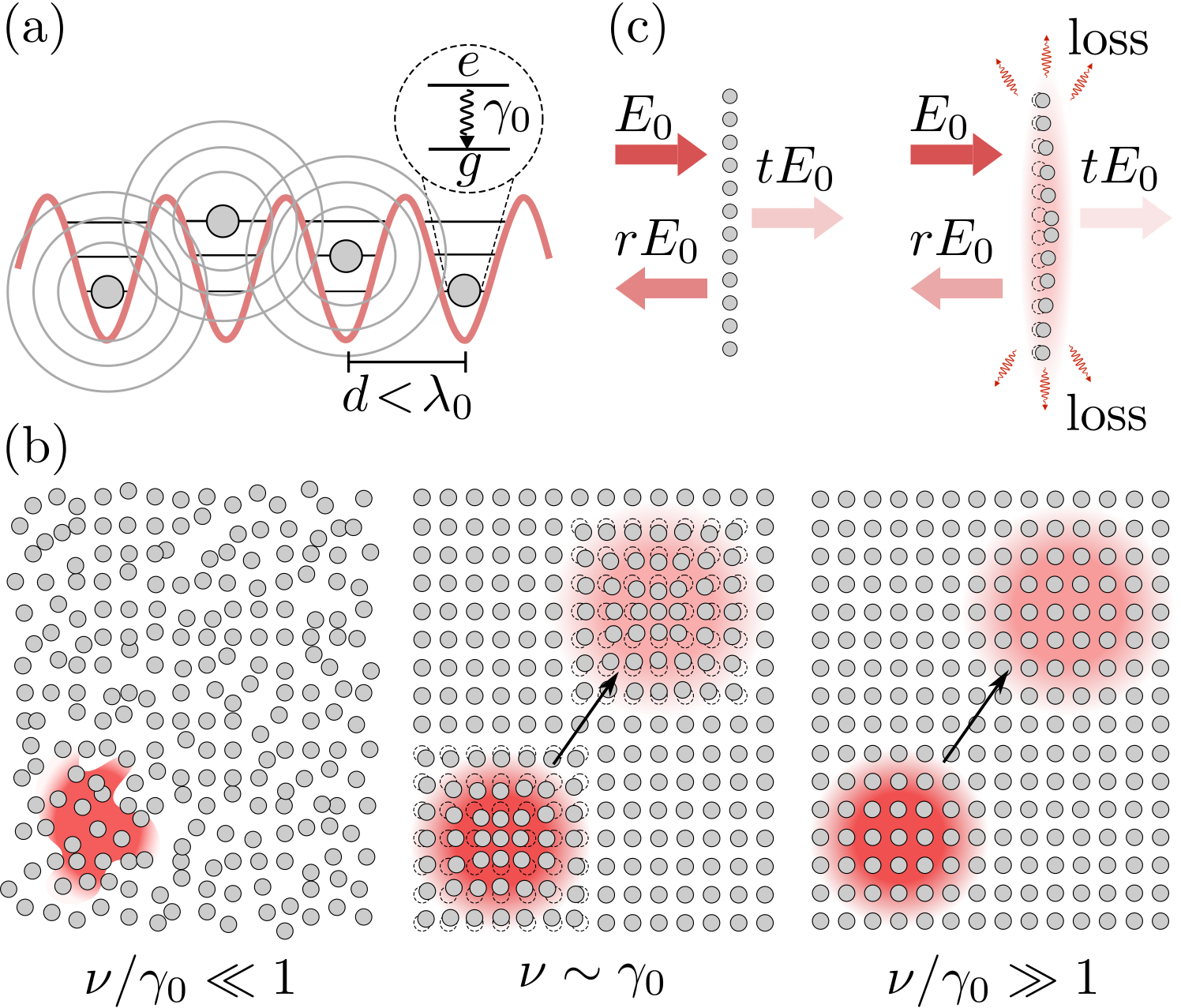}
    \caption{{\bf Motion in subwavelength arrays.} (a) Subwavelength arrays of trapped atoms: recoil due to emission and absorption couples internal excitations to atomic motion. (b) Propagation of a subradiant excitation in a 2D array. The frozen motion approximation may predict a breakdown of transport in the limit $\nu \ll \gamma_0$ (left panel). We identify conditions where the propagation of a polaron-polariton remains robust over a broad range of trap frequencies $\nu\sim \gamma_0$ (central panel). In these cases, transport is similar to the case of fast motion $\nu\gg\gamma_0$ (right panel).
    (c) Atomic mirrors exhibit near-perfect reflectivity when atoms are pinned (left). For trapped atoms, phonon-assisted scattering processes can significantly reduce reflectivity (right). Nevertheless, high reflectivities can be obtained by suppressing resonant scattering to motional sidebands through suitable experimental design.}
    \label{fig:fig1}
\end{figure}

We apply the polaron-polariton theory to address the following open problems.

(1) How does motion affect subradiance? We show that atomic motion opens up additional decay pathways and enables scattering processes between polaritons and phonons: Off-resonant scattering processes induce a dressing of the bare polariton, while resonant scattering induces a non-radiative decay into motional sidebands at a rate corresponding to Fermi's golden rule.

(2) How does motion affect the propagation of subradiant excitations in subwavelength atomic arrays? In the absence of atomic motion, such excitations propagate losslessly through the system, akin to guided modes in nanophotonic devices. In the slow motion limit, the frozen motion approximation models the effect of motion as static disorder and could predict a breakdown of transport~\cite{Deng2018,Needham2019,GutierrezJauregui2022}, analogous to Anderson localization in closed systems. Within the polaron-polariton picture, the suppression of transport arises from strong resonant phonon-assisted scattering. Building on this insight, we identify conditions under which such resonant scattering effects are weak even at small trapping frequencies and show that transport remains remarkably robust as long as resonant scattering is suppressed [Fig.~\ref{fig:fig1}(b)].

(3) How does atomic motion modify selective radiance in a 2D atomic mirror? We find that resonant scattering into sidebands is the dominant process leading to emission into unwanted channels, thereby degrading the mirror's reflectivity [Fig.~\ref{fig:fig1}(c)]. At the same time, we show that the impact of resonances can be mitigated by appropriate choices of the trap frequency and lattice geometry, allowing reflectivities at the 99\% level to be achieved under realistic experimental conditions.

(4) Can motional degrees of freedom be harnessed as a resource? 
We demonstrate that resonant phonon scattering enables the direct excitation of subradiant states of a 2D subwavelength array with an external laser, a process otherwise disallowed by quasimomentum conservation when atoms are pinned.

The paper is organized as follows: in Sec.~\ref{sec:Model}, we present the general model and introduce the polaron–polariton framework. We apply the theory to the polaron-polariton quasiparticle properties in free-space subwavelength arrays in Sec.~\ref{sec:subwavelength_arrays}. We analyze excitation transport in Sec.~\ref{sec:transport}, light-scattering from a 2D array in Sec.~\ref{sec:input-output}, including the collective nature of phonon-assisted light-scattering processes, and discuss how to harness phonons to directly excite subradiant states in Sec.~\ref{sec:excitation_of_subradiant_states}.
In Section~\ref{sec:Discussion}, we discuss experimental implementations and generalizations of our approach. We draw our conclusions in Sec.~\ref{sec:Conclusions}.

\section{Model}\label{sec:Model}

We consider an array of atoms interacting with the surrounding electromagnetic field. We model the atoms as two-level systems with ground state $\ket{g}$, excited state $\ket{e}$, resonance frequency $\w_0$, and natural linewidth $\gamma_0$. We assume that the ground and excited states experience the same trapping potential, which we model as a harmonic potential for each site of the array with frequencies $\nu_\alpha$ along the spatial directions $\alpha = x, y, z$. We denote atomic positions as $\hat{\vec r}_\vec{n} = d \, \vec n+\hat{\vec R}_\vec{n}$, where $d$ is the lattice constant and $\hat{\vec R}_\vec{n}$ describes deviations from the trap center of site $\nn$, and the positions of the trap centers as ${\vec r}_\vec{n} = d \, \vec n$. Tracing out the photonic field within the Born-Markov approximation, the dynamics of the atoms
is described by ($\hbar=1$)
\be\label{eq:ME}
\begin{split}
	\partial_t\rhop &= -i\pare{\Hop\rhop - \rhop\Hop^\dag} + \mathcal{R}\rhop,
\end{split}	
\ee
with the non-Hermitian Hamiltonian 
\be\label{eq:H}
\begin{split}
	\Hop = \w_0 \!\sum_\nn  \spl_\nn\smi_\nn \!+\! \sum_{\alpha,\nn}\nu_\alpha \bdop_{\alpha,\nn}\bop_{\alpha,\nn} \!+\! \sum_{\nn,\mm} G(\rrop_\nn \!-\! \rrop_\mm)\spl_\mm\smi_\nn,
\end{split}	
\ee
and recycling term 
\begin{equation}
    \mathcal R\rho=\sum_{\vec m,\vec n}\left[e^{\hat{\vec R}_{\vec n}\cdot\nabla_\vec{n}}\hat\sigma^-_\vec{n}\rho\hat\sigma^+_\vec{m}e^{\hat{\vec R}_{\vec m}\cdot\nabla_\vec{m}}\right]\Im[G(\vec{n}d-\vec{m}d)],
\end{equation}
where $\mathbf{e}_\alpha\cdot\nabla_{\nn} = \partial/\partial(d n_\alpha)$.
The first and second terms in \cref{eq:H} represent, respectively, the internal and mechanical energy of the atoms where the atomic internal excitations are described in terms of spin excitations by the operator $\spl_\nn = (\smi_\nn)^\dag \equiv \ketbra{e_\nn}{g_\nn}$. In this description, we will interchangeably refer to collective dipolar excitations as polaritons and spin waves. In Eq.~\eqref{eq:H}, the bosonic operator $\bdop_{\alpha,\nn}$ ($\bop_{\alpha,\nn}$) creates (annihilates) an atomic motional excitation (a phonon) at the site $\nn$ along a direction $\alpha=x,y,z$. 
The last term in \cref{eq:H} describes dipole--dipole interactions. Since the dipole--dipole interactions depend on interatomic distances, this term induces a coupling between the internal and motional degrees of freedom. The complex coupling rate $G(\rr)$, describing both coherent and dissipative interactions, is proportional to the electromagnetic Green's function. \Cref{eq:H} applies to the case of atoms along a waveguide, in a cavity, or in free space depending on the form of $G(\rr)$~\cite{AsenjoGarcia2017PRA}. 

Solving \cref{eq:ME} in full generality is intractable. In the following, we restrict our attention to the linear-response regime with at most one photon and one phonon excitation. This is permissible far from saturation, provided that we work in the Lamb-Dicke regime with low phonon occupations,
\be\label{eq:LambDicke_condition}
	\eta_\alpha \equiv \sqrt{\frac{\nu_R}{\nu_\alpha}} \ll 1 {\;\rm and\;} n_{\mathrm{th},\alpha} \ll 1,
\ee
by which phonons have a perturbative effect. Here, $\nu_R = k_0^2/2M$ is the resonant recoil frequency of the atoms with mass $M$, $k_0\equiv\w_0/c$, and $n_{\mathrm{th},\alpha}$ is the average number of thermal excitations in the atomic center of mass motion.

We can then expand \cref{eq:H} in powers of the Lamb-Dicke parameters $\eta_\alpha$ as 
\begin{align} \label{eq:H_expansion}
\Hop \simeq \Hop_0 + \sum_\alpha \eta_\alpha \Hop^{\alpha}_1 + \frac{1}{2}\sum_{\alpha,\beta} \eta_\alpha\eta_\beta \Hop^{\alpha\beta}_{2}.
\end{align}
As shown in Ref.~\cite{AsenjoGarcia2017PRX}, for arrays containing many atoms ($N\gg1$), $\Hop_0$ can be well approximated in Fourier space by
\be\label{eq:H0_diagonal}
	\Hop_0 = \sum_{\alpha,\pp}\nu_\alpha\bdop_{\alpha,\pp}\bop_{\alpha,\pp} + \sum_\pp \left(\w_0 + \varepsilon_\pp^{(0)}\right)\sdop_\pp \sop_\pp,
\ee
where $\sdop_\pp=N^{-1/2}\sum_\mm e^{i\pp\cdot\rr_\mm}\spl_\mm$ creates a polariton with quasimomentum $\pp$, the dispersion of the polaritons $\varepsilon_\pp^{(0)} =\sum_{\vec n}G(\vec r_\vec{n})e^{-i{\vec p} \cdot\vec r_\vec{n}}$ is the Fourier transform of $G$, and we introduced $\bdop_{\alpha,\pp}\equiv N^{-1/2}\sum_\mm e^{i \pp\cdot \rr_\mm}\bdop_{\alpha,\mm}$.
The dispersion and decay rate of the collective modes are given by $J_\pp^{(0)} = {\rm Re}(\varepsilon_\pp^{(0)})$ and $\Gamma_\pp^{(0)} = -2{\rm Im}(\varepsilon_\pp^{(0)})$, respectively. 
To first order in $\eta_\alpha$, we obtain the spin--phonon coupling
\be\label{eq:H1_crystal_momenta}
	\Hop_1^\alpha = \frac{1}{\sqrt{N}}\sum_{\pp,\qq} g^\alpha_{\qq,\pp}\pare{\bdop_{\alpha,\pp-\qq}\sdop_{\qq}\sop_{\pp} - \sdop_{\pp}\sop_{\qq} \bop_{\alpha,\pp-\qq}},
\ee
where we define
\be\label{eq:H1_vertex}
	g^\alpha_{\qq,\pp} = \sum_\nn G^\alpha_{\nn{\mathbf 0}} \pare{e^{i\qq\cdot\rr_\nn} - e^{i\pp\cdot\rr_\nn}},
\ee 
with $G^\alpha(\rr) \equiv k_0^{-1} \partial_\alpha G$. We note that the coupling, $g^\alpha_{\qq,\pp}$, between subradiant modes ($|\qq|,|\pp| > k_0$) is purely imaginary, whereby the corresponding term in \cref{eq:H1_crystal_momenta} is Hermitian as one might expect.

If the phonon population is negligible, the term to second order in $\eta_{\alpha}$ modifies the dispersion relation as (see \cref{app:Zero_point_motion})
\begin{equation}\label{eq:epsilon_infinity_general}
    \varepsilon^{(0)}_\pp \longrightarrow \varepsilon^{\infty}_\pp \equiv \varepsilon^{(0)}_\pp + \Delta\varepsilon_\pp.
\end{equation}
In this manner, the coupling is explicitly written as a Fr{\"o}hlich interaction \cite{Frohlich1952} of the polariton excitation with phononic mechanical modes. 

This suggests to think of the motion of the spin wave in terms of \emph{polarons}, describing how the spin excitation becomes dressed by quantum mechanical lattice distortions in direct analogy to the original work by Landau and Pekar \cite{Landau1933,Landau1948} on the motion of electrons in dielectric media. One should note, however, that the interaction is not due to an inherent attraction or repulsion of ground state atoms to the excited atom, but rather arises due to atomic displacement affecting the hopping \cite{Barisic1970,Zhang2021}, which is here mediated by the exchange of photons. It is the photon-mediated hopping of the excitation that leads to emission and absorption of phonons. In this sense, the photonic coupling is both responsible for the formation of the polariton and its coupling to vibrational modes: that is, without the light, there would neither be a polariton \emph{nor} a polaron.

\subsection{Polaron-polariton description} \label{subsec:polaron_description}

As a result of the displacement-mediated hopping in \cref{eq:H1_crystal_momenta}, the polaritons in \cref{eq:H0_diagonal} couple to the mechanical oscillations in the array and should thus lead to the formation of phonon-dressed polaritons, i.e. polaron-polaritons. 

We quantify these motional effects by calculating their influence on the spin-wave retarded Green's function 
$iG_\vec{p}^R(t)=\theta(t)\tr\{\hat s_\vec p(t)\hat s_\vec{p}\dagg\rho_0\}$, where $\rho_0 = \ket{0}\bra{0}$ is the vacuum state. Computing this using \cref{eq:ME}, we obtain
\be \label{eq:Greens_function}
\!\!iG_\pp^{\rm R}(t) = \theta(t)\bra{0} \sop_\pp e^{-i\hat{H} t}  \sdop_\pp\ket{0} \equiv \theta(t)\bra0 \hat s_{\vec p}\ket{\Psi_\vec{p}(t)},\!\!
\ee
where $\hat H$ is the non-Hermitian Hamiltonian in \cref{eq:H} and, consequently, 
$\ket{\Psi_\vec{p}(t)}$ is not normalized, but rather its norm decays with time $t$.
In the Lamb-Dicke regime, we treat the phonons perturbatively and restrict our attention to the subspace with at most one phonon excitation.
This corresponds to expressing the evolved state in terms of the Chevy ansatz~\cite{Chevy2006}
\be \label{eq:Chevy}
\ket{\Psi_\pp(t)} \simeq B_\pp(t) \sdop_\pp \ket{0} + \sum_{\qq,\alpha} C_{\pp,\qq}^{\alpha}(t) \sdop_\qq \bdop_{\alpha,\pp-\qq} \ket{0}.
\ee 
Evolving this state with the initial condition $B_\pp(0) = 1$, $C_{\pp,\qq}^{\alpha}(0) = 0$, we obtain the retarded Green's function as $G_\pp^{\rm R}(t) = -i\theta(t) B_\pp(t)$.

To derive the equation of motion for $G^{\rm R}_\pp(t)$, we first use the Schr{\"o}dinger equation $i\partial_t\ket{\Psi_\pp(t)} = \Hop\ket{\Psi_\pp(t)}$ and project it onto the single phonon subspace. We then integrate out the phononic contributions, $C_{\pp,\qq}^\alpha(t)$, to get 
\begin{align}
i\partial_tG^{\rm R}_\pp(t) ={}& \delta(t) + \varepsilon_\pp^{\infty} G^{\rm R}_\pp(t)+ \frac{i}{N}\sum_{\qq,\alpha} (\eta_\alpha g^\alpha_{\qq,\pp})^2  \nonumber \\
\times&\int_0^{\infty} \!\! d\tau \, e^{-i(\varepsilon_\qq + \nu_\alpha - i0^+)\tau}G^{\rm R}_\pp(t\!-\!\tau),\!\!
\label{eq:G_R_equation_of_motion}
\end{align}
where $\delta(t) = \partial_t \theta(t)$ is the Dirac delta function, and the positive infinitesimal $0^+$ is used to regularize the temporal integral. Equation \eqref{eq:G_R_equation_of_motion} is solved by a Fourier transformation, $G^{\rm R}_\pp(t)\! =\! \int d\omega \, \exp(-i\omega t) G^{\rm R}_\pp(\omega) / (2\pi)$. This yields
\be
G^{\rm R}_\pp(\omega) = \frac{1}{\omega - \varepsilon_\pp^{\infty} - \Sigma_\pp(\omega) + i0^+},
\label{eq:Greens_function_frequency_space}
\ee
in which we identify the self-energy for the emission and absorption of phonons
\be\label{eq:Sigma_p_omega}
\Sigma_\pp(\omega) = - \frac{1}{N} \sum_{\qq,\alpha}\frac{(\eta_\alpha g^\alpha_{\qq,\pp})^2}{\omega - \varepsilon_\qq^{\infty} - \nu_\alpha + i0^+}.
\ee
The poles of $G^{\rm R}_\pp(\omega)$ define the modified eigenenergies, which to second order in $\eta_\alpha$ read 
\begin{align} \label{eq:second_order_energy_general}
E_\pp = \varepsilon_\pp^{\infty} - \frac{1}{N} \sum_{\qq,\alpha}\frac{(\eta_\alpha g^\alpha_{\qq,\pp})^2}{\varepsilon_\pp^{(0)} - \varepsilon_\qq^{(0)} - \nu_\alpha + i0^+}.
\end{align}
Equation~\eqref{eq:second_order_energy_general} represents the excitation energy and linewidth of the emergent polaron-polaritons. It contains two contributions at order $\eta_\alpha^2$. The first comes from the zero-point motion of the atoms and is contained in $\varepsilon_\pp^{\infty}$, while the second arises due to the Fr{\"o}hlich-type coupling to mechanical modes when the photon hops from one spin to another. Indeed, in analogy to the dynamical Chevy ansatz in \cref{eq:Chevy}, we find a perturbed quasiparticle eigenstate~\cite{Chevy2006} 
\begin{align}
\ket{\psi_\pp} =Z_\pp^{1/2}\!\left[\sdop_\pp \!+\! \frac{1}{\sqrt{N}}\sum_{\qq,\alpha} \frac{\eta_\alpha g^\alpha_{\pp,\qq}}{\varepsilon^{(0)}_\pp \!-\! \varepsilon^{(0)}_\qq \!-\! \nu_\alpha} \sdop_\qq \bdop_{\alpha\pp - \qq}\right]\!\ket{0} \label{eq.polaron_state}
\end{align}
of the non-Hermitian Hamiltonian in \cref{eq:H_expansion}. The prefactor $Z_\pp = (\bra{0}\hat{s}_\pp \ket{\psi_\pp})^2$ is the quasiparticle spectral weight. It defines the amplitude of the long-time behavior of the retarded Green's function, $iG^{\rm R}_\pp(t) \to Z_\pp e^{-iE_\pp t}$, and may be calculated from the self-energy as
\begin{align} \label{eq:residue}
Z_\pp &= \frac{1}{1-\partial_\omega \Sigma_\pp(\omega)|_{\omega = E_\pp}},
\end{align}
using \cref{eq:Sigma_p_omega}. Note that unlike Hermitian systems, \cref{eq:residue} can be complex and its norm can be \emph{larger than one}~\cite{Scarlatella2019}. 
As a result, the probabilistic interpretation of the spectral weight does not fully apply in the non-Hermitian case (see Appendix~\ref{app:sum_rule}). In particular, it means that a breakdown in the quasiparticle theory developed here occurs for both $|Z_\pp| \ll 1$ and $|Z_\pp| \gg 1$. 

Equations \eqref{eq:Chevy} and \eqref{eq.polaron_state} describe the dressing of polaritons with phonons, in contrast to the usual scenario of dressing with matter-matter excitations in the context of exciton-polaritons \cite{Sidler2017,Tan2020} and impurities in quantum gases \cite{Nielsen2020,Camacho2020}. 

In the following, we treat the ratio $\nu_\alpha/\gamma_0$ and the Lamb-Dicke parameters $\eta_\alpha$ as independent. We do so to keep our discussion as general as possible in light of the different experimental regimes currently available by using different atomic species and different atomic transitions within the same species. A summary of the most commonly used atomic species and the corresponding values of the relevant parameters are given below in Table~\ref{TAB:parameters}.

\section{Polaron-polaritons in free-space subwavelength arrays} \label{sec:subwavelength_arrays}
In the following, we analyze the impact of quantum motion on the energy, decay rate and spectral weight of the phonon-dressed polaritons described by \cref{eq:second_order_energy_general,eq:residue}. We consider one- and two-dimensional atomic arrays in free space. The main takeaway messages of our analysis are: (i) The dispersion relation is only weakly perturbed. 
(ii) The decay rate and spectral weight of the subradiant states can be strongly modified, due to singularities in the density of states at saddle points (in 2D arrays) or maxima (in 1D arrays) of the polariton's dispersion relation.

The dipole coupling is obtained from the free-space electromagnetic Green's tensor as specified in \cref{app:Free_Space}. For convenience, we assume the case of isotropic trap frequencies, $\eta_\alpha=\eta$, for all $\alpha$. We assume that the array contains a large number of atoms ($N\gg 1$), such that the bulk physics is well captured by \cref{eq:second_order_energy_general,eq:residue}. For $N \gg 1$, we may convert the discrete sum in \cref{eq:second_order_energy_general} to an integral over the first Brillouin zone (BZ)
\begin{equation} \label{eq:second_order_energy_D_dimension}
\begin{split}
E_\pp = \varepsilon^{\infty}_\pp
- \eta^2 d^D \int_\text{BZ} \frac{\text{d}^D\qq}{(2\pi)^D}\frac{\sum_\alpha(g^\alpha_{\qq,\pp})^2}{\varepsilon_\pp^{(0)} - \varepsilon_\qq^{(0)} - \nu + i0^+}.
\end{split}
\end{equation}
For atoms in free space and isotropic traps, the first term in \cref{eq:second_order_energy_D_dimension} takes the simple form~\cite{Guimond2019,Rusconi2021}
\be\label{eq:E_p_fast_motion}
\varepsilon_\pp^{\infty} = (1-\eta^2)\varepsilon^{(0)}_\pp - i \frac{\eta^2\gamma_0}{2}.
\ee
To understand the different contributions in Eq.~\eqref{eq:second_order_energy_D_dimension} it is instructive to first consider the fast-motion limit ($\nu\gg \gamma_0$). In this limit, the trap frequency is much faster than the dipole interaction rate and the second term in \cref{eq:second_order_energy_D_dimension}, which originates from the scattering between polariton modes with the creation of a phonon, is suppressed.
The sole motional correction is thus Eq.~\eqref{eq:E_p_fast_motion} which coincides with the one found by other authors in this regime~\cite{Guimond2019, Rusconi2021, CastellsGraells2024, olmos2025, eltohfa2025}. It modifies the polariton dispersion and linewidth as $J^{(0)}_\pp \!\rightarrow\! (1-\eta^2)J_\pp^{(0)}$ and $\Gamma^{(0)}_\pp \!\rightarrow\! (1-\eta^2)\Gamma^{(0)}_\pp \!+ \eta^2\gamma_0$ respectively, where $\eta^2$ quantifies the probability of an atom to recoil when emitting a photon into the far field. 
The modified linewidth can then be simply understood as the incoherent sum of two decay channels.
In the absence of recoil, happening with a probability $1-\eta^2$, a photon is scattered collectively at the rate $\Gamma^{(0)}_\pp$. With probability $\eta^2$, an atom recoils and the photon is independently emitted by that atom at a rate $\gamma_0$ along directions dictated by the atom's dipole transition.
For atoms initially prepared in their motional ground state, the presence of a phonon at a particular site provides which-path information about which atom has
radiated the photon, thus suppressing collective scattering~\cite{Fedoseev2025, Zhang2025, Scully1991, Itano1998}.
Notably, subradiant modes acquire a nonzero decay rate, $\eta^2 \gamma_0$, due to this process. 

For finite trap frequencies, a polariton with momentum $\pp$ can scatter to a polariton with momentum $\qq$ by creating a lattice phonon. This leads to a change in the polariton dispersion described by 
the second term in \cref{eq:second_order_energy_D_dimension}.
This scattering process is suppressed across most of the Brillouin zone. For a polariton with energy $\varepsilon_{\pp}^{(0)}$, the energy mismatch in scattering to a polariton-phonon pair with energy $\varepsilon_\qq^{(0)} + \nu$ is typically much larger than the coupling $\eta_\alpha g^\alpha_{\qq,\pp}$. Nonetheless, in most situations there exist specific momenta $\qq$ for which the process becomes resonant, satisfying $\varepsilon_{\qq}^{(0)} + \nu = \varepsilon_{\pp}^{(0)}$.
In such cases, the integrand in \cref{eq:second_order_energy_D_dimension} has a pole~\footnote{There are also poles in both the interaction vertex $g^\alpha_{\qq,\pp}$ and the energy $\varepsilon_\qq^{(0)}$ at $|\qq| = k_0$. These poles, however, are integrable as we show in \cref{app:calculation_self_energy}. The integrability of these poles is important for the consistency of our approach because the description of Eq.~\eqref{eq:ME} breaks down at the light cone.}. We isolate the contribution from this pole by writing the self-energy as
\be\label{eq:Sigma_two_terms}
	\Sigma_\pp = \Sigma_\pp^{\text{res}} + \Sigma_\pp^{\text{off-res}}.
\ee
Here, $\Sigma_\pp^{\text{res}}$ is the contribution from the pole that arises due to the resonant scattering processes. Coupling resonantly to a band of states induces a decay of the polariton at a rate $-2 \Im(\Sigma_\pp^{\text{res}})$. This decay does not describe scattering to free space, but non-radiative losses where excitations are transferred to other modes in the array. 
The term $\Sigma_\pp^{\text{off-res}}$, instead, corresponds to the off-resonant phonon-mediated interaction with all the other (bright and dark) modes of the array. In particular, $-2{\rm Im}(\Sigma_{\pp}^\text{off-res})$ can be interpreted as a renormalization of the free-space scattering rate due to the dressing of the bare polariton mode $\pp$ with all the non-resonant modes.

\subsection{2D arrays}\label{sec:applications-2D}

\begin{figure*}[t!]
\begin{center}
\includegraphics[width=1\textwidth]{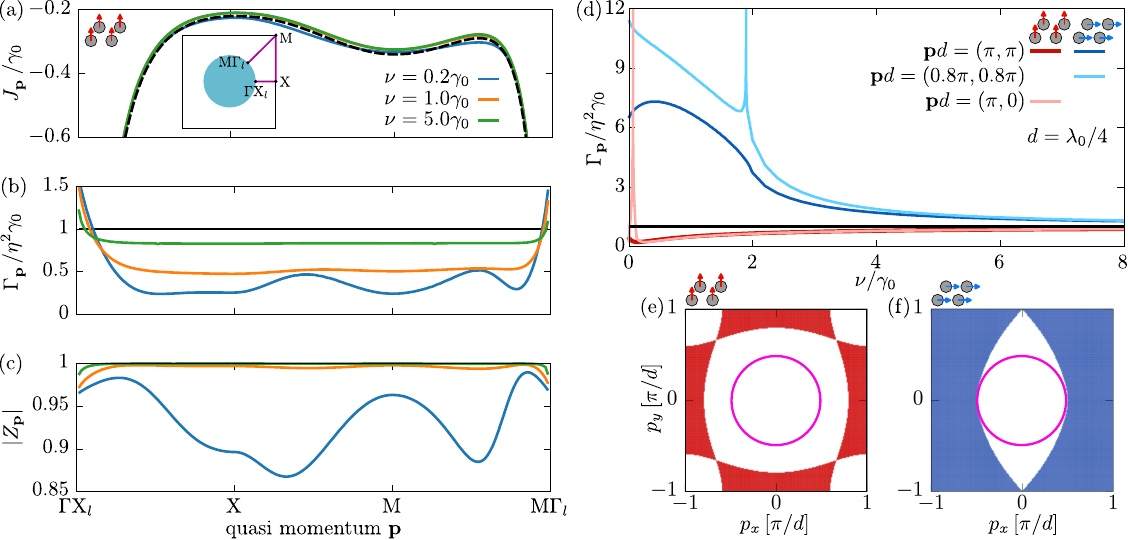}
\end{center}\vspace{-0.5cm}
\caption{{\bf Polaron-polariton properties in 2D arrays.} (a) Dispersion relation, (b) decay rate and (c) quasiparticle weight for a polaron-polariton in a 2D array for dipole moments polarized perpendicular to the array [see schematic array representation in the top left corner of (a)] and for the indicated values of the trapping frequency, $\nu$. The inset in (a) shows the symmetric lines in the first Brillouin zone outside of the light cone (blue disk region).
The black dashed lines in (a) and (b) show, respectively, the unperturbed value $J_\pp^{(0)}$ and the fast motion correction $\eta^2\gamma_0$ to the decay rate. (d) Decay rate vs. trap frequency for indicated quasimomenta and dipole polarizations. These are further compared to the contribution from the zero-point motion, $\eta^2\gamma_0$ (black line), which they approach for $\nu \gg \gamma_0$. The divergences for $\pp = (0.8\pi,0.8\pi)$ and $\pp = (\pi,0)$ are logarithmic, appearing due to a resonant coupling to a saddle-point in the dispersion, $J_\qq$. For (e) perpendicular and (f) parallel polarization, we indicate (colors) the modes which admit a logarithmic divergence at a particular value $\nu/\gamma_0$.
The pink circles show the light cone at $|\pp| = k_0$. Other parameters: Lattice spacing $d = \lambda_0/4$ and in panels (a) and (c), we use $\eta = 0.2$.}
\label{fig.quasiparticle_properties_perp} 
\vspace{-0.25cm}
\end{figure*}

For brevity, we focus our detailed numerical analysis on 2D arrays. Qualitatively similar results are found for 1D arrays (see~\cref{app:Polaron-Polariton_1D}). For 2D arrays, the resonant and off-resonant contributions in \cref{eq:Sigma_two_terms} read 
\bea
    \!\!\!\!\!\Sigma_\pp^\text{res} \!&=&\! \frac{i \eta^2 d^2 }{2}\int_0^{2\pi} \frac{{\rm d}\varphi}{2\pi} \sum_{\qq \in \mathcal{R}(\varphi)} \frac{q \, \sum_\alpha(g^\alpha_{\qq,\pp})^2}{|\partial_k \varepsilon^{(0)}_\kk|_{\kk = \qq}}, \label{eq:2D_resonant}\\
    \!\!\!\!\!\Sigma_\pp^\text{off-res} \!&=&\! -\eta^2 d^2 \! \int_0^{2\pi}\!\frac{d\varphi}{2\pi} \, {\rm P}\!\int_0^{q_{\max}}\!\frac{dq}{2\pi} \frac{q\, \sum_\alpha (g^\alpha_{\qq,\pp})^2}{\varepsilon^{(0)}_\pp \!-\! \varepsilon^{(0)}_\qq \!-\! \nu}. \label{eq:2D_off_resonant}
\eea
Here, $\mathcal{R}(\varphi)$ denotes the set of resonances $\qq$ at polar angle $\varphi$, i.e. $\varepsilon^{(0)}_\pp = \varepsilon^{(0)}_\qq + \nu$. Moreover, P denotes the Cauchy principal value, while $q_{\max}(\varphi)$ is the maximum crystal momentum at $\varphi$ within the first Brillouin zone. We note that the resonant contribution depends on the density of states, $\propto 1/|\partial_k \varepsilon_\kk^{(0)}|_{\kk=\qq}$. 

From Equations \eqref{eq:2D_resonant}, \eqref{eq:2D_off_resonant}, \eqref{eq:second_order_energy_D_dimension} and \eqref{eq:residue}, we numerically calculate the perturbed energy and spectral weight of the polaron-polariton (see \cref{app:pole_contribution_2D} for details). The results are plotted in Fig. \ref{fig.quasiparticle_properties_perp}(a)-\ref{fig.quasiparticle_properties_perp}(c) for a few indicated values of the trapping frequency $\nu/\gamma_0$ and in the case of atoms linearly polarized along a direction perpendicular to the array plane. We note that even as the trap frequency decreases below the natural linewidth, $\nu \lesssim \gamma_0$, the dispersion $J_\pp = \Re(E_\pp)$ remains robust and is only weakly affected by atomic motion [\cref{fig.quasiparticle_properties_perp}(a)]. Moreover, while a diminishing trap frequency eventually leads to a stronger generation of phonons and smaller values of the spectral weight in Fig. \ref{fig.quasiparticle_properties_perp}(c), it generally remains close to unity, $|Z_\pp| \simeq 1$. 

We attribute the robustness of the dispersion relation, $J_\pp$, to the fact that scattering to almost all crystal momenta $\qq$ is suppressed even for small trap frequencies $\nu < \gamma_0$, because the energy mismatch $\sim \gamma_0$ is larger than the coupling $\eta g_{\qq,\pp}^\alpha$. In addition, differences in the decay rates ($\Gamma^{(0)}_\pp = -2 \Im(\varepsilon^{(0)}_\pp)$) generally suppress scattering between subradiant and superradiant modes. Finally, when resonant scattering channels exist, they contribute to the non-radiative decay of the polariton via Eq.~\eqref{eq:2D_resonant} but do not modify the dispersion. Mathematically, this follows from the fact that resonant scattering channels exist only from one subradiant state to another, for which the spin-phonon coupling in \cref{eq:H1_vertex} is Hermitian. Therefore, the resonance only contributes to the imaginary part of the self-energy and, thereby, to the collective (non-radiative) decay rate of that particular polaron-polariton state.

On the other hand, we find that the collective decay rate can be strongly modified by atomic motion [\cref{fig.quasiparticle_properties_perp}(b)], and can \emph{both} be higher or lower than the fast-motion limit of $\eta^2 \gamma_0$ described in \cref{eq:E_p_fast_motion}. This enhancement or suppression of decay can be viewed as the result of constructive or destructive interference, respectively, of two processes. The first, as discussed above, contributes with $\eta^2 \gamma_0$ and is associated with the creation of a phonon during photon emission into the far field. The second arises from Eq.~\eqref{eq:2D_off_resonant} and corresponds to the emission of a photon to free space through the off-resonant scattering of a polariton into a bright mode with $|\qq|<k_0$ and a phonon with quasimomentum $\pp-\qq$. In both instances, the final state of the system contains one phonon and, therefore, the two processes can interfere. 
Note that the resonant scattering between dark polaritons, described by Eq.~\eqref{eq:2D_resonant}, does not interfere with the previous two processes, since it does not generate radiation. 

To analyze this behavior further, we plot in \cref{fig.quasiparticle_properties_perp}(d) the collective decay rate at the indicated crystal momenta and for both parallel and perpendicular polarizations. When the trap frequency is large, $\nu \gg \gamma_0$, the off-resonant contribution in Eq.~\eqref{eq:2D_off_resonant} becomes highly detuned. Moreover, only modes near the light cone contribute to the resonant correction in Eq.~\eqref{eq:2D_resonant}. Since the density of states is small here, it has little impact on the collective decay. As a result of both of these effects, the decay rate for states outside the light cone converges to the fast motion limit $\eta^2\gamma_0$ in Eq.~\eqref{eq:E_p_fast_motion} for $\nu \gg \gamma_0$. On the other hand, when the trap frequency is lowered, perpendicular polarization generally leads to a considerable suppression of the collective decay rate, below $\eta^2 \gamma_0$, associated with destructively interfering processes, whereas parallel polarization leads to a large enhancement above $\eta^2 \gamma_0$, associated with constructive interference. 

Moreover, we find that for some momenta, there is a \emph{logarithmic} divergence in the decay rate,
\begin{align}
\Gamma_\pp \simeq -\eta^2 c_\pp \gamma_0 \ln\left[\frac{|\nu - \nu_c|}{\nu_c}\right], 
\end{align}
with a momentum-dependent prefactor $c_\pp$. This happens for one particular critical value of the trap frequency, $\nu_c$, at which the mode $\pp$ scatters resonantly and non-radiatively to a \emph{saddle-point} mode of the dispersion $\varepsilon_\qq^{(0)}$, $\varepsilon_\pp^{(0)} = \varepsilon_\qq^{(0)} + \nu_c$, as described by Eq.~\eqref{eq:2D_resonant}. For $x$-polarized dipoles, this saddle point appears at $\qq d = (0,\pi)$, while for perpendicularly polarized dipoles it depends on the chosen inter-particle distance. For the one considered here, $d = \lambda_0 / 4$, it occurs around $\qq d \simeq (0.65\pi,0.65\pi)$. This enhanced decay rate, thereby, corresponds to an enhanced rate of non-radiative scattering into specific subradiant modes. In Sec.~\ref{sec:excitation_of_subradiant_states}, we show that this mechanism can be used to excite subradiant modes via a phononic sideband. 

At these saddle points, it follows from \cref{eq:residue} that the residue in second-order perturbation theory vanishes linearly,
\begin{equation}
|Z_\pp| = \frac{1}{|(1 + \partial_\nu J_\pp)^2 + ( \partial_\nu \Gamma_\pp)^2/4|^{1/2}} \simeq \frac{|\nu - \nu_c|}{2\eta^2 c_\pp\gamma_0}.
\end{equation}
This is a clear indication of the breakdown of our perturbative treatment. 
When scattering to saddle points with a diverging density of states, it is thus necessary to include higher-order processes to quantitatively predict the modified decay rate of subradiant states. This calculation is beyond the scope of this work, though we expect it to merely lead to a softening of the appearing resonances. To further characterize these instabilities, we plot in Figs. \ref{fig.quasiparticle_properties_perp}(e) and \ref{fig.quasiparticle_properties_perp}(f) the momenta which become unstable at some critical trapping frequency, $\nu_c$. These are defined by points for which the mode at $\pp$ is energetically above the saddle-point at $\qq$, $J_\pp > J_\qq$. Note, however, that for parallel polarization, the momenta at the corners of the BZ, $(\pm\pi,\pi),(\pm\pi,-\pi)$, as well as $(\pm \pi,0)$ are actually stable, because the coupling to $\qq = (0,\pi)$ vanishes faster than the divergence occurs. As a result, the associated collective decay rate, as exemplified by the dark blue curve in Fig. \ref{fig.quasiparticle_properties_perp}(d), is well-behaved. 

So far, we focused solely on the subradiant states outside the light cone. For the superradiant states inside the light cone, the resonant contribution in Eq.~\eqref{eq:2D_resonant} is slightly modified, as we describe in \cref{app:pole_contribution_2D}. However, for the cases investigated here, the dispersion of the collective states has saddle points within the light cone. Therefore, there are \emph{no} logarithmic divergences in the decay rate, whereby its modification will be of order $\eta^2$. As a result, their superradiant behavior is robust.

In the following sections, we apply the polaron description to investigate the effects of atomic motion on the properties of subwavelength atomic arrays.

\section{Transport} \label{sec:transport}

In this section, we study the effect of atomic motion on the transport of a subradiant excitation through both 1D and 2D arrays of atoms.
When a single excitation is present, a subwavelength array behaves like a dielectric material where subradiant modes correspond to guided modes propagating through the system with minimal to no loss~\cite{AsenjoGarcia2017PRX,Masson2020PRR,CastellsGraells2021,Patti2021,GutierrezJauregui2022,RubiesBigorda2022PRR}. 
Previous works have modeled the effect of atomic motion as classical disorder within the frozen motion approximation, predicting suppression of propagation in 1D arrays~\cite{Olmos2013,GutierrezJauregui2022,Peter2024}.
In contrast with these predictions, we showed in \cref{sec:subwavelength_arrays} that a polaron-polariton has a well-defined quasimomentum, a well-defined quasiparticle weight, $|Z_\pp| \simeq 1$, and its dispersion relation shows a behavior close to the case of pinned atoms over a wide range of values of $\nu/\gamma_0$ (see~\cref{fig.quasiparticle_properties_perp} and \cref{fig:Polaron_1D} in Appendix~\ref{app:Polaron-Polariton_1D} for 1D arrays). Here, we demonstrate that this translates into robust transport properties for polaron-polaritons, whenever resonant phonon scattering is weak.

In the following, we simulate transport of an excitation in real space in both 1D and 2D arrays.
Specifically, we solve the Schr{\"o}dinger equation, $\im \pa{t}\ket{\psi} = \Hop\ket{\psi}$, where $\Hop$ is the non-Hermitian Hamiltonian in \cref{eq:H_expansion} truncated to contain at most one spin and one phonon excitation (Appendix~\ref{app:Simulation_Transport}). 
We consider an initial Gaussian wave packet, 
\be\label{eq:Gaussian_WP}
	\ket{\psi} = \inv{\sqrt{2\pi \sigma^2}}\sum_{\nn} e^{-(\rr_\nn-\rr_0)^2/4\sigma^2}e^{\im \kk_s\cdot \rr_\nn} \spl_\nn\ket{g}\otimes \ket{0},
\ee
where the initial polariton excitation is centered at $\rr_0$ and delocalized over a width $\sigma$, chosen so that in momentum space the wave-packet is concentrated within the subradiant sector around the carrier momentum $\kk_s$.
We compare these results with the numerical simulations for pinned atoms and for the frozen motion approximation. The latter is obtained by treating the atomic position in \cref{eq:H} as a variable sampled from a Gaussian distribution at each lattice site with standard deviation $\sigma = \eta/k_0$.

\subsection{1D arrays}
\begin{figure*}
	\includegraphics[width=2\columnwidth]{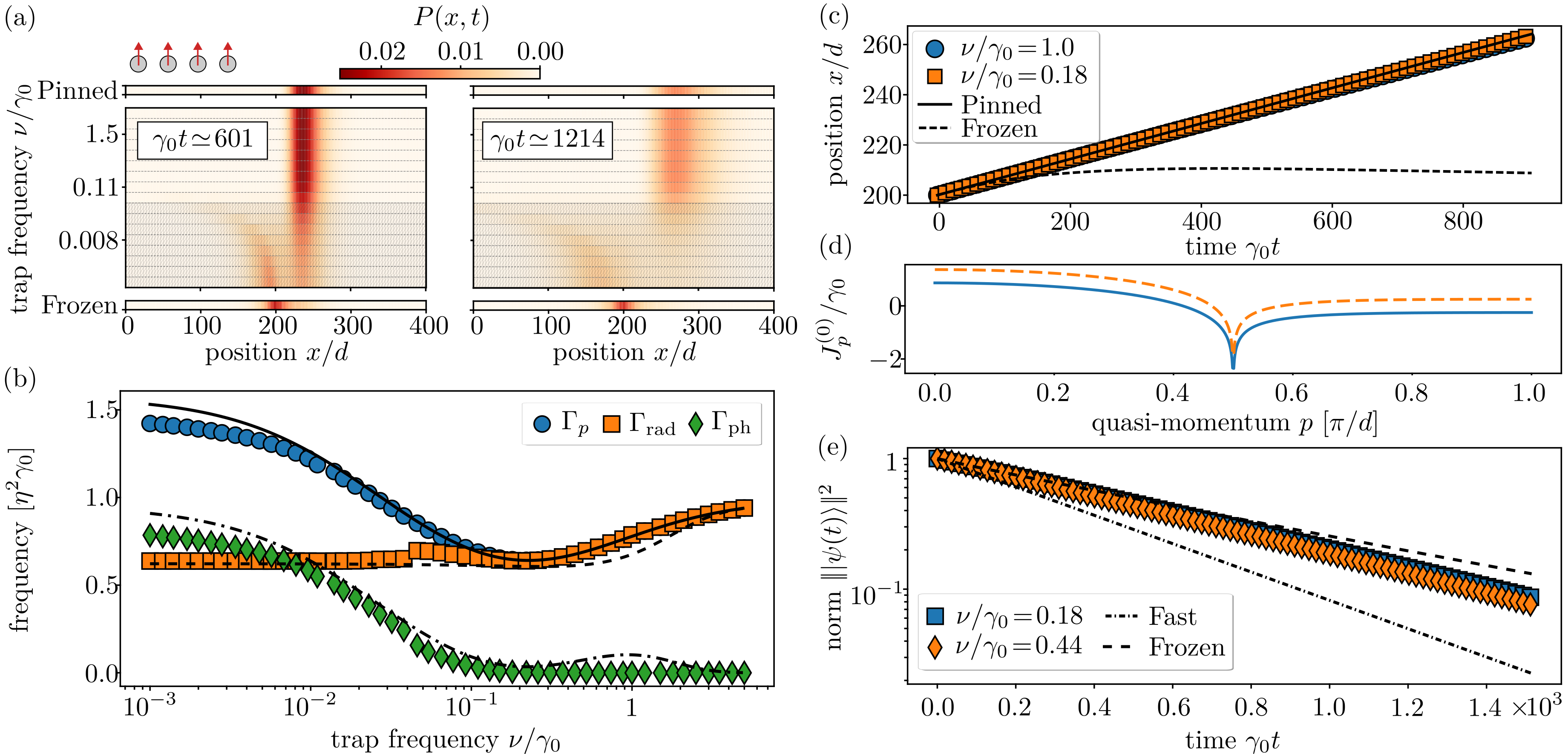}
	\caption{{\bf Transport in a 1D array.} We consider $N=400$ atoms with lattice constant $d=\lambda_0/4$, perpendicular polarization, and $\eta=0.05$. (a) Propagation of the normalized spin population along the array for different values of the trap frequency $\nu/\gamma_0$ at indicated times. Each horizontal line corresponds to a particular trap frequency: From top to bottom, $\nu/\gamma_0 = 4.2$, 2.5, 1.5, 0.88, 0.52, 0.31, 0.18, 0.11, $6.5\!\times\! 10^{-2}$, $3.8\!\times\! 10^{-2}$, $2.3\!\times\! 10^{-2}$, $1.4\!\times\! 10^{-2}$, $8.1\!\times\! 10^{-3}$, $4.8\!\times\! 10^{-3}$, $2.8\!\times\! 10^{-3}$, $1.7\!\times\! 10^{-3}$, $10^{-3}$.
    The hatched region indicates the regime where the phonon population is significant, which can render the numerical simulations inaccurate. 
    (b) Polariton damping rate ($\Gamma_p$), radiative decay rate ($\Gamma_\text{rad}$) and phonon-scattering rate ($\Gamma_\text{ph}$) as a function of the trap frequency. The values extracted from the numerical simulations (colored markers) show good agreement with the theoretical prediction of the polaron ansatz (Appendix~\ref{app:Polaron-Polariton_1D}) for $\Gamma_p$ (black solid line), $\Gamma_\text{rad}$ (dashed line), and $\Gamma_\text{ph}$ (dot-dashed line). 
    (c) Evolution of the center of the wave-packet compared to the case of pinned atoms (black solid line) and to the frozen motion approximation (black dashed line).
    (d) Dispersion relation for the zero-phonon band (solid blue) and for the one-phonon sideband (dashed orange) with $\nu/\gamma_0=0.5$.
    (e) Decay of the excitation as a function of time for different values of the trap frequency (colored markers) and for the fast (dash-dotted line) and frozen motion approximations (dashed line). Other parameters: $k_s=0.8\pi/d$, and $\sigma_k = (\pi/d-k_s)/2$. The frozen motion simulations are averaged over 1000 realizations for Gaussian disorder with standard deviation $\eta/k_0$.}\label{fig:Transport_1D_perpendicular}
\end{figure*}
We study propagation of an excitation in a 1D array of $N=400$ perpendicularly polarized atoms with interatomic distance $d=\lambda_0/4$. To avoid boundary effects, we initialize a Gaussian wave packet centered in the middle of the array.
In Fig.~\ref{fig:Transport_1D_perpendicular}(a), we plot the probability distribution of the spin excitation as a function of time, normalized by the probability of the excitation remaining in the system,
\begin{equation}\label{eq:Spin_Distribution}
    P(x_j,t) = \bra{\psi(t)}\hat \sigma_j^+\hat \sigma_j^-\ket{\psi(t)}/\norm{\ket{\psi(t)}}^2.
\end{equation} 
The left and right panels show $P(x_j,t)$ at two points in time as a function of the trap frequency $\nu/\gamma_0$ and of the position $x_j=j d$. We identify two distinct dynamical regimes: For $\nu / \gamma_0 \gtrsim 0.1$, propagation remains robust and closely resembles the pinned-atom case. In contrast, when $\nu / \gamma_0 \lesssim 0.1$, propagation is suppressed and a backward-moving component appears. As we explain in the following, robust transport coincides with the regime in which resonant phonon scattering is weak, while the suppression of transport at lower trap frequencies originates from strong resonant phonon scattering. 

To elucidate this connection, we plot the different contributions to the polaron-polariton damping rate in Fig.~\ref{fig:Transport_1D_perpendicular}(b). Specifically, we show the decay of the bare polariton component $\Gamma_p$ (blue circles) obtained by fitting the decay of population in the zero-phonon sector (see Appendix~\ref{app:Simulation_Transport} for details on the fitting procedure), the rate of radiative decay $\Gamma_\text{rad}$ (orange squares) extracted from the norm decay of the state $\norm{\ket{\psi(t)}}^2$, and the rate of resonant phonon scattering $\Gamma_\text{ph} \equiv \Gamma_p - \Gamma_\text{rad}$ (green diamonds). The latter quantity corresponds to the rate of resonant scattering as defined in Eq.~\eqref{eq:Sigma_two_terms}. We find good agreement between the numerical results and the corresponding predictions for a polaron-polariton with quasi-momentum $k_s$ based on Eqs.~\eqref{eq:second_order_energy_D_dimension}, \eqref{eq:main_Self_Energy_1D_res} and~\eqref{eq:main_Self_Energy_1D_offres}. 

For $\nu/\gamma_0 \gtrsim 0.1$, we observe uniform behavior for the transport of an excitation [Fig.~\ref{fig:Transport_1D_perpendicular}(a)]. The wave packet propagates ballistically at a group velocity $v_g$, which closely matches the corresponding value for pinned atoms [Figs.~\ref{fig:Transport_1D_perpendicular}(a) and (c)]. In this regime, the phonon population grows slowly, in agreement with the prediction of polaron theory [Fig.~\ref{fig:Transport_1D_perpendicular}(b)]:
For $\nu/\gamma_0 \gtrsim 0.1$, the phonon scattering rate is small and largely insensitive to the trap frequency $\nu/\gamma_0$. This originates from the shape of the dispersion $J_p^{(0)}$, which is nearly flat throughout most of the subradiant sector ($|p|>k_0$). Thus, even for relatively small values of $\nu/\gamma_0$, resonant scattering only occurs to modes close to the light cone, where the density of states is low.
This is illustrated in Fig.~\ref{fig:Transport_1D_perpendicular}(d) for a zero and one-phonon polariton band with $\nu/\gamma_0=0.5$. Accordingly, the phonon population during the dynamics remains well below one, which justifies the truncation of the Hilbert space to states with at most one phonon.

For smaller trap frequencies $\nu / \gamma_0 \lesssim 0.1$, the phonon scattering rate increases because a resonance occurs between modes in the flat region of the dispersion relation where the density of states is large [Fig.~\ref{fig:Transport_1D_perpendicular}(b)]. Consequently, resonant phonon scattering depletes the propagating wave packet. Additionally, we observe the appearance of a backwards moving component in Fig.~\ref{fig:Transport_1D_perpendicular}(a), which arises from phonon-assisted scattering to polariton modes with wavevector $q < 0$. In Sec.~\ref{sec:conditions_transport}, we discuss in detail how these mechanisms induce a suppression of transport. 
We remark that when $\Gamma_\text{ph}/\Gamma_\text{rad}$ is large, the phonon excitation number quickly grows to order unity (for details see \cref{app:Simulation_Transport}) and thus the results of the numerical simulation cannot be used to quantitatively predict the dynamics in this case.
Interestingly, we still find $|Z_p|\simeq 1$, which suggests that the polaron description may still be valid approximately.

In \cref{app:Simulation_Transport}, we analyze the case of parallel polarization, which yields broadly similar results as for perpendicular polarization. However, due to the different shape of the dispersion relation $J_p$ for parallel polarization (discussed in \cref{app:Polaron-Polariton_1D}), resonant phonon scattering plays a strong role already at higher trap frequencies and transport is suppressed for $\nu / \gamma_0 \lesssim 1.2$, assuming all other parameters chosen as in this subsection.

In general, the shape of the polariton dispersion relation determines when resonant phonon scattering is significant and thus directly affects the robustness of transport. By engineering the dispersion relation such that resonant phonon scattering is suppressed for the relevant modes, excitation transport can be achieved even at relatively low trap frequencies.

\subsection{2D arrays}
\begin{figure}
    \centering
    \includegraphics[width=\linewidth]{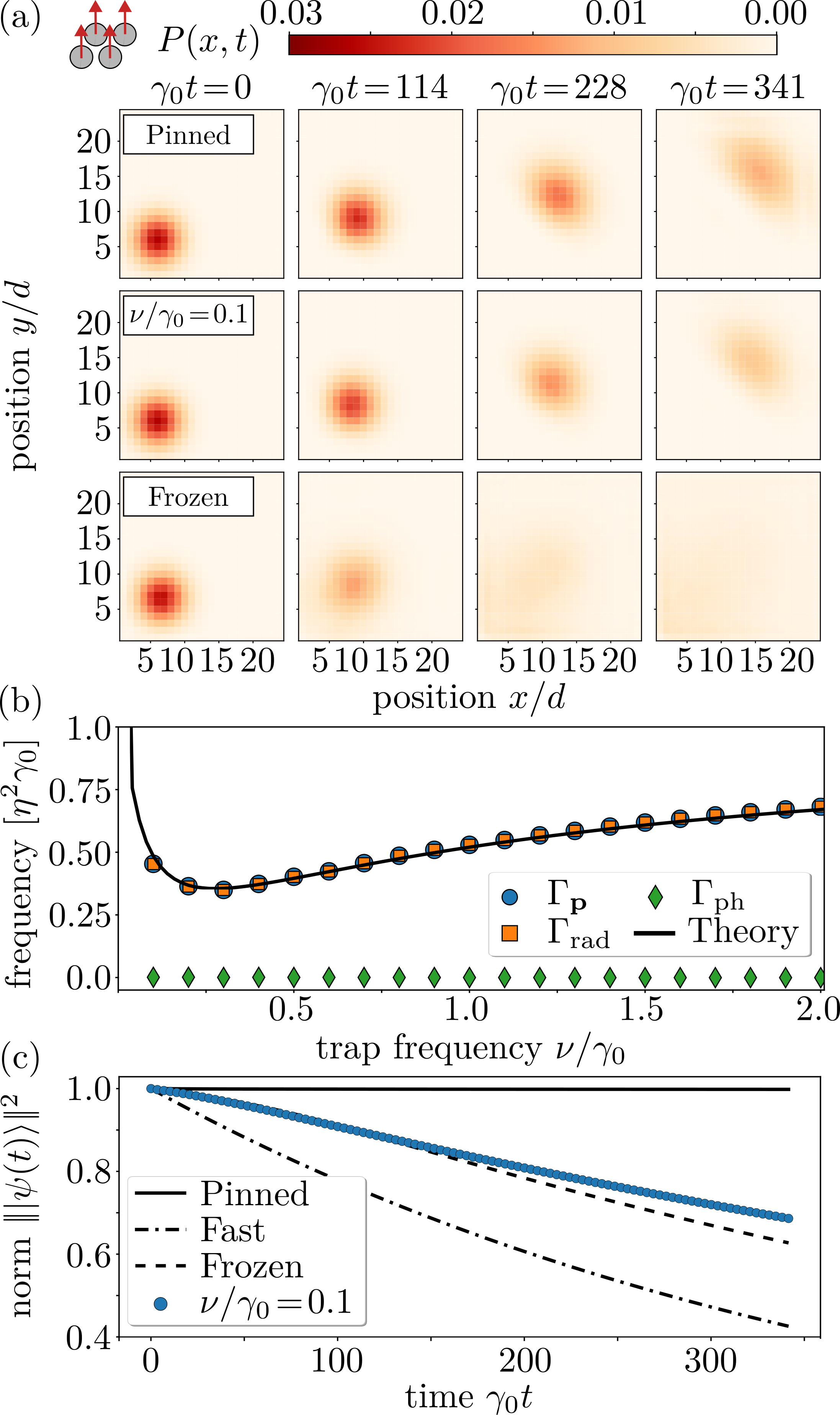}
    \caption{{\bf Transport in a 2D array.} (a) Snapshots of the spin population distribution at different times (increasing from left to right starting at $\gamma_0 t=0$) for an array of $24\times 24$ atoms polarized perpendicular to the array plane. (b) Trap frequency dependence of the polariton damping $\Gamma_\pp$ (blue circles) and of the radiative scattering rate $\Gamma_\text{rad}$ (orange squares) compared to the polaron theory in \cref{eq:second_order_energy_D_dimension} for $\mathbf{p}=(0.8,0.8)\pi/d$. (c) Decay of $\vert \vert\ket{\psi(t)}\vert\vert^2$ as a function of time for: pinned atoms (solid line), fast motion (dash-dotted line), frozen motion (dashed line) and $\nu=0.1\gamma_0$ (blue dots). Other parameters: $\eta=0.05$, $d=\lambda_0 / 4$, $\kk_s = (0.8,0.8)\pi/d$. For the frozen motion simulation, we average over 1000 realizations.}
    \label{fig:Transport_2D}
\end{figure}
Let us now consider transport in a 2D array of $24 \times 24$ atoms polarized perpendicular to the array plane and with lattice constant $d=\lambda_0/4$.
As for the case of 1D arrays, we numerically simulate the propagation of the initial Gaussian excitation in Eq.~(\ref{eq:Gaussian_WP}) by integrating the Schr{\"o}dinger equation. In Fig.~\ref{fig:Transport_2D}(a), we compare the results obtained for pinned atoms (top), trapped atoms with $\nu=0.1\gamma_0$ and $\eta=0.05$ (middle), and with the frozen motion approximation (bottom). We see that even at trap frequencies as small as $\nu=0.1\gamma_0$, the propagation is highly similar to the case of pinned atoms, while it differs qualitatively from the frozen motion approximation. 
The robustness of the excitation transport has the same explanation as for 1D arrays with perpendicular polarization. Since the dispersion for 2D arrays is approximately flat in the subradiant sector, resonant scattering occurs close to the light cone for a broad range of trap frequencies. The small density of states in this regime then leads to a suppressed phonon scattering rate for these trap frequencies [see Fig.~\ref{fig:Transport_2D}(b)]. 
Finally, the decay rate $\Gamma_\text{rad}$ is smaller than the fast motion limit leading to a slow radiative decay of the excitation, in agreement with the theoretical calculations, as shown in Fig.~\ref{fig:Transport_2D}(c).

\subsection{Conditions for transport} \label{sec:conditions_transport}

Finally, we discuss under which conditions we expect to observe transport of a polaron-polariton. 
An initial excitation with momentum $\pp$ propagates along the array as long as its quasiparticle character is well defined: the bare polariton component, $|Z_\pp|^2\exp(-\Gamma_\pp t)$, must be larger than the phonon population.
This sets a maximum propagation length $L\simeq v_g/\Gamma_\pp$ and effectively constrains transport over length scales $\sim v_g/(\eta^2 \gamma_0)$. 
It is important, however, to distinguish between two different contributions to the polaron decay $\Gamma_\pp$ as, depending on which one dominates, we observe significantly different dynamical behavior. The first contribution is the radiative decay to free space that limits the propagation of an excitation to length scales $L \simeq v_g/\Gamma_\text{rad}$ as the probability to observe an excitation at times $t>1/\Gamma_\text{rad}$ is exponentially suppressed [Fig.~\ref{fig:Transport_1D_perpendicular}(e)]. Note that, when this process dominates, the excitation clearly propagates along the array as shown in Fig.~\ref{fig:Transport_1D_perpendicular}(a,b).
The second contribution is the resonant phonon scattering rate $\Gamma_\text{ph}$. 
This process is comparable to the former at small values of the trap frequency $\nu/\gamma_0$ [see Fig.~\ref{fig:Transport_1D_perpendicular}(b)]. For $\nu/\gamma_0\rightarrow 0$, scattering to modes with quasimomentum $\approx -\pp$ is resonantly enhanced (scattering to $\pp$ is suppressed as $g_{\pp,\pp}^\alpha = 0$). Scattering a phonon thus reverses the direction of propagation, effectively suppressing transport. This qualitatively recovers results from the frozen motion approximation which also predicts absence of transport [Fig.~\ref{fig:Transport_1D_perpendicular}(a)]. 
It is tempting to make an analogy with localization in disordered systems where the localization length scales as the inverse of the square of the disorder strength $\sim\eta^{-2}$, here quantified by the Lamb-Dicke parameter. This analogy is, however, not entirely correct: true localization cannot occur because any initial excitation will eventually radiate away.

Our approach identifies under which conditions the frozen motion description captures the behavior of the system at least qualitatively ($\Gamma_\text{ph}\gtrsim\Gamma_\text{rad}$) and suggests a dynamical mechanism for the suppression of transport: an excitation gets trapped by phonon scattering processes in which its propagation direction is reversed.

\section{Optical properties}\label{sec:input-output}

Here, we analyze how atomic motion modifies light scattering from an array, with a particular focus on its impact on selective radiance. For concreteness, we discuss these effects for a 2D subwavelength array in free space driven at normal incidence, although many of our conclusions extend more broadly. In the chosen setting, and for an infinite array, light is scattered only along the direction of the incoming field or in the back-reflected direction. When driven at the collective resonance, the system behaves like a mirror~\cite{BettlesPRL2016,Shahmoon2017,Rui2020,Srakaew2023}.
Recent theoretical work \cite{Solomons2024,Mann2024} suggested that the mirror reflectivity is generally related to the efficiency for performing various quantum tasks with the system, such as storing or entangling photons.

In this section, we develop an input-output theory for light scattering from an atomic array in the presence of motion and apply it to study the optical response of a 2D atomic array. We analyze the reflectivity and the fraction of photons that are scattered off-axis for the cases of arrays of $^{87}$Rb and $^{88}$Sr atoms.
We reveal that selective radiance is particularly sensitive to resonant scattering into phonon sidebands, leading to emission into other channels than the target one. We also demonstrate that such detrimental effects can be mitigated by \emph{shifting} the sideband scattering out of resonance. 

Finally, we analyze the emitted light in the event of atomic recoil. We show that when resonant phonon scattering is suppressed, the emission of light mostly occurs as independent scattering from the recoiled atom. Instead, when a phonon sideband is resonant, light emission is collective, and it retains a directional character, albeit not necessarily in the direction of the reflected light from the mirror.

\subsection{Input-output theory}
We consider an atomic array driven by a classical field and model the scenario by adding a driving term $\hat{H}_{\rm d}$ to the master equation \cref{eq:ME}. Within the rotating wave approximation the driving term reads $ \hat{H}_{\rm d} = -\sum_{\nn} \left(\spl_\nn \dpp^\dagger \EE^+_0(\hat{\rr}_\nn, t) + {\rm H.c.}\right)$, where $\dpp$ is the dipole matrix element, $\EE^+_0$ denotes the positive frequency component of the input field.
While the master equation captures the dynamics of the atomic degrees of freedom, the electric field follows from the input-output relation~\cite{BettlesPRL2016,Shahmoon2017,AsenjoGarcia2017PRA,Shahmoon2020}
\begin{align}
\!\!\!\!\!\hat{\EE}^+(\rr, t) = \hat{\EE}^+_0(\rr, t) \!+\! \mu_0\w_0^2\!\sum_{\nn} \!\bar{\bar{G}}(\rr \!-\! \hat{\rr}_\nn(t),\w_0) \dpp \smi_\nn (t),\!\!
\label{eq:input_output_formula_1}
\end{align}
which describes how the total field depends on the input field and the field scattered by the atoms. The latter contribution depends on the dynamics of the atoms via the electromagnetic Green's tensor $\bar{\bar{G}}(\rr \!-\! \hat{\rr}_\nn(t), \w_0)$ evaluated at the resonance frequency of the dipole transition, $\omega_0$. In the following we drop the explicit dependence on $\w_0$ in the Green tensor for notational simplicity.
Equation~\eqref{eq:input_output_formula_1} includes motional effects through the atomic position operator $\hat{\rr}_\nn(t)$.

We consider a 2D atomic array located in the $xy$-plane, with in-plane and out-of-plane motion characterized by the Lamb-Dicke parameters $\eta_\parallel$ and $\eta_z$.
In the same spirit as in Sec. \ref{sec:Model}, we expand $\bar{\bar{G}}(\rr - \hat{\rr}_\nn(t))$ to second order around the equilibrium position, $\rr_{\nn} = d \, \nn$, of atom $\nn$, and exploit that phononic populations are negligible. We obtain
\begin{align}\label{eq:General_Input_Ouput_Formula}
    &\hat{\EE}^+(\rr,t)\! = \EE^+_0(\rr,t)\!+\!\mu_0\w_0^2\bigg\{ -\sum_{\nn,\alpha} \eta_\alpha\bar{\bar{G}}^\alpha(\rr-\rr_\nn)\Rop_{\alpha,\nn}(t) \nonumber \\
    & +\sum_\nn\!\! \spare{\pare{1 \!-\!\frac{\eta_\parallel^2}{2}}\!+\!\frac{\eta_z^2 \!-\!\eta_\parallel^2}{2k_0^2}\partial^2_z}\!\bar{\bar{G}}(\rr \!-\! \rr_\nn)\bigg\} \dpp \smi_\nn(t),
\end{align}
where we defined $\bar{\bar{G}}^\alpha(\rr) = k_0^{-1} \partial_\alpha \bar{\bar{G}}(\rr)$. In Equation~\eqref{eq:General_Input_Ouput_Formula}, we recognize the same contributions to the modified scattering that we described in Sec.~\ref{sec:Model}. In particular, the terms proportional to $\eta_z^2, \eta_\parallel^2$ (second line) describe the modification to the scattering from zero-point motion of the ground state, whereas the terms linear in $\eta_\alpha$ (first line) come from the absorption/emission of phonons. 

We assume the array is driven with a plane-wave of the form $\EE^+_0(\rr, t) = \EE_0 e^{i \pp \cdot \rr_{xy} + i p_z z - i \omega_{\rm d} t}$, where the wavevector has in-plane components $\pp = (p_x, p_y)$ and an out-of-plane component $p_z = \sqrt{(\omega_d/c)^2 - \pp^2} \simeq \sqrt{k_0^2 - \pp^2}$, describing a wave propagating in the positive $z$ direction. We transform to a frame rotating at the frequency of the drive, $\omega_{\rm d}$, and 
approximate $\Hop_d$ to second order in the Lamb-Dicke parameter. Assuming negligible phonon populations $\avg{\bdop_{\alpha\nn}\bop_{\alpha\nn}}\approx 0$, we obtain
\begin{align}
\hat{H}_{\rm d} =&\, - \sum_\nn \dpp^\dagger \EE_0^+ \, e^{i \pp \cdot \rr_\nn} \Bigg[ \left(1 - \sum_\alpha \frac{\eta_\alpha^2 p_\alpha^2}{2 k_0^2}\right) \spl_{\nn} \nonumber \\ &+ i \sum_\alpha \frac{\eta_\alpha p_\alpha}{k_0}  \left(\bdop_{\alpha,\nn} + \bop_{\alpha,\nn}\right) \spl_{\nn}\Bigg] + {\rm H.c.}
\label{eq:driving_Lamb_Dicke_expansion}
\end{align}
Crucially, the recoil of an absorbed or emitted photon can lead to the creation or annihilation of a phonon, as described by the second line in Eq.~\eqref{eq:driving_Lamb_Dicke_expansion}.

\subsection{Optical response at normal incidence}\label{sec:Normal_Incidence}

We now analyze the optical properties in the particular case of driving a 2D array at normal incidence. In this case, the recoil can only excite phonons in the out-of-plane direction.\footnote{We assume no cross coupling in the trapping potential between different motional directions.} Transforming the driving Hamiltonian \cref{eq:driving_Lamb_Dicke_expansion} to momentum space, we obtain
\begin{align}
\!\!\!\hat{H}_{\rm d}
&= -\Omega_{\mathbf{0}} \Big[\!\Big(1 \!-\! \frac{\eta_z^2}{2}\Big) \sdop_{\bf 0} \!+\! \frac{i\eta_z}{\sqrt{N}} \sum_{\qq}\!\sdop_{\qq} (\hat{b}^\dagger_{z,-\qq} \!+\! \bop_{z,\qq} )\Big] \!+\! {\rm H.c.}
\label{eq:driving_Hamiltonian_normal_incidence}
\end{align}
Here, we defined the collective Rabi frequency for the target mode $\kk=\mathbf{0}$ as
$\Omega_\mathbf{0} \equiv \dpp^\dagger \EE_0^+(\kk=\mathbf{0},z = 0)/(\sqrt{N}d^2)$ where $\EE_0^+(\kk,z) = \int d^2\rr_{xy}~ e^{-i \kk \cdot \rr_{xy}} \EE_0^+(\rr)$.\footnote{Note that for a plane wave $\dpp^\dagger \EE_0^+(\kk={\bf 0},0)\propto N$ so that $\Omega_\mathbf{0} \propto \sqrt{N}$ as expected when driving the collective symmetric mode.} The first term in \cref{eq:driving_Hamiltonian_normal_incidence} describes the direct excitation of the $\kk={\bf 0}$ mode of the system, whereas the second term drives motional sideband excitations.

To describe reflection and transmission from an infinite array, it is more convenient to express the input-output formula in Fourier space for the $xy$-plane. Due to momentum conservation, only the $\kk = {\bf 0}$ component of the output field contributes and \cref{eq:General_Input_Ouput_Formula} reads (Appendix~\ref{app:input_output})
\begin{align}
&\!\!\!\!\!\!\!\!\braket{\hat{\EE}^+({\bf 0}, z)} = \EE^+_0e^{i k_0 z} + \mu_0 \omega_0^2 \sqrt{N} \bar{\bar{G}}(\kk = {\bf 0},z)\dpp \nonumber \\
&\!\!\!\!\!\!\!\!\times \!\Big[ \Big(1 \!-\! \frac{\eta_z^2}{2}\Big) \braket{\sop_{\bf 0}} \!-\! \frac{i\,{\rm sgn}(z)\eta_z}{\sqrt{N}} \!\sum_{\qq}\braket{\sop_{\qq}(\bdop_{z, \qq} \!+\! \bop_{z, -\qq})} \Big],
\label{eq:input_output_with_phonons_momentum_space}
\end{align}
where we assumed the input field to be a classical field propagating along the $z$-direction and left the temporal dependency implicit for notational simplicity. Moreover, $\bar{\bar{G}}(\kk, z)$ is the electromagnetic Green's tensor Fourier transformed in the plane of the array and here evaluated in $\kk = {\bf 0}$. Compared to the case of pinned atoms, there are two major changes in Eq.~\eqref{eq:input_output_with_phonons_momentum_space}. The first term shows that the scattering from the zero-momentum transition operator $\sop_{\bf 0}$ is reduced by a factor of $1 - \eta_z^2/2$ due to the atomic zero-point fluctuations. 
The second term describes the scattering of a photon by the simultaneous creation or absorption of a phonon. 

We note that the optical response is largest when the incoming electric field is polarized in the same direction as the dipoles: $\EE_0^+ \parallel \dpp$, which is assumed to be in the plane of the array. In this case, the scattered field has the same polarization as the dipoles. This can be seen by the fact that the Fourier transformed Green's tensor is proportional to the identity in the plane of the array at normal incidence for $z \neq 0$: $\bar{\bar{G}}(\kk = {\bf 0}, z) \propto \hat{\bold{x}}\hat{\bold{x}}^\dagger + \hat{\bold{y}}\hat{\bold{y}}^\dagger$, where $\hat{\bold{x}}$ and $\hat{\bold{y}}$ are unit vectors along the $x$- and $y$-axes respectively [see Eq.~\eqref{eq:Green_Function_Mixed_Rep}].

The output field in \cref{eq:input_output_with_phonons_momentum_space} depends on the dynamics of the atoms, which obey the equations of motion (Appendix~\ref{app:input_output}) 
\begin{align} \label{eq.EOMS_mirror}
\!\!\!\!i\partial_t\!\braket{\sop_{\bf 0}} &= -\!\left(1 \!-\! \frac{\eta_z^2}{2}\right) \Omega_{\mathbf{0}} +\!(\varepsilon_{\bf 0}^\infty\!-\!\Delta)  \braket{\sop_{\bf 0}} \nonumber \\
&\;\;\;\;- \frac{1}{\sqrt{N}}\!\!\sum_{\qq,\alpha = x,y} \!\!\eta_\alpha g^\alpha_{\qq,{\bf 0}}\braket{\bop_{\alpha,-\qq}\sop_\qq}, \nonumber \\
\!\!\!\!i\partial_t\!\braket{\hat{b}_{z,-\qq} \hat{s}_\qq} &= (\nu_z \!+\! \varepsilon_\qq^\infty \!-\! \Delta)\braket{\hat{b}_{z,-\qq} \hat{s}_\qq} \!-\! \frac{i\eta_z\Omega_{\mathbf{0}}}{\sqrt{N}}, \nonumber\\
\!\!\!\!\!\!i\partial_t\!\braket{\bop_{\alpha,-\qq}\sop_\qq} &= (\nu_\alpha \!+\! \varepsilon_\qq^\infty \!-\! \Delta )  \braket{\bop_{\alpha,-\qq}\sop_\qq} \!+\! \frac{\eta_\alpha g^\alpha_{\qq,\bf 0}}{\sqrt{N}}  \braket{\sop_{\bf 0}}\!, \!\!\!
\end{align}
where $\alpha = x,y$ for the bottom equation, and with detuning $\Delta \equiv \omega_d - \omega_0$. Note that, at normal incidence, it is solely the Fr{\"o}hlich phonon coupling $g^\alpha$ that leads to nonzero spin-phonon coherences  $\braket{\bop_{\alpha,-\qq}\sop_\qq}$ for $\alpha = x,y$, while there is a direct driving of the $z$-component, i.e. a term $\propto \Omega_{\mathbf{0}}$. To obtain \cref{eq.EOMS_mirror}, we neglect expectation values of the form $\braket{\sdop \sop}$, consistent with our linear response description, as well as purely phononic expectation values like $\braket{\bop}, \braket{\bdop \bop}$ etc. This is fulfilled when the array is initially close to its motional ground state and on a timescale such that the drive has not yet heated the array appreciably. In Sec. \ref{sec:Discussion}, we estimate how many photons can be scattered before this occurs.
Assuming that the internal dynamics reaches its steady state well before heating effects become significant, we can extract the optical properties from the steady state of \cref{eq.EOMS_mirror}
\begin{align} \label{eq.steady_state}
\braket{\sop_{\bf 0}} &= -\left(1 - \frac{\eta_z^2}{2}\right)\Omega_{\mathbf{0}} \times G_{\bf 0}^{\rm R}(\Delta), \nonumber \\
\braket{\bop_{z,-\qq} \sop_{\qq}} &= -\frac{i\eta_z\Omega_{\mathbf{0}}}{\sqrt{N}} \times \frac{1}{\Delta - \varepsilon_{\qq}^\infty - \nu_z}.
\end{align}
The retarded Green's function $[G_{\bf 0}^{\rm R}(\omega)]^{-1} = \omega - \varepsilon_{\bf 0}^\infty - \Sigma_{\bf 0}(\omega)$ indicates that the response of the system is described by collective polaron-polariton excitations. Combining \cref{eq.steady_state} with the input-output formula in \cref{eq:input_output_with_phonons_momentum_space} and taking the continuum limit yields 
\begin{align}
\frac{\braket{\hat{E}^+({\bf 0}, z)}}{E^+_0({\bf 0}, 0)} =\; &  e^{ik_0z} \!-\! e^{ik_0|z|} \frac{i\Gamma_{\bf 0}^{(0)}}{2}\bigg[\left(1 \!-\! \eta_z^2\right) G_{\bf 0}^{\rm R}(\Delta)\nonumber\\
&+\int\!\!\frac{{\rm d}^2\qq}{(2\pi)^2} \frac{{\rm sgn}(z)\eta_z^2 d^2}{\Delta \!-\! \varepsilon_\qq^\infty \!-\! \nu_z} \bigg],
\label{eq:input_output_with_phonons_momentum_space_2}
\end{align}
with $E^+ = \dpp^\dagger \EE^+ / |\dpp|$. This gives a direct link between the input field, $E^+_0e^{i k_0 z}$, and the expected output field, $\braket{\hat{E}^+({\bf 0}, z)}$, in linear response. In Equation~\eqref{eq:input_output_with_phonons_momentum_space_2}, we use that $\mu_0\omega_0^2 \sqrt{N} \Omega_{\mathbf{0}} \dpp^\dagger \bar{\bar{G}}(\kk = {\bf 0},z)\dpp /|\dpp| = - i E_0^+({\bf 0}, 0) e^{ik_0 |z|} \Gamma^{(0)}_{\bf 0} / 2$, where $\Gamma_{\bf 0}^{(0)} = 3\pi \gamma_0 / (k_0d)^2$ is the  collective decay for the $\kk = {\bf 0}$ mode. 

Equation \eqref{eq:input_output_with_phonons_momentum_space_2} gives direct access to the optical properties of the array. 
We define $\braket{\hat{E}^+({\bf 0}, z)}/E^+_0({\bf 0}, 0)\equiv t(\Delta)\theta(+z) e^{+i k_0 z} + r(\Delta) \theta(-z) e^{-i k_0 z}$ with transmission and reflection amplitudes 
\begin{align}
\!\!\!\!t(\Delta) &= 1 \!-\! \frac{i\Gamma_{\bf 0}^{(0)}}{2}\left[\left(1\!-\!\eta_z^2\right)\! G_{\bf 0}^{\rm R}(\Delta) \!+\!\!\int\!\!\!\frac{{\rm d}^2\qq}{(2\pi)^2} \frac{\eta_z^2d^2}{\Delta \!-\! \varepsilon_\qq^\infty \!-\! \nu_z} \right]\!\!,\!\nonumber \\
\!\!\!\!r(\Delta) &= \!- \frac{i\Gamma_{\bf 0}^{(0)}}{2}\!\left[\!\left(1 \!-\! \eta_z^2\right) \! G_{\bf 0}^{\rm R}(\Delta) \!-\!\!\int\!\!\!\frac{{\rm d}^2\qq}{(2\pi)^2} \frac{\eta_z^2d^2}{\Delta \!-\! \varepsilon_\qq^\infty \!-\! \nu_z} \right]\!\!.\!\!
\label{eq:reflection_amplitude_normal_incidence}
\end{align}
Equation \eqref{eq:reflection_amplitude_normal_incidence} gives the coherent sum of two processes which bring the system to the same final state and can, therefore, interfere. More precisely, the first process is the direct excitation of the zero-momentum polaron-polariton mode, as can be seen by the appearance of the Green's function $G_{\bf 0}^{\rm R}(\Delta)$ in place of the bare optical response, $[\Delta - (J_{\bf 0}^{(0)} - i\Gamma_{\bf 0}^{(0)}/2)]^{-1}$~\cite{Shahmoon2017}. The amplitude of this first process is suppressed by a factor of $1-\eta_z^2$. This factor arises from the reduction of the coupling to the driving field, Eq.~\eqref{eq:driving_Hamiltonian_normal_incidence}, as well as to the scattering amplitude into the back-reflected mode, Eq.~\eqref{eq:input_output_with_phonons_momentum_space}. Each contributes with a factor of $(1-\eta_z^2/2)$, leading to a total suppression of $(1-\eta_z^2/2)^2\simeq 1-\eta_z^2$. In the limit of $\nu_z\gg\gamma_0$, this process is the only modification to the reflection and transmission amplitude from the pinned atoms case. 

The second process describes the situation where a phonon is generated and subsequently reabsorbed during the scattering event. This is resonant when $\Delta = {\rm Re}(\varepsilon_{\qq}^\infty) + \nu_z$. 
Resonant coupling to subradiant sidebands, which have a small linewidth, $-2{\rm Im}(\varepsilon_{\qq}^\infty)\ll \gamma_0$, can have a strong effect on the reflection and transmission of the array.
Crucially, this also means that the drive can directly excite subradiant modes via the excitation of an additional mechanical oscillation, a task that is impossible for pinned atoms. We will explore this possibility further in Sec.~\ref{sec:excitation_of_subradiant_states}.

\subsection{Reflection and transmission}

The reflectance ($R = |r|^2$) and transmittance ($T = |t|^2$) obtained from \cref{eq:reflection_amplitude_normal_incidence} do not sum to one, i.e., $R + T < 1$. 
This occurs because \cref{eq:input_output_with_phonons_momentum_space_2} only describes the coherent part of the output field, which arises from processes that return the array to its internal and motional ground state at the end of the scattering process. 
The resulting loss $L \equiv 1 - (R + T)$, therefore, describes the incoherent part of the output field and is generally dominated by the coupling to the phononic sideband. In particular, larger losses occur when the drive is resonant to a phononic sideband, $\Delta = {\rm Re}(\varepsilon_{\qq}^\infty) + \nu_z$, of collective subradiant modes $\qq$ which are located where the dispersion, ${\rm Re}(\varepsilon_{\qq}^\infty)$, is locally flat---i.e., where there is a large density of states. In the following, we illustrate these features by considering the experimentally relevant examples of 2D atomic mirrors based on $^{87}$Rb~\cite{Rui2020,Srakaew2023} and $^{88}$Sr atoms.

\begin{table}
\caption{{\bf Experimental parameters for different atomic species.} ${}^{87}$Rb, ${}^{162}$Dy and ${}^{168}$Er atoms are assumed trapped in optical lattices with lattice spacing $d$ and driven on the D2-line, a narrow cycling transition~\cite{Lu2011}, and the $\lambda_0 = 841 \, {\rm nm}$ transition~\cite{GreinerSuperradiance}, respectively. Trap frequencies for ${}^{87}$Rb are obtained from trap depths of $300 \nu_R$ ($2000\nu_R$), where $\nu_R= k_0^2/2M$, perpendicular to (in the plane of) the array~\cite{Priv_Comm_Bloch}. 
The remaining parameters for ${}^{168}$Er and ${}^{162}$Dy are taken from Refs.~\cite{GreinerSuperradiance} and, partially from Ref.~\cite{Hofer2024}. As compared to Ref.~\cite{Hofer2024}, we consider the collective transition wavelength $\lambda_0$, which has a magic trapping condition for the $440$nm trap wavelength~\cite{Priv_Comm_Igor}. ${}^{88}$Sr atoms are assumed trapped in a tweezer array using metasurface traps and driven on the ${}^3\text{D}_3-{}^3\text{P}_2$ transition ~\cite{Holman2026,Holman2024,Priv_Comm_Will}. As compared to Ref.~\cite{Holman2026}, we assume a different ratio of radial to axial tweezer trap frequencies, which is, however, within experimental reach~\cite{Priv_Comm_Will}. 
}\label{TAB:parameters}
\begin{ruledtabular}
\begin{tabular}{lccccc}
Parameter & ${}^{87}$Rb & ${}^{88}$Sr & ${}^{162}$Dy & ${}^{168}$Er\\
\hline
$\lambda_0$ & 780~nm &  2.9~$\mu$m & 741~nm & 841~nm\\
$d/\lambda_0$ & 0.47 & 0.34 & 0.36 & 0.32\\
$\gamma_0/2\pi$ & 6.06~MHz & 57~kHz & 1.8~kHz & 8~kHz\\
$\nu_\parallel/2\pi$ & 335~kHz & 100~kHz & 50~kHz & 40~kHz\\
$\nu_z/2\pi$ & 129~kHz& 20~kHz & 35~kHz & 18~kHz\\
$\eta_\parallel$ & 0.10 & 0.05 & 0.21 & 0.2\\
$\eta_z$ & 0.17 & 0.11 & 0.25 & 0.3 \\
\end{tabular}
\end{ruledtabular}
\end{table}

\begin{figure}[t!]
\begin{center}
\includegraphics[width=1.0\columnwidth]{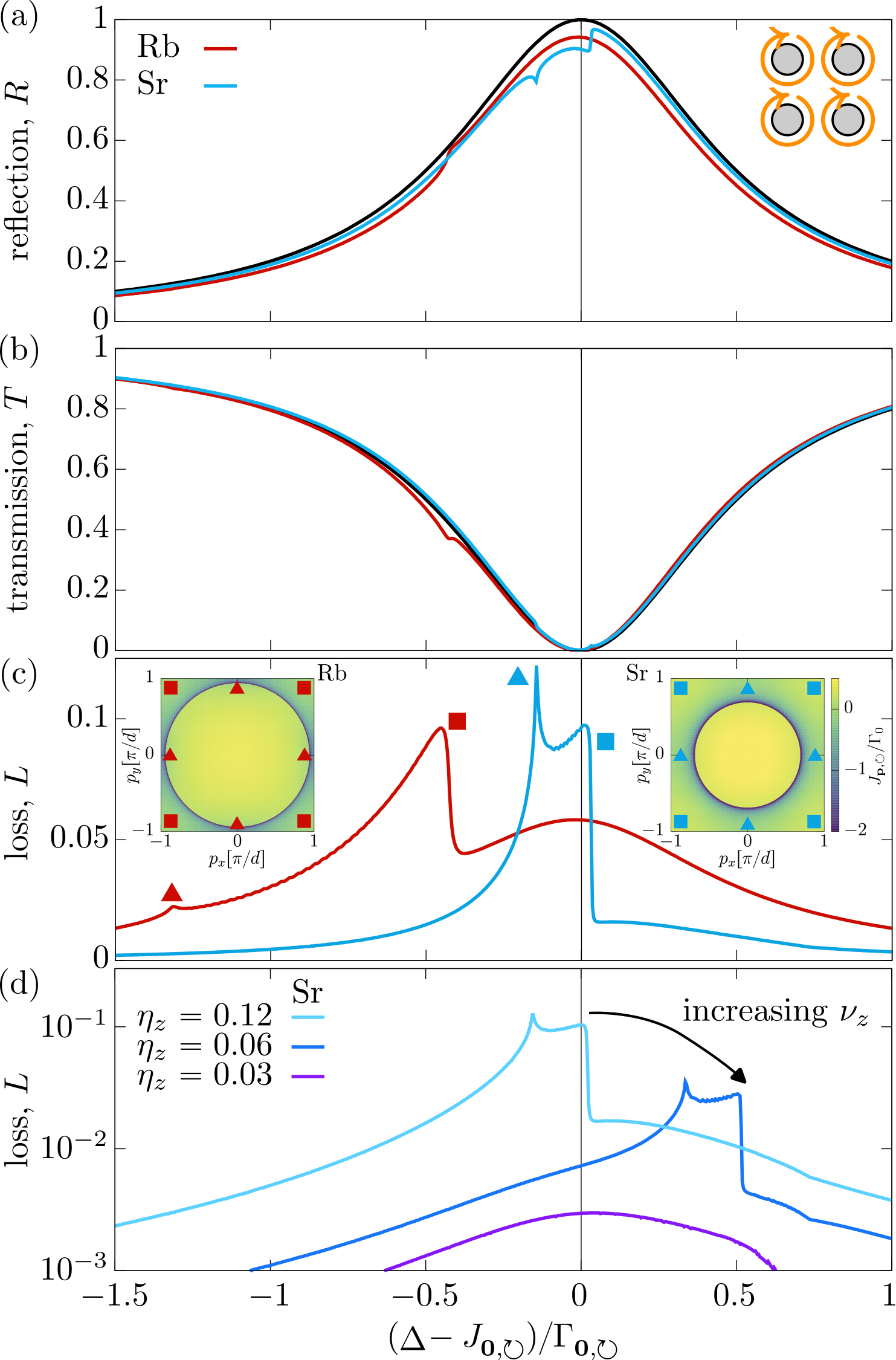}
\end{center}\vspace{-0.5cm}
\caption{{\bf Effect of motion on a 2D atomic mirror.} Reflection $R$ (a), transmission $T$ (b), and loss $L$ (c, d) as a function of detuning for two experimental setups based on $^{87}$Rb in an optical lattice (red) and $^{88}$Sr in a tweezer array (blue) compared to the perfectly pinned case (black). In both cases the atomic dipole moments are circularly polarized in the plane of the array (inset in (a)). The detuning is set relative to the resonance frequency, $J_{\bf 0}$, and in units of the collective decay rate, $\Gamma_{\bf 0}$, for pinned atoms. The insets in (c) show the dispersion relation in case of Rubidium (left inset) and Strontium (right inset). The peaks appearing in the loss spectrum are directly related to energy ranges where the density of states of the motional sidebands is large, corresponding to flat dispersions (triangles and squares). (d) Loss coefficients for the $^{88}$Sr setup and indicated out-of-plane Lamb-Dicke parameters by varying the trap frequency $\nu_z$. These correspond to (top to bottom) $\nu_z/2\pi \simeq 19 {\rm\, kHz}, 75 {\rm\, kHz}, 300 {\rm \,kHz}$. The main loss suppression results from moving the phononic sideband out of resonance (black arrow).} 
\label{fig:RTL_plot} 
\vspace{-0.25cm}
\end{figure} 

Figure~\ref{fig:RTL_plot} quantifies the reflectance, transmittance and loss for two realizations of an atomic mirror with ${{}^{87}}$Rb and ${}^{88}$Sr atoms, respectively. For Rubidium (indicated by red lines in the plot), we use the parameters in \cref{TAB:parameters}, which are inspired by the recent realization of an atomic mirror with atoms trapped in an optical lattice~\cite{Rui2020,Srakaew2023}. Due to the broad linewidth of the D2-transition in ${{}^{87}}$Rb, the trap frequencies are small compared to $\gamma_0$. Arrays of tweezer-trapped ${}^{88}$Sr atoms, whose loss is shown by the blue line in Fig.~\ref{fig:RTL_plot}(c), have instead a much smaller natural linewidth resulting in a larger ratio $\nu/\gamma_0$ (see \cref{TAB:parameters}). 

The Strontium setup additionally displays smaller values of the Lamb-Dicke parameters~\cite{Holman2026,Holman2024,Priv_Comm_Will}. Larger values of $\nu_\alpha / \gamma_0$ and smaller Lamb-Dicke parameters should in principle lead to a better performance of the atomic mirror. However, as seen in Fig. \ref{fig:RTL_plot}(c), driving the Strontium array on resonance also couples resonantly to a phononic sideband. Therefore, even though the overall scale of the loss coefficient for ${}^{88}$Sr is suppressed in comparison to the ${^{87}}$Rb setup, the performance on the collective resonance $\Delta = J_{\bf 0}$ is actually worse. 
These higher losses arise from resonant scattering between the target $\kk = {\bf 0}$ mode and points of the Brillouin zone with large densities of states, indicated by the squares and triangles in Fig.~\ref{fig:RTL_plot}(c).
We note, however, that the maximal reflection for the Strontium array, which occurs slightly above resonance, away from the phononic sideband, \emph{is} slightly higher than for Rubidium [\cref{fig:RTL_plot}(a)].

These results also highlight how losses may be alleviated by avoiding resonances to phononic sidebands. Indeed, Fig.~\ref{fig:RTL_plot}(d) shows directly that by increasing the out-of-plane trapping frequency, $\nu_z$, the losses on resonance can be suppressed by an order of magnitude. 
This happens as a combined effect of suppressing the overall scale of the losses, due to decreasing $\eta_z$, and shifting the phononic sideband out of resonance. When driving on resonance, $\Delta = J_{\bf 0}$, we additionally find that the transmission remains vanishingly small, $T \approx 0$. As a result, the reflectance $R = 1 - L - T$ for $\eta_z \leq 0.06$ exceeds $99.2\%$ on resonance. Changing the lattice geometry (e.g. to a triangular structure) or the lattice constant provides further tuning knobs for suppressing the loss. Since the dispersion $\varepsilon_\pp$ depends on $k_0 d$ and the lattice symmetry, one can engineer the dispersion such that scattering resonances are shifted away from the driving frequency. While the requirement of trapping the ground and excited state at equal strength can impose constraints on the lattice constant, experiments with atoms trapped in tweezers or in an accordion lattice allow $k_0 d$ to be varied.

Reducing losses is important because high reflectivity is a key parameter that determines the fidelity and efficiency of entangled photon generation based on 2D atomic arrays~\cite{Bekenstein2020, Moreno-Cardoner2021, Srakaew2023}. Alleviating losses is also fundamental for storing and retrieving photons efficiently in quantum memory protocols based on 2D atomic arrays~\cite{Manzoni2018,Guimond2019,Solomons2024, Mann2024}.
By indicating ways to circumvent resonant scattering, our results thus pave the way for establishing arrays of trapped atoms as highly efficient light-matter interfaces. 

We end this analysis with a comparison to previous approaches for modeling the impact of motion on the mirror response. Specifically, we consider the case of frozen motion and the effective model presented in Refs.~\cite{Shahmoon2019,Shahmoon2020}.

\begin{figure}[t!]
\begin{center}
\includegraphics[width=1.0\columnwidth]{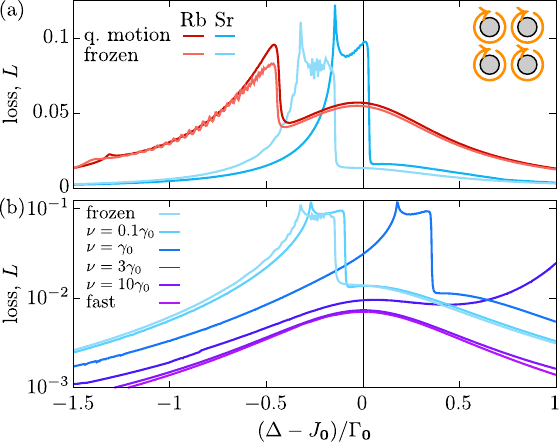}
\end{center}\vspace{-0.5cm}
\caption{{\bf Dependence of the loss coefficient on the trap frequency.} (a) Loss coefficient for the ${}^{87}$Rb (red line) and ${}^{88}$Sr (blue line) experimental setups, repeated from Fig. \ref{fig:RTL_plot}, compared to the frozen motion approximation (light red and blue lines). (b) Loss coefficients for the ${}^{88}$Sr setup and the indicated isotropic trap frequencies. Frozen and fast motion results are obtained for an array of $40\times 40$ atoms.} 
\label{fig:loss_fast_and_frozen_motion} 
\vspace{-0.25cm}
\end{figure}

The frozen motion approximation is often used to model the effects of motion in the experiments~\cite{Rui2020,Srakaew2023}.
For the same realization using ${}^{87}$Rb and ${}^{88}$Sr, Fig. \ref{fig:loss_fast_and_frozen_motion}(a) shows good agreement between the full quantum motion results and the frozen motion approximation, which is obtained by simulating the response of a finite array of $40\times 40$ atoms.
Note, however, that a finite out-of-plane trap frequency, $\nu_z$, can lead to shifts in the resonance positions between the two approaches as evidenced by the case of $^{88}$Sr in Fig.~\ref{fig:loss_fast_and_frozen_motion}(a). More specifically, in Appendix~\ref{app:frozen_motion_proof} we prove that for arrays driven at normal incidence and when atoms can move only in the out-of-plane direction, in the limit $\nu_z\ll \gamma_0$ frozen motion yields \emph{exactly} the same result as \cref{eq:reflection_amplitude_normal_incidence}. In this limit, the two approaches agree \emph{quantitatively} as shown in Fig.~\ref{fig:loss_fast_and_frozen_motion}(a) for the case of $^{87}$Rb. 
This agreement highlights that although for some aspects---like transport---it is difficult to make a quantitative connection to sampling static disorder configurations, for the mirror properties this connection can be exactly derived. 
Indeed in Fig. \ref{fig:loss_fast_and_frozen_motion}(b), we see how our results for the loss coefficient seamlessly interpolate between the limits of the frozen and fast motion regimes for the case of isotropic trap frequencies $\nu = \nu_z = \nu_\parallel$. For the fast and frozen motion results, we simulated a $40\times40$ atomic array. 
Specifically, for the fast-motion approximation dipole–dipole interactions are averaged over the spatial wavefunction, while in the frozen-motion approximation, atomic positions are sampled from frozen disorder configurations~\cite{Porras2008}. 

We also compare our results for the transmission and reflection amplitudes in Eqs.~\eqref{eq:reflection_amplitude_normal_incidence} to the results presented in Refs.~\cite{Shahmoon2019,Shahmoon2020} where in-plane motion was neglected ($\eta_\parallel = 0$) and the optical response of the array was approximated by the one of the $\kk = {\bf 0}$ mode only. Indeed, if we set $\eta_\parallel=0$, $\nu_z\to 0$, and neglect the momentum dependency of the collective modes, $\varepsilon_\qq^\infty \to \varepsilon_{\bf 0}$ in Eqs.~\eqref{eq:reflection_amplitude_normal_incidence}, we recover the results of Refs.~\cite{Shahmoon2019,Shahmoon2020}: to leading order the reflection amplitude is suppressed by a factor of $1 - 2\eta_z^2$ from the fully pinned case, whereas the transmission amplitude is unaffected. 
The extra factor of two emerges from the second term in the square bracket of Eq.~\eqref{eq:reflection_amplitude_normal_incidence} within this approximation. We emphasize that to correctly compute the effects of the phononic sidebands in Eq.~\eqref{eq:reflection_amplitude_normal_incidence}, one needs to take into account the full dispersion relation. This is most crucially important for capturing the impact of the scattering resonances.

\subsection{Output field under recoil and which-path information}\label{sec:output_field_recoil}

As the dominant contribution to the loss coefficient $L$ comes from the (resonant) scattering of a phonon, it is useful to investigate the properties of the scattered light in the event of a recoil. We are particularly interested in the conditions for which the scattered light retains a directional (\ie collective) character. Below we demonstrate that if the array is driven far off resonance from a phonon sideband, the profile of the emitted light looks like that of a recoiling point-dipole positioned at the place of the recoil. On the other hand, no matter how large $\nu_\alpha / \gamma_0$, if the array is driven nearly resonant with spin-phonon pairs (\ie on a phononic sideband), the collective effects are essential and cannot be ignored.

\begin{figure*}[t!]
\begin{center}
\includegraphics[width=1\textwidth]{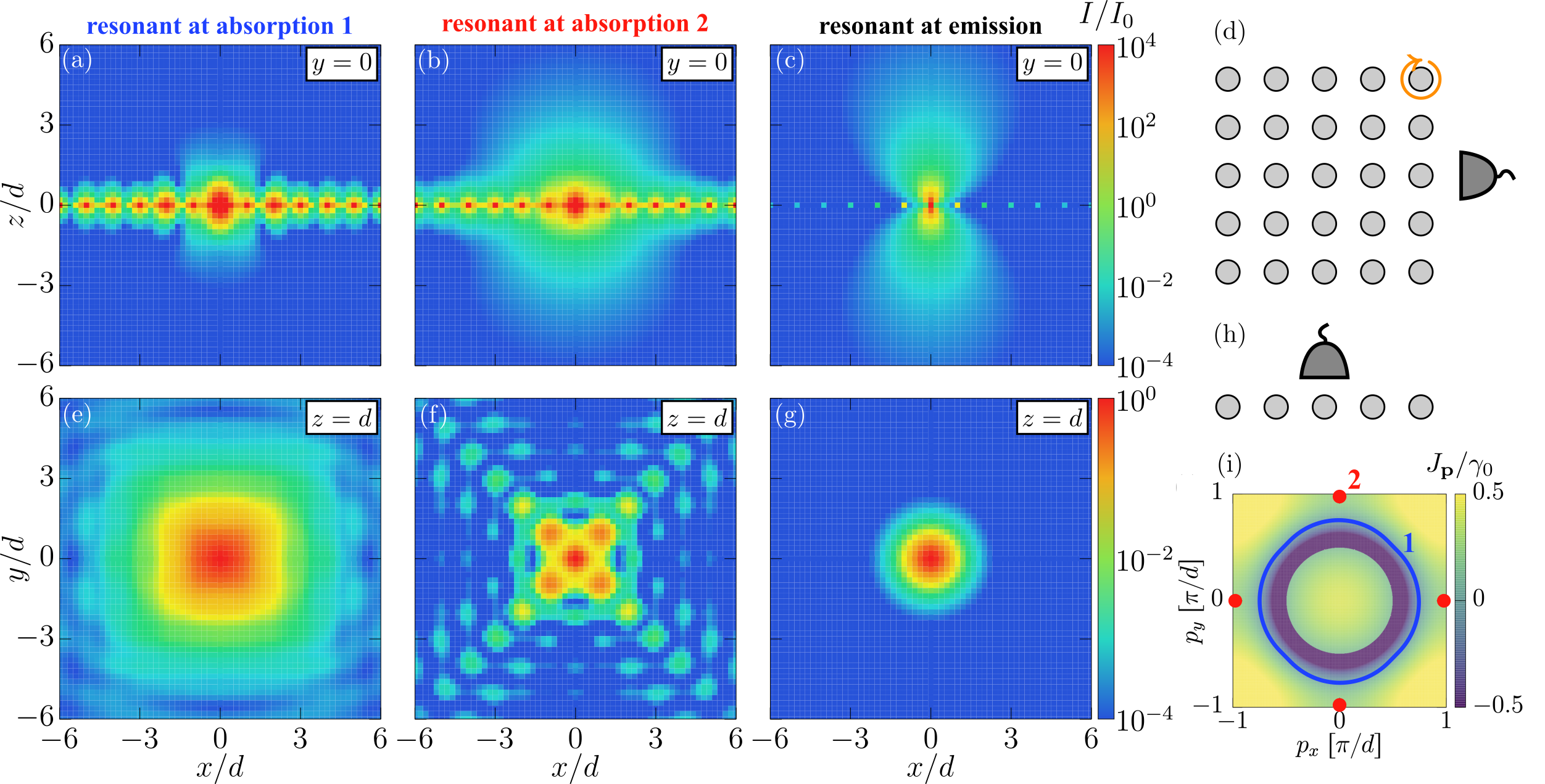}
\end{center}\vspace{-0.5cm}
\caption{{\bf Effects of motion on the emission pattern.} Radiated intensity under recoil, $I = |\braket{\hat{b}_{z,{\bf n} = {\bf 0}}\hat{E}^+({\bf r})}|^2$, in the $y = 0$ plane perpendicular to the array (a),(b),(c) for circularly polarized dipole moments in the plane of the array [as indicated in (d)], as well as at a fixed distance of $z = d$ above the array (e), (f), (g) [as indicated in (h)]. For better relative comparison, the field in each panel is normalized by $I_0 = I({\bf r} = (0,0,d))$. In (a),(e) the drive is resonant with producing a phonon and collective eigenmodes along the blue line in (i) as the photon is \emph{absorbed}, $\Delta = {\rm Re}(\varepsilon^\infty_{\bf p}) + \nu$, described by the first term in \cref{eq:field_with_recoil}. In panels (b) and (f), it is resonant with a phonon and the collective modes at the red dots in (i). The latter are saddle points in the dispersion and lead to characteristic \emph{ridges} in the $xy$-plane as evident in (f). In panels (c) and (g) the drive is instead resonant with the target $\kk=\bold{0}$ mode, $\Delta = {\rm Re}(\varepsilon^\infty_{\bf 0})$. In this case, the second term in \cref{eq:field_with_recoil}, proportional to $\partial_z \bar{\bar{G}}$, dominates, yielding a localized emission pattern with minor corrections from the other atoms. We assume a lattice spacing $d = \lambda_0 / 4$ and trap frequency $\nu = 10\gamma_0$.} 
\label{fig.field_with_recoil} 
\vspace{-0.25cm}
\end{figure*} 

We study the output field amplitude subject to a recoil event at the center of the array, ${\bf n}={\bf 0}$. We compute the coherence $\braket{\bop_{\alpha,\nn={\bf 0}}\hat{\EE}^+(\rr)}$ from \cref{eq:General_Input_Ouput_Formula}, by using the steady state solutions in Eq.~\eqref{eq.steady_state} under the assumption that the array is driven at normal incidence and the atomic polarization $\hat{\dpp}$ is constant across the array. The general result, and the details of the derivation are given in \cref{app:output_field_with_phonon}. In the following, we focus on the case where the recoil occurs out of the plane of the array, \ie along $\alpha = z$, in which case the result is most easily interpreted.
We may then express the scattered field concisely as the interference of two dipole responses which correspond to the recoil of the atom at $\nn = {\bf 0}$ due to scattering into a collective phononic sideband at absorption, and the recoil of the same atom at emission. These two processes are analogous to the ones in Eq.~\eqref{eq:reflection_amplitude_normal_incidence}, but where we now consider the dynamics in the one phonon sector. Explicitly, we find that the output field in the presence of a phonon at $\nn  = {\bf 0}$ is
\begin{align} \label{eq:field_with_recoil}
\!\!\braket{\bop_{z,\nn={\bf 0}}\hat{\EE}^+(\rr)} =\;& \frac{\mu_0 \omega_0^2\eta_z}{\sqrt{N}}\Bigg[i\!\int\!\!\frac{{\rm d}^2{\bf q}}{(2\pi)^2} e^{i\qq\cdot\rr} \bar{\bar{G}}(\qq,z) \hat{\dpp}\hat{\dpp}^\dagger \PP_{z,\qq}^{(1)}(\Delta)  \nonumber \\
&-\frac{\partial_z \bar{\bar{G}}(\rr)}{k_0} \hat{\dpp}\hat{\dpp}^\dagger\PP^{(2)}(\Delta) \Bigg],
\end{align}
where the overall prefactor correctly scales with the amplitude for producing a phonon, the Lamb-Dicke parameter $\eta_z$, as well as the amplitude for observing that phonon at site $\vec{n} = \mathbf{0}$, given by $1/\sqrt{N}$. We note that, in the event of a recoil, the polarization of the scattered field can change so that the scattered field does not necessarily have the same in-plane polarization as the atoms, unlike what we found in Sec.~\ref{sec:Normal_Incidence} for the zero phonon sector. Mathematically, this comes about, because there are contributions from $\bar{\bar{G}}(\qq,z)$ at nonzero $\qq$, which produce contributions with different in-plane polarizations.

The first term in Eq.~\eqref{eq:field_with_recoil}, containing an integral over all in-plane momenta, describes collective polarization fields
\be
\PP_{z,\qq}^{(1)}(\Delta) = -\frac{|\dpp|^2}{\Delta - \varepsilon_\qq^\infty - \nu_z} \frac{1}{\sqrt{N}}\sum_\nn \EE_0^+(\rr_\nn,0),
\ee
which explicitly depend on the momentum $\qq$ through the collective energy $\varepsilon_\qq$. This describes the situation in which the driving field $\EE_0^+(\rr)$ directly excites a spin-phonon pair $\sdop_\qq \bdop_{z,-\qq}$ in the array. Hence, the phonon is produced as the \emph{photon is absorbed} into the array---through the excitation of a collective sideband~\cite{RubiesBigorda2024}.

The second term in Eq.~\eqref{eq:field_with_recoil} has a polarization field
\be
\PP^{(2)}(\Delta) = -|\dpp|^2G_{\bf 0}^{\rm R}(\Delta) \frac{1}{\sqrt{N}}\sum_{\nn} \EE_0^+(\rr_\nn,0),
\ee
and describes the situation, where the array absorbs the photon \emph{without} a recoil, whereby the dipole moment scales with the retarded Green's function $G_{\bf 0}^{\rm R}(\Delta)$ of the $\qq = {\bf 0}$ mode from \cref{eq:Greens_function_frequency_space}. In this case, the recoil (and thereby the production of the phonon) happens as the photon is re-emitted into free space. The associated motion of the dipole leads to the spatial derivative $\partial_z \bar{\bar{G}}(\rr)$ of the electromagnetic Green's tensor. The same correction appears for a classical dipole undergoing periodic motion, as we elucidate in Appendix~\ref{app:classical_dipole_field}. The emitted field subject to a recoil is, thus, given by the interference of this localized response and the collective response described by the first term in \cref{eq:field_with_recoil}.  

To understand the behavior and relative importance of the two responses, we plot the squared norm of \cref{eq:field_with_recoil} in Fig. \ref{fig.field_with_recoil} under different driving conditions. We only keep the \emph{far-field} component of the emission stemming from the terms in $\bar{\bar{G}}({\bf r}) \propto 1/r$. Figures \ref{fig.field_with_recoil}(a), \ref{fig.field_with_recoil}(e) as well as \ref{fig.field_with_recoil}(b), \ref{fig.field_with_recoil}(f) describe situations in which the drive is resonant with collective sidebands associated to the modes indicated by the blue line and red circles in Fig.~\ref{fig.field_with_recoil}(i) respectively. The first term in \cref{eq:field_with_recoil} is thus resonantly enhanced, leading to a strong \emph{collective} response with a radiation pattern showing a clear and sensitive dependence on which collective modes are resonant with the drive. While the emitted field is naturally strongest closest to the recoiling atom, the emission patterns at intermediate distances to the array [$r \sim d$] look nothing like the emission from a single recoiling dipole. In particular, the characteristic ridges in Fig. \ref{fig.field_with_recoil}(f) come about, because the drive resonantly excites the modes at the saddle points ${\bf p}d = (\pm \pi,0),(0,\pm \pi)$. Locally, these saddle points are flat along the diagonals $\Delta q_x = \pm \Delta q_y$ leading to the ridges in real space along $x = \pm y$ [Fig. \ref{fig.field_with_recoil}(f)]. 

Figures.~\ref{fig.field_with_recoil}(c) and \ref{fig.field_with_recoil}(g) describe, instead, a situation, in which the driving is off-resonant with any phonon-sideband. In this case the second term in \cref{eq:field_with_recoil} dominates and the field's emission pattern is well described by that of a single recoiling dipole, whereas all the other dipoles in the array have only a minuscule contribution. This is the situation that occurs when driving an array on resonance with a collective mode (\eg~the $\kk = {\bf 0}$ mode) in the limit of large trap frequencies, $\nu_\alpha/\gamma_0 \gg 1$. In this regime, when an atom recoils the photon is emitted with frequency $\omega_0-\nu_\alpha$ which is far off resonant with the polariton band, thus suppressing interference with other polariton modes. Accordingly, the recoiled atom retains which-path information~\cite{Scully1991, Zhang2025, Fedoseev2025, Itano1998} while radiating a photon along the direction prescribed by its dipole-emission pattern, as we confirm below.

The central insight from this calculation is that, regardless of the trap frequency, the output field exhibits collective behavior rather than that of a single localized dipole if the drive is nearly resonant with a phonon sideband, that is for either $\Delta \simeq {\rm Re}(\varepsilon_\qq^\infty) + \nu_z$ or $\Delta \simeq {\rm Re}(\varepsilon_\qq^\infty) + \nu_\parallel$ for some $\qq$. On the other hand, in the other extreme where the array is far detuned from such resonances, we obtain
\begin{equation}
\!\!\!\!\braket{\bop_{z,\nn={\bf 0}}\hat{\EE}^+(\rr)}\! \simeq \!\frac{\mu_0\omega_0^2\eta_z}{\sqrt{N}}\!\!\left(\!i\bar{\bar{G}}(\rr) \!-\! \frac{\partial_z \bar{\bar{G}}(\rr)}{k_0}\!\right)\!\!\hat{\dpp}\hat{\dpp}^\dagger\PP(\Delta),\!\!\!
\end{equation}
where $\PP(\Delta) = -|\dpp|^2 \Delta^{-1} N^{-1/2}\sum_\nn \EE^+_0(\rr_\nn,0)$. The total field in this limit is, thus, simply the sum of a standard dipolar field $\sim \bar{\bar{G}}$ and the linear correction $\sim \partial_z \bar{\bar{G}}$.

\section{Motion-induced excitation of subradiant states} \label{sec:excitation_of_subradiant_states}

In the previous sections, we identified the leading modifications in the array's optical properties as coming from (resonant) scattering of a phonon by a polariton. This process modifies the polariton's momentum potentially hindering transport (Sec.~\ref{sec:transport}) or reducing selective radiance by scattering light into uncontrolled directions (Sec.~\ref{sec:input-output}).
In this section, we show how this very same process can be controlled to enhance the scattering of polaritons into subradiant collective modes of the system. This technique can be immediately implemented with current experiments to spectroscopically investigate subradiant excitations, prepare a propagating excitation in the array, such as those studied in Sec.~\ref{sec:transport}, or observe topological edge states in arrays of multilevel atoms~\cite{Perczel2017, Perczel2017PRA}.

Single-photon excitation of subradiant modes in arrays of immobile atoms is disallowed by momentum conservation. Hence, previous work has proposed to excite multi-photon transitions~\cite{Rusconi2021,Cech2023}, apply field gradients~\cite{Plankensteiner2015}, or drive auxiliary transitions~\cite{He2020}. 
However, as discussed in Sec.~\ref{sec:input-output}, a subradiant polariton can be excited by scattering a phonon via two distinct processes.
(1) A direct excitation of a motional sideband by the laser, as expressed by the second line of Eq.~\eqref{eq:driving_Lamb_Dicke_expansion}, as discussed in Ref.~\cite{RubiesBigorda2024}; (2) Scattering of a spin excitation \emph{after} the photonic excitation is stored in the array, by producing a phonon. This process is described by Eq.~\eqref{eq:H1_crystal_momenta}. 
These two processes allow us to resonantly excite subradiant states via a more straightforward protocol than previous schemes. Moreover, we show in the following that the second scattering channel can be significantly enhanced by driving the array at an angle, providing a more efficient way to produce subradiant excitations. 
A related mechanism for populating subradiant states via coupling to phonons was proposed in molecular aggregates in Ref.~\cite{Holzinger2022}, where vibronic coupling modulates the emitters’ resonance frequencies. While similar in spirit, our protocol differs in that here the phonon coupling modifies the excitation hopping rate rather than the on-site resonance frequency. 

\begin{figure}[t!]
\begin{center}
\includegraphics[width=1.0\columnwidth]{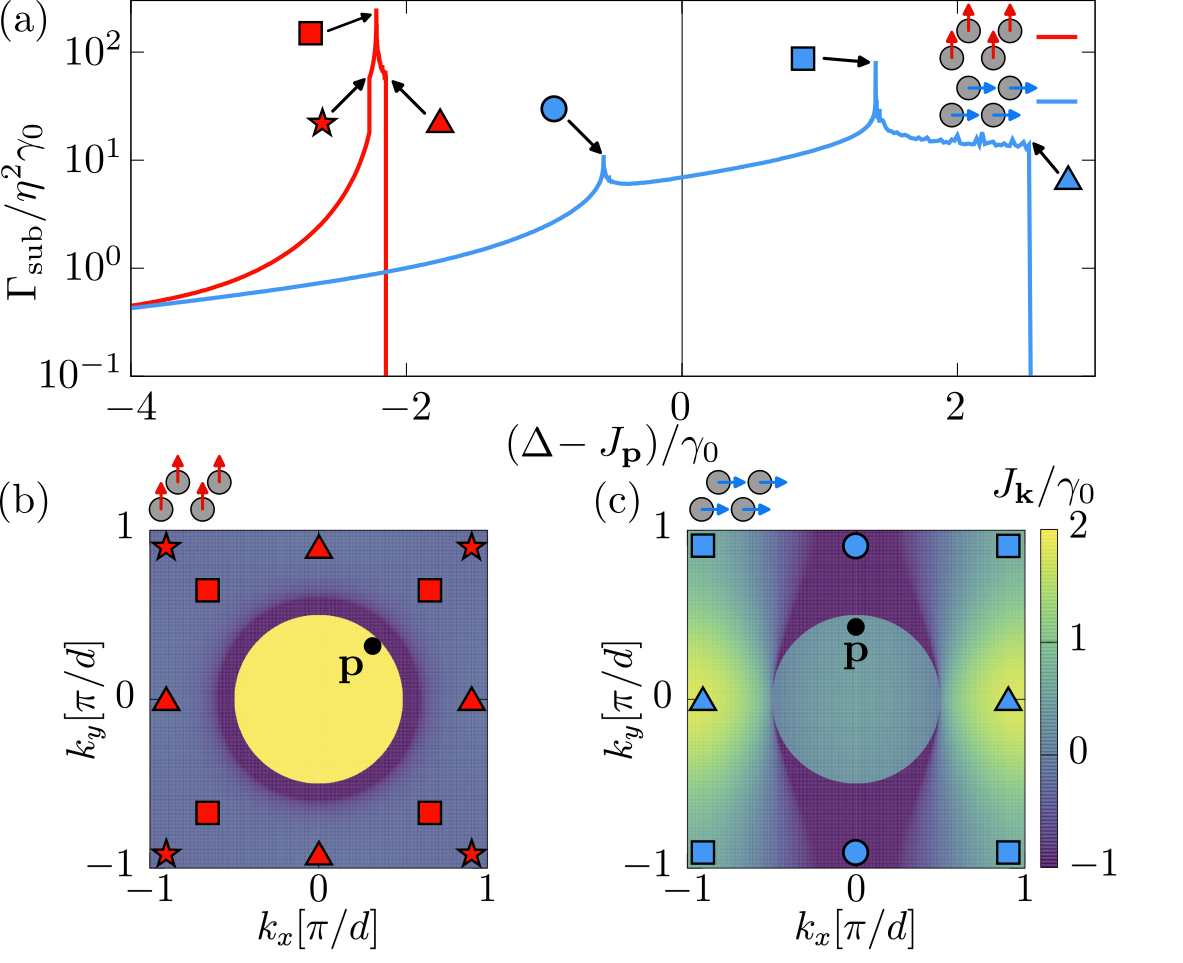}
\end{center}\vspace{-0.5cm}
\caption{{\bf Rate of excitation of subradiant states.} (a) Scattering rate to the subradiant sector for atoms polarized perpendicular ($\perp$) and parallel ($\parallel$) to the array as a function of detuning relative to the energy of the driven mode $J_\pp$ for $\nu = \gamma_0$. The input light field is coupled to highly superradiant modes $\pp = (k_0 - \Delta k, k_0 - \Delta k)/\sqrt{2}$ ($\pp = (0, k_0 - \Delta k)$) for perpendicular (parallel) polarization, just inside the light cone ($\Delta k = 0.1/d$). Note that we cannot describe driving the mode exactly at the light cone, $\Delta k=0$, because at those points both our perturbative approach as well as the Markovian assumption break down. Dispersion relations $J_{\kk}$ for perpendicular (b) and parallel [along $x$] (c) polarization. The black points show the driving modes $\pp$. The colored points indicate which subradiant states come into resonance with the driving field at the indicated features in (a).} 
\label{fig.subradiant_scattering_rate_2D} 
\vspace{-0.25cm}
\end{figure}

We consider a 2D array driven with a general in-plane momentum $\pp$. The driving term \cref{eq:driving_Lamb_Dicke_expansion} reads in momentum space
\begin{align}
&\hat{H}_{\rm d} = \!-\Omega_\pp\Bigg[\!\left(1 \!-\frac{1}{2}\! \sum_\alpha\left[\frac{\eta_\alpha p_\alpha}{k_0}\right]^2\right) \sop^{\dagger}_{\pp} \nonumber \\
&\!+\! \frac{i}{\sqrt{N}} \sum_\alpha \frac{\eta_\alpha p_\alpha}{k_0} \sum_{\qq} \sop^{\dagger}_{\qq}\left(\hat{b}^\dagger_{\alpha,\pp - \qq} \!+\! \bop_{\alpha,\qq - \pp}\right)\!\Bigg] \!+\! {\rm H.c.},
\label{eq:driving_Hamiltonian_general}
\end{align}
with the collective Rabi frequency $\Omega_\pp = \dpp^\dagger \EE^+_0(\pp, z = 0) / (\sqrt{N} d^2)$. 
The idea is then to drive the superradiant state at $\pp$ in resonance with a \emph{subradiant} state. The fast decay of $\pp$ then quickly leads to a steady state value of $\braket{\sop_\pp} = -\Omega_\pp / (\Delta - \varepsilon_\pp)$. The phonon coupling along with the direct driving of phononic sidebands in the second line of \cref{eq:driving_Hamiltonian_general} then leads to the equation
\begin{align}\label{eq:EoM_coherent_subrad_scattering}
\!\!\!\!\! i\partial_t \!\braket{\bop_{\alpha,\pp - \qq}\sop_\qq} &= (\nu_\alpha \!+\! \varepsilon_\qq \!-\! \Delta)\braket{\bop_{\alpha,\pp - \qq}\sop_\qq} \!-\! \frac{\Omega_\pp M^\alpha_{\qq,\pp}}{\sqrt{N}} ,\!\!\!
\end{align}
with coupling $M^\alpha_{\qq,\pp} = \eta_\alpha [ip_\alpha/k_0 + (1 - \delta_{\alpha,3}) g^\alpha_{\qq,\pp}/(\Delta - \varepsilon_\pp)]$ leading to a continuous production of scattered excitations. We directly integrate \cref{eq:EoM_coherent_subrad_scattering} to obtain $\braket{\bop_{\alpha,\pp - \qq}\sop_\qq}$. 
At intermediate timescales---long compared to the superradiant dynamics, short compared to the decay of subradiant states---the rate of scattering an excitation into the subradiant sector by producing a phonon is given by Fermi's golden rule,
\begin{align} \label{Eq:subradiant_scattering_rate}
\!\!\!\!\Gamma_{\rm sub} &\equiv \partial_t \sum_{\qq \in {\rm sub},\alpha} |\braket{\bop_{\alpha, \pp - \qq}\sop_\qq}|^2 \nonumber \\
\!\!\!\!&= |\Omega_\pp|^2 d^2  \sum_\alpha \!\! \int_{\rm sub}\!\! \frac{{\rm d}^2q}{2\pi}  |M^\alpha_{\qq,\pp}|^2 \delta(\Delta \!-\! J_\qq^{(0)} \!-\! \nu_\alpha).\!\!
\end{align}
In the last passage, we assume an infinite array and convert the sum to a two-dimensional integral over the subradiant sector of the Brillouin zone. In Fig. \ref{fig.subradiant_scattering_rate_2D}(a), we show the resulting linear scattering rate both for parallel and perpendicular polarizations, when we couple to highly superradiant modes just inside the light cone at $|\pp|=k_0$. Here, we use $\Omega_\pp = 0.3\Gamma_\pp$ to remain in the linear response regime.

These results show several striking features. First, the scattering rate is around two orders of magnitude larger than the inherent lifetime of the subradiant state $\sim \eta^2 \gamma_0$. This comes about as a consequence of the divergence of the interaction $g^{\alpha}_{\qq,\pp}$ at the light cone, $p = k_0$, along with large densities of states (DoS). Indeed, inspecting \cref{Eq:subradiant_scattering_rate}, the subradiant scattering rate closely mirrors the DoS, as expected from Fermi's golden rule. 
Second, since a state with energy $J_\qq^{(0)} + \nu$ resonant with the driving frequency must be present for a nonzero scattering to occur, there is a threshold, indicated by triangular markers in Fig.~(\ref{fig.subradiant_scattering_rate_2D}), above which the scattering rate vanishes.
Third, atoms polarized perpendicular (parallel) to the array show one (two) peak(s) in $\Gamma_{\rm sub}$ as the detuning is lowered. These strong responses correspond to drives at resonance with the subradiant states at the square and circular markers in the Brillouin zone, which are all saddle points in the dispersion. As a result, the density of states diverges logarithmically here, and the ability to excite these states is considerably enhanced. Note that in Eq.~\eqref{Eq:subradiant_scattering_rate}, we have approximated the dispersion in the integral with its unperturbed value, $J_\qq^{(0)}$, leading to the observed sharp logarithmic divergences. In a fully self-consistent calculation, the finite lifetime of the subradiant states induced by motion should smoothen these sharp divergences. Finally, for perpendicular polarization, there is a local minimum of the dispersion at the corner of the Brillouin zone, $(\pi, \pi)$. When the detuning decreases below the corresponding energy of that subradiant state, the scattering rate suddenly decreases by an order of magnitude (star point). For parallel polarization, on the other hand, there is \emph{no} local minimum in the dispersion, and the scattering, therefore, smoothly decreases once the detuning is below the energy of the lowest saddle point (blue circular point).

\section{Relevance for experiments} \label{sec:Discussion}

We complement our results by reviewing the main assumptions behind our formalism and their validity for experiments.

The description of recoil effects rests on a perturbative treatment. This is valid when the coupling to atomic motional fluctuations, as quantified by the Lamb-Dicke parameter $\eta=\sqrt{\nu_R/\nu}$, is weak. We also choose $\eta$ independently from $\nu/\gamma_0$ to reflect the diverse regimes that could be achieved in experiments using different atomic species or atomic transitions within the same species and for which $\gamma_0$ and $\nu_R$ might have significantly different values.
Indeed, several atomic species host long-wavelength laser-compatible dipole transitions suitable for realizing subwavelength arrays~\cite{Masson2024}. 
In this article, we focused on ${}^{87}$Rb~\cite{Glicenstein2020,Rui2020,Srakaew2023} and ${}^{88}$Sr~\cite{Huang2023,Holman2024,Holman2026}, however, our method applies in general to different atomic species. As an example, we summarize in Table~\ref{TAB:parameters} the relevant parameters used in recent experiments on subwavelength atomic arrays for different atomic species.
We see that the Lamb-Dicke parameters have similar values, and all are small enough to justify our perturbative approach. Importantly, $\nu/\gamma_0$ varies across several orders of magnitude.

We also assumed the atoms to be trapped at a magic wavelength such that the center-of-mass potential is insensitive to the internal electronic states ($\ket{g},\ket{e}$). 
While this condition has been shown in experiments with tweezer-trapped ${}^{88}$Sr~\cite{Priv_Comm_Will} and ${}^{162}$Dy~\cite{Bloch2023,Bloch2024} atoms, experiments with ${}^{87}$Rb atoms in Refs.~\cite{Rui2020,Srakaew2023} and with ${}^{168}$Er in Ref.~\cite{GreinerSuperradiance} currently lack this capability. However, in Table~\ref{TAB:parameters} we assume a trapping laser with wavelength $\approx 740~\text{nm}$ for ${}^{87}$Rb, which has been predicted to be magic~\cite{Priv_Comm_Bloch}.
If the trap is not magic, the potential suddenly changes when atoms are excited on the dipole transition $\ket{g}\leftrightarrow \ket{e}$ thus leading to additional heating of the atomic motional state~\cite{Karanikolaou2024}. Future work should investigate effects of non-magic traps as they are relevant for current experiments.

We described recoil effects in subwavelength arrays for atoms prepared in their motional ground state. This assumption is well justified, as recent experiments with optical lattices are initialized by cooling atoms deeply into the Mott-insulator regime~\cite{Rui2020,Srakaew2023}. 
Additionally, there are ongoing efforts to cool tweezer-trapped atoms into the ground state~\cite{Kaufman2012, Cooper2018, Saskin2019, Hofer2024, Manetsch2025}. However, phonon-assisted scattering processes heat up the atoms during the dynamics. 
Heating occurs on a timescale set by the phonon-scattering rate. For transport, we showed in Sec.~\ref{sec:transport} that these effects are negligible if resonant phonon scattering is weak. For the case of an atomic mirror, we assumed that the internal dynamics reaches a steady state before the array heats up. It is important to estimate how many photons can be scattered before this occurs. The associated phonon population is produced at the inelastic scattering rate $R_\text{in}= \eta^2 d^2 \int\!\! \text{d}\qq\,\W^2/(\Delta -\varepsilon_\qq-\nu)$ mainly via the recoil of an atom during the absorption of a laser photon. This should be compared with the elastic scattering rate $R_\text{el} \equiv \Gamma_\mathbf{0} |\avg{s_\mathbf{0}}|^2$ leading to a ratio $R_\text{in}/R_\text{el} \simeq \eta^2 d^2\int\!\!\text{d}\qq\, \Gamma_\mathbf{0}/(J_\mathbf{0} -\varepsilon_\qq-\nu)$ of the order of the mirror's losses.
For the cases studied in Fig.~\ref{fig:RTL_plot}, a recoil only occurs somewhere between once every ten to a hundred scattering events. 
Accordingly, our theory correctly captures the behavior of the system at early times when the number of recoil events is small. For an array of $N\sim 10^3$ Sr atoms initially prepared in their motional ground state, we estimate that a phonon density of $0.1$ is reached after $10^3-10^4$ photons have been reflected off the mirror if $R_\text{in}/R_\text{el} \simeq 0.01-0.10$. 
In case of mirrors with larger losses, it is still possible to reduce phonon scattering over the timescale of the internal dynamics by reducing the driving strength, $\Omega$. This, however, comes at the cost of a longer duration of the experiment.

Thermal effects due to a finite atomic motional temperature~\footnote{This can be either due to initial state preparation, where the atomic motional state cannot be ground-state cooled, or to heating of the array and subsequent thermalization of its phonon population.} lead to enhanced phonon scattering and to an additional scattering channel where a polariton is scattered by \emph{absorbing} a phonon. Assuming thermal equilibrium, we can describe finite temperature effects by modifying the self-energy as
\be\label{eq:Sigma_Thermal}
	\!\!\!\Sigma_\pp = -\frac{1}{N}\sum_{\qq,\alpha}\!\spare{\frac{(\bar{n}_\qq+1)(\eta_\alpha g^{\alpha}_{\qq,\pp})^2 }{\varepsilon^{(0)}_\pp \!-\! \varepsilon^{(0)}_\qq \!-\! \nu_\alpha} \!+\! \frac{\bar{n}_\qq(\eta_\alpha g^{\alpha}_{\qq,\pp})^2 }{\varepsilon^{(0)}_\pp \!-\! \varepsilon^{(0)}_\qq \!+\! \nu_\alpha}}\!,\!\!
\ee
where $\bar{n}_\qq$ is the average thermal phonon population in the mode $\qq$. We remark that while \cref{eq:Sigma_Thermal} correctly captures the thermal corrections to the self-energy in perturbation theory, a precise description of how the array heats up during the dynamics requires more sophisticated methods that go beyond the scope of this article. 

\section{Conclusions and Outlook} \label{sec:Conclusions}

In this work, we studied the optical properties of a subwavelength array of trapped atoms, where atomic vibrations couple to collective light-matter excitations through optical forces. 
We introduced a framework that treats atomic motion as an intrinsic part of the system's collective response, leading to the concept of polaron-polariton quasiparticles, which we identify as the elementary optical excitations in a subwavelength array of trapped atoms.
Our method provides a fully quantum mechanical treatment of motion that goes beyond the commonly considered extreme limits of frozen ($\nu/\gamma_0 \ll 1$) and fast ($\nu/\gamma_0 \gg 1$) motion by capturing correlations between internal and external degrees of freedom across a broad range of trap frequencies, Lamb-Dicke parameters, and atomic polarizations.

Using our framework, we determined how atomic motion affects selective radiance, subradiance, and near-lossless propagation of atomic excitations, which form the essential ingredients for virtually all known applications of subwavelength atomic arrays. We found that the most detrimental effects arise from resonant scattering processes into motional sidebands.
This insight points toward trap frequencies, lattice geometries, and dipole polarizations that minimize such resonant scattering as promising routes for mitigating the adverse effects of atomic motion.

Our results provide a path to quantify motion-induced errors in protocols for quantum memories and generating entangled photons based on atomic arrays~\cite{AsenjoGarcia2017PRX,Manzoni2018,Guimond2019,Solomons2024,Moreno-Cardoner2021,Porras2008,Bekenstein2020,Zhang2022,Ballantine2021,RubiesBigorda2022PRR}.
These protocols have been theoretically predicted to achieve high efficiencies and fidelities, and to outperform comparable protocols with disordered atomic ensembles. These predictions, however, rely on the assumption of immobile atoms. Our work enables the assessment of their performance under realistic experimental conditions.

Furthermore, our results have direct implications for all-atomic nanophotonic devices based on subwavelength atomic arrays, such as atomic mirrors~\cite{Bettles2015,Shahmoon2017,Rui2020} or atomic waveguides~\cite{Masson2020PRR, Patti2021, CastellsGraells2021}: For atomic mirrors, our analysis suggests that experiments designed to avoid resonant scattering into motional sidebands could reach reflectances approaching $\sim 99 \%$ in the near future. For all-atomic waveguides, near-lossless transport of excitations along the array is essential. While transport generically breaks down in the limit of very small trap frequencies, we identify regimes where transport remains remarkably robust even at relatively small trap frequencies, opening prospects for realizing all-atomic waveguide QED with current experimental capabilities.

Beyond these specific applications, our work opens the door to exploring new phenomena arising from the interplay of internal excitations and motion in subwavelength arrays. For instance, we demonstrated how phonon scattering into subradiant modes can be harnessed to generate subradiant excitations without resorting to complex multi-photon schemes~\cite{Rusconi2021,Cech2023}.
Future work may investigate phonon-mediated interactions between multiple polaron-polaritons, as a novel type of nonlinearity capable of inducing effective photon-photon interactions. 

We identify the influence of atomic motion in highly excited systems as an important avenue for future research. In the unsaturated regime explored in this work, motion primarily affects subradiant modes, while bright states experience negligible corrections. This suggests that many-body superradiance in atomic arrays~\cite{Masson2020PRL,Masson2022,Robicheaux2021,Mok2024} might be resilient to the effects of atomic motion, whereas many-body subradiance~\cite{GreinerSuperradiance} might be generically fragile.

Our results provide the foundation for understanding collective optical phenomena in the presence of motion, an essential aspect in the rapidly emerging experiments on light-matter interactions with subwavelength atomic arrays. Our insights help to guide the design of future experiments and open up possibilities for harnessing atomic motion as a resource.

\acknowledgments{We are grateful to Daniel~Adler, Immanuel~Bloch, Darrick E.~Chang, Igor Ferrier-Barbut, Aaron~Holman, Charlie-Ray~Mann, Simon~P.~Pedersen, Thomas~Pohl, Oriol~Rubies-Bigorda, Ephraim~Shahmoon, Ximo~Sun, Panagiotis~Tselifis, Sebastian~Will, Pascal~Weckesser, and Johannes~Zeiher for insightful discussions. 
We are particularly thankful to I.~Bloch, C.-R.~Mann, P.~Weckesser, and J.~Zeiher for valuable feedback on the manuscript.
K.K.N. acknowledges support from the Carlsberg Foundation through a Carlsberg Reintegration Fellowship (Grant No. CF24-1214). L.W. acknowledges support from the “Secretaria d’Universitats i Recerca del Departament de Recerca i Universitats de la Generalitat de Catalunya” under grants 2023 FI-1 00716, 2024 FI-2 00716 and 2025 FI-3 00716, the European Social Fund Plus, Generalitat de Catalunya (CERCA program), Fundaci\'{o} Cellex, and Fundaci\'{o} Mir-Puig. D.C.G. acknowledges support from the Munich Quantum Valley (Bavarian State Government, Hightech-Agenda Bayern Plus) and the Munich Center for Quantum Science and Technology (DFG, EXC2111–390814868). D.\ M.\ acknowledges support from the Novo Nordisk Foundation under grant numbers NNF22OC0071934 and NNF20OC0059939. A.\ A.-G. acknowledges support by the National Science Foundation through the CAREER Award (No. 2047380), the David and Lucile Packard Foundation, and the Chu Family Foundation. C.\ C.\ R. acknowledges support from the European Union’s Horizon Europe program under the Marie Sklodowska Curie Action LIME (Grant No. 101105916). This research was supported in part by the grant NSF PHY-2309135 to the Kavli Institute for Theoretical Physics (KITP).}

\appendix

\section{Lamb-Dicke expansion of the non-Hermitian Hamiltonian} \label{app:Zero_point_motion}

In this Appendix, we present the derivation of the different orders in the Lamb-Dicke expansion of the non-Hermitian Hamiltonian in \cref{eq:H_expansion}.

Let us first consider the first-order correction and derive the Fröhlich-like interaction in \cref{eq:H1_crystal_momenta}. Starting from the Taylor expansion in real space in \cref{eq:H_expansion} and using the Fourier modes $\bop_{\alpha,\nn} = \sum_\kk b_{\alpha,\kk} e^{i\kk\cdot\rr_\nn}/\sqrt{N}$, we get 
\begin{align}\label{eq:H1_appendix}
\Hop_1^\alpha &= \sum_{\nn,\mm} G^\alpha_{\nn\mm}\pare{\bop_{\alpha,\nn} + \bdop_{\alpha,\nn} -\bop_{\alpha,\mm}-\bdop_{\alpha,\mm}}\spl_{\mm}\smi_{\nn} \nonumber \\
&= \frac{1}{N^{3/2}}\sum_{\substack{\pp,\qq,\kk\\\nn,\mm}} G^\alpha_{\nn{\bf 0}}\pare{\bop_{\alpha,\kk} + \bdop_{\alpha,-\kk}}\sdop_{\qq}\sop_{\pp} \nonumber\\
&\times \pare{e^{i(\pp+\kk)\cdot\rr_\nn} - e^{i\pp\cdot\rr_\nn}} e^{+i(\pp+\kk-\qq)\cdot\rr_\mm} \nonumber \\
&= \frac{1}{\sqrt{N}}\sum_{\substack{\pp,\qq,\nn}} G^\alpha_{\nn{\bf 0}}\pare{\bop_{\alpha,\qq-\pp} + \bdop_{\alpha,\pp-\qq}}\sdop_{\qq}\sop_{\pp} \nonumber\\
&\times \pare{e^{i\qq\cdot\rr_\nn} - e^{i\pp\cdot\rr_\nn}} \nonumber\\
&= \frac{1}{\sqrt{N}}\sum_{\substack{\pp,\qq}} g^\alpha_{\qq,\pp}\pare{\bop_{\alpha,\qq-\pp} + \bdop_{\alpha,\pp-\qq}}\sdop_{\qq}\sop_{\pp}.
\end{align}
Identifying like terms in the last two lines leads to the interaction vertex in \cref{eq:H1_vertex}.

Let us now consider the second-order Lamb-Dicke correction in \cref{eq:H_expansion}. It reads
\begin{align}\label{eq:H2_appendix}
\Hop_2 \!=\! \sum_{\substack{\alpha,\beta\\ \nn,\mm }}&\!\frac{\eta_\alpha\eta_\beta}{2} G^{\alpha\beta}_{\nn\mm}(\hat{\widetilde{R}}_{\alpha,\nn} \!-\! \hat{\widetilde{R}}_{\alpha,\mm}) ( \hat{\widetilde{R}}_{\beta,\nn} \!-\! \hat{\widetilde{R}}_{\beta,\mm}) \sdop_\nn \sop_\mm
\end{align}
where $\hat{\widetilde{R}}_{\alpha,\nn} = \bdop_{\alpha,\nn}+\bop_{\alpha,\nn}$ is the displacement operator in units of the zero-point fluctuation.
For low-temperature states, $k_BT \ll \nu$, which have a vanishingly small phonon occupation, $\braket{\bdop_{\alpha,\nn}\bop_{\alpha,\nn}} \simeq 0$, we can approximate 
\begin{equation}\label{eq:Approx_Position_Op_2}
    (\hat{\widetilde{R}}_{\alpha,\nn} \!-\! \hat{\widetilde{R}}_{\alpha,\mm}) ( \hat{\widetilde{R}}_{\beta,\nn} \!-\! \hat{\widetilde{R}}_{\beta,\mm}) \simeq 2\delta_{\alpha,\beta}(1 - \delta_{\nn,\mm}).
\end{equation}
All other terms proportional to $\bdop_{\alpha,\nn}\bop_{\beta,\mm}$, $\bdop_{\alpha,\nn}\bdop_{\beta,\mm}$, or $\bop_{\alpha,\nn}\bop_{\beta,\mm}$ contribute to higher-order corrections than $\eta^2$ for the effects that we study in this work.
Substituting \cref{eq:Approx_Position_Op_2} into \cref{eq:H2_appendix}, we obtain
\begin{equation}\label{eq.app_H2_general_expression}
\begin{split}
\Hop_2 &\simeq \sum_{\nn,\mm,\alpha} \eta_\alpha^2 G^{\alpha\alpha}_{\nn{\mm}}(1 - \delta_{\nn,\mm})\sdop_\nn \sop_\mm \\
&= \sum_{\pp,\nn,\alpha} \eta_\alpha^2G^{\alpha\alpha}_{\nn{\bf 0}} (1 - \delta_{\nn,{\bf 0}})\cdot e^{-i\pp\cdot\rr_\nn} \sdop_\pp \sop_\pp,
\end{split}
\end{equation}
where in the second step we assumed translational invariance, which is known to correctly capture the spectrum of large atomic arrays. It is convenient to define the correction coming from the contribution in \cref{eq.app_H2_general_expression} as
\begin{equation}\label{eq:Correction_fast_motion_General}
    \Delta \varepsilon_\pp \equiv \sum_{\nn,\alpha} \eta_\alpha^2G^{\alpha\alpha}_{\nn{\bf 0}} (1 - \delta_{\nn,{\bf 0}})e^{-i\pp\cdot\rr_\nn}.
\end{equation}
The impact of the second-order correction is thus to modify the dispersion relation of the bare polariton $\varepsilon_\pp^{(0)}$ as
\begin{equation}\label{eq:epsilon_infinity_general_appendix}
    \varepsilon^{(0)}_\pp \longrightarrow \varepsilon^{\infty}_\pp \equiv \varepsilon^{(0)}_\pp + \Delta\varepsilon_\pp.
\end{equation}

We emphasize that the result derived here is generally applicable to atoms in a generic electromagnetic environment described by the Green's function $G(r)$.

\section{Sum rule of the spectral function} \label{app:sum_rule}

Within the Chevy ansatz, the solution of the effective Schr{\"o}dinger equation $\Hop\ket{\psi^r_{\pp}} = E_\pp\ket{\psi^r_{\pp}}$ gives the right perturbed polaronic eigenstates~\cite{Chevy2006}
\begin{align} \label{eq:right_polaron_eigenvectors}
\ket{\psi^r_{\pp}} = \sqrt{Z_\pp}\left[\sdop_\pp + \frac{1}{\sqrt{N}}\!\sum_{\qq,\alpha} \frac{\eta_\alpha g^\alpha_{\qq,\pp}}{\varepsilon^{(0)}_\pp - \varepsilon^{(0)}_\qq - \nu_\alpha}\sdop_\qq \bdop_{\alpha,\pp-\qq} \right]\ket{0}, 
\end{align}
whereas the left polaronic eigenstates $\ket{\psi^l_{\pp}}$ have complex conjugate amplitudes. This implies that the overlap condition for the residue is $Z_\pp = (\bra{0} s_\pp\ket{\psi^r_\pp})^2$, which does not involve an absolute value. This subtle difference from Hermitian systems means that the residue is no longer guaranteed to lie in the interval $[0,1]$, nor to be real~\cite{Scarlatella2019}. 

In fact, resolving the real-time Green's function $G^{\rm R}_\pp(t) = -i\theta(t)\bra{0}\sop_\pp e^{-i\Hop t} \sdop_\pp\ket{0}$ using the left-right eigenvector identity $\mathbb{1} = \sum_n \ket{\varphi_n^r}\bra{\varphi_n^l}$ shows that its Fourier transform may be expressed as
\begin{align}
G^{\rm R}_\pp(\omega) = \sum_n \frac{\bra{0}\sop_\pp\ket{\varphi^r_n}\bra{\varphi^l_n}\sdop_\pp\ket{0}}{\omega - E_n},
\end{align}
where $\Hop\ket{\varphi^r_n} = E_n\ket{\varphi^r_n}$ defines the complex energy eigenvalues $E_n$. Using the integral identity $\int_{-\infty}^{\infty} d\omega (\omega-E_n)^{-1} = -i\pi$ finally shows that the spectral function, $A_\pp(\omega) = -2{\rm Im}(G_\pp(\omega))$, obeys the same normalization condition as for a Hermitian system,
\begin{align}
\int_{-\infty}^{\infty} \frac{d\omega}{2\pi} A_\pp(\omega) = 1.
\end{align}
From this, it follows that if ${\rm Re}[Z_\pp] = {\rm Re}[(\bra{0}\sop_\pp\ket{\psi^r_{\pp}})^2] > 1$, there must be frequencies for which $A_\pp(\omega)$ is negative.

\section{Master equation for atoms in free space} \label{app:master_equation}

In this appendix, we derive the Lamb-Dicke expansion for the master equation describing dipole-interacting atoms in free space, including the effects of recoil. This result was originally derived in Refs.~\cite{Palmer2010,RubiesBigorda2024} and is a generalization of the case of a single atom first presented in Ref.~\cite{Stenholm1986}. Here, we sketch the derivation for completeness.

The starting point is the master equation \eqref{eq:ME} where, for the case of atoms in free space, the population recycling term has the form
\be\label{eq:Recycling_FreeSpace}
    \mathcal{R}\rhop = \gamma_0 \sum_{\nn,\mm} \int\!\!\text{d}\uu\,\mathcal{D}(\uu) e^{-i k_0 \uu \cdot \rrop_\nn}\smi_\nn \rhop e^{i k_0 \uu \cdot \rrop_\mm}\spl_\mm.
\ee
Here, $\uu$ is a unit vector describing the direction of the emitted photon, and the integral is taken over all directions in space weighted by the dipole emission pattern $\mathcal{D}(\uu)= (3/8\pi) \hat{\dpp}^\dag (\id -\uu\otimes\uu )\hat{\dpp}$, normalized such that $\int\!\text{d} \uu\,\mathcal{D}(\uu)=1$. 

We now proceed to expand the master equation in powers of the atomic position fluctuations up to second order in the Lamb-Dicke parameter. The expansion of the non-Hermitian Hamiltonian is given in \cref{eq:H_expansion}. For the population recycling term, we have 
\be\label{eq:Recycling_Lamb-Dicke}
    \mathcal{R}\rhop \simeq \bigg(\mathcal{R}_0+\sum_\alpha \eta_\alpha \mathcal{R}^\alpha_1+\sum_{\alpha,\beta}\eta_\alpha\eta_\beta\mathcal{R}_2^{\alpha,\beta}\bigg)\rhop,
\ee
where $\mathcal{R}_n$ contains contributions to order $n=0,1,2$ in the Lamb-Dicke parameter in \cref{eq:LambDicke_condition}.
The leading-order term reads
\be\label{eq:R_0}
    \mathcal{R}_0\rhop \equiv  \sum_{\nn,\mm}\Gamma_{\nn\mm}\smi_\nn \rhop\spl_\mm,
\ee
where we defined the dissipative coupling
\be
    \Gamma_{\nn\mm} \equiv \gamma_0\int\!\!\text{d}\uu \,\mathcal{D}(\uu) e^{-i k_0 \uu\cdot \rr_{\nn \mm}} =  -2\Im[G_{\nn\mm}],
\ee
where $\rr_{\nn \mm} \equiv \rr_\nn - \rr_\mm = d (\nn - \mm)$.
The first-order Lamb-Dicke correction reads
\be\label{eq:R_1}
    \mathcal{R}^\alpha_1 \rhop \equiv \sum_{\nn\neq\mm} \Gamma_{\nn\mm}^\alpha \smi_\nn\pare{\hat{\widetilde{R}}_{\alpha,\nn}\rhop -\rhop \hat{\widetilde{R}}_{\alpha,\mm}}\spl_\mm,
\ee
where we defined
\be
    \Gamma_{\nn\mm}^\alpha \equiv \gamma_0 \int\!\!\text{d}\uu\,\mathcal{D}(\uu)\frac{\partial}{\partial(k_0 r_{\alpha})}e^{-i k_0\uu\cdot\rr}\biggr\rvert_{\rr=\rr_{\nn \mm}},
\ee
where $r_{\alpha} \equiv \mathbf{e}_\alpha\cdot\rr$ and $\mathbf{e}_\alpha$ is the unit vector along the direction $\alpha=x,y,z$. The summation in \cref{eq:R_1} does not include the diagonal case $\nn=\mm$ because $\Gamma_{\nn\nn}^\alpha = 0$ due to symmetry.

Finally, the second-order Lamb-Dicke correction in \cref{eq:Recycling_Lamb-Dicke} reads
\be\label{eq:R_2}
\begin{split}
    \mathcal{R}_2^{\alpha\beta}\rhop &= \sum_{\nn,\mm} \Gamma_{\nn\mm}^{\alpha\beta}\smi_\nn\Big[\inv{2}\hat{\widetilde{R}}_{\alpha,\nn}\hat{\widetilde{R}}_{\beta,\nn}\rhop + \inv{2}\rhop\hat{\widetilde{R}}_{\alpha,\mm}\hat{\widetilde{R}}_{\beta,\mm} \\
    &-\hat{\widetilde{R}}_{\alpha,\nn}\rhop\hat{\widetilde{R}}_{\beta,\mm}\Big]\spl_\mm,
\end{split}
\ee
where we defined
\be
    \Gamma_{\nn\mm}^{\alpha\beta}\equiv \gamma_0 \int\!\!\text{d}\uu\,\mathcal{D}(\uu) \frac{\partial^2}{k_0^2 \partial r_{\alpha}\partial r_{\beta}} e^{-i k_0\uu\cdot\rr}\biggr\rvert_{\rr = \rr_{\nn \mm}}.
\ee
Note that in \cref{eq:R_2}, the summation includes the case $\nn=\mm$, which corresponds to the recoil heating of an atom at a rate $\Gamma_{\nn\nn}^{\alpha\alpha}$ due to the scattering of photons.

Similar results have been obtained in Ref.~\cite{olmos2025} following a slightly different approach. In this latter case, however, the contributions to first order in the Lamb-Dicke parameter, namely \cref{eq:H1_crystal_momenta} and \cref{eq:R_1}, were neglected under the assumption of a large trap frequency ($\nu\gg\gamma_0$). Here, instead, we do not make this assumption.

\section{Spin-phonon coupling in free-space arrays}\label{app:Free_Space}
In this appendix, we consider the case of atomic arrays in a three-dimensional electromagnetic vacuum and derive explicit expressions for the spin-phonon coupling used to obtain the results presented in the main text.

The dipole-dipole coupling strength between atoms $\nn$ and $\mm$ in free space reads~\cite{Lehmberg1970a} 
\be\label{eq:G_coupling}
    G(\rrop_\nn -\rrop_\mm) \equiv  -\frac{3\pi \gamma_0}{k_0} \hat{\dpp}^\dagger \bar{\bar{G}}(\rrop_\nn-\rrop_\mm,\omega_0)  \hat{\dpp},
\ee
where $\gamma_0=\mu_0\w_0^3|\dpp|^2/(3\pi c)$ is the free-space single-atom spontaneous emission rate, $\dpp$ is the dipole matrix element of the atomic transition, and $\hat{\dpp} = \dpp / |\dpp|$ is the associated unit vector. 
The free-space electromagnetic Green's tensor $\bar{\bar{G}}(\rr,\omega_0)$ is obtained as a solution of the Helmholtz equation
\be\label{eq:Helmholtz_Eq}
    \nabla\times\nabla\times \bar{\bar{G}} (\rr-\rr',\w_0) - \frac{\w_0^2}{c^2}\bar{\bar{G}} (\rr-\rr',\w_0)  = \delta(\rr-\rr')\id.
\ee
Setting $\rr'=0$, the general expression for the Green's function everywhere except at the origin reads
\be\label{eq:green}
\begin{split}
    \bar{\bar{G}} (\rr,\w_0) \equiv & \frac{e^{\im k_0 r}}{4\pi k_0^2r^3}\Big[(k_0^2r^2+\im k_0 r-1)\id\\
    &+(-k_0^2r^2-3\im k_0r+3)\frac{\rr\otimes\rr}{r^2}\Big].
\end{split}
\ee
At the origin, we define $G({\bf 0})=-i\gamma_0/2$, thereby including the vacuum Lamb shift, proportional to the real part of \cref{eq:G_coupling}, in the definition of the dipole resonance frequency $\w_0$. This corresponds to defining the dispersion relation as $\varepsilon_\pp^{(0)} \equiv \tilde{G}(\pp) - {\rm Re}(G({\bf 0}))$, where $\tilde{G}(\pp)$ is the Fourier transform of \cref{eq:green}.

In the following, we present explicit expressions for the spin-phonon coupling to first and second order in the Lamb-Dicke expansion.

\subsection{First-order spin-phonon coupling}

Using the specific properties of the free-space Green's function, we obtain explicit expressions for the spin-phonon coupling in \cref{eq:H1_vertex} to first order in the Lamb-Dicke approximation. It is convenient to consider the cases of 1D and 2D arrays separately.

\begin{figure*}
    \centering
    \includegraphics[width=2\columnwidth]{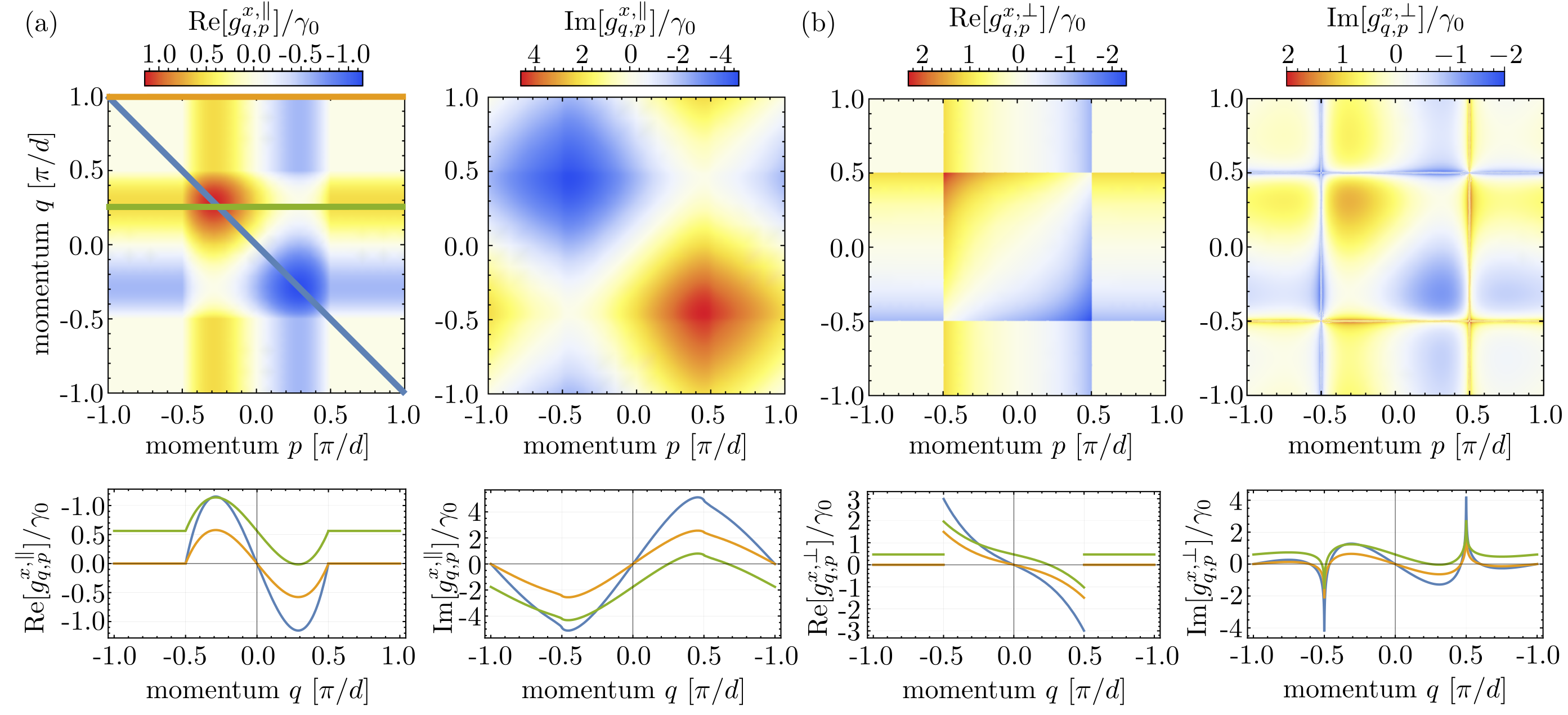}
    \caption{{\bf Spin-phonon coupling in 1D arrays.} Coupling amplitude for atoms polarized (a) parallel and (b) perpendicular to the array axis. In each panel, the top row shows the real (left) and imaginary (right) parts of $g_{p,q}^x$ as a function of $p$ and $q$. Each line in the bottom panels shows cuts along particular directions as specified by the correspondingly colored lines in the top-left plot of panel (a). Note that for perpendicular polarization, $\Im[g_{p,q}^{x,\perp}]$ diverges at the light cone (bottom right panel of (b)). In the top right panel, we capped the plot range, which makes the divergences appear as thin gray regions. All figures are obtained for $d=\lambda_0/4$.}
    \label{fig:Coupling_1D}
\end{figure*}

{\bf 1D arrays.} We consider a one-dimensional atomic array aligned along the $x$-axis. We can obtain analytical expressions using the definition in \cref{eq:H1_vertex} and the expression in \cref{eq:green}. We obtain 
\begin{equation}\label{eq:g_x}
    g_{q,p}^x =\sum_n G^x_{n0}\pare{e^{iq dn}-e^{ip dn}} = G^x(q)-G^x(p),
\end{equation}
and $g^z_{q,p}=g^y_{q,p}=0$, because corrections along directions perpendicular to the array are of second order in the displacement or higher. 
The explicit expression for $G^x(q)$ depends on whether the atoms are polarized perpendicular or parallel to the atomic array. In particular, for perpendicular polarization:
\be \label{eq:G_x_q_perp}
\begin{split}
    G_\perp^x(q)&= \frac{3\gamma_0}{4}\sum_{s=\pm} s\Big[\frac{i\log(1-\lambda_s)}{k_0 d}+ \frac{2\Li_2(\lambda_s)}{(k_0 d)^2}\\
    &+\frac{3 i \Li_3(\lambda_s)}{(k_0d)^3} - \frac{3 \Li_4(\lambda_s)}{(k_0d)^4}\Big],
\end{split}
\ee
while for parallel polarization:
\begin{equation}\label{eq:G_x_q_parallel}
\begin{split}
    G_\parallel^x(q)&= -\frac{3\gamma_0}{2}\sum_{s=\pm} s\Big[\frac{\Li_2(\lambda_s)}{(k_0 d)^2}+3i \frac{\Li_3(\lambda_s)}{(k_0d)^3}-3\frac{\Li_4(\lambda_s)}{(k_0d)^4}\Big],
\end{split}
\end{equation}
where $\lambda_\pm \equiv \exp[i d(k_0\pm q)]$ and $\Li_n(x)$ is the polylogarithm function of order $n$.
In Fig.~\ref{fig:Coupling_1D}(a,b), we plot the real and imaginary parts of \cref{eq:g_x} for both polarization choices.

In both cases, $\Re(g_{p,q}^x)=0$ for $|q|,|p|>k_0$. This is why the resonant contribution to the self-energy in \cref{eq:Sigma_two_terms} is purely imaginary. We also observe that while the coupling is continuous and finite for parallel polarization, it has a discontinuity for perpendicular polarization. Furthermore, in the latter case, $\Im[g_{p,q}^x]$ diverges at the light cone.

{\bf 2D arrays.}
For atomic arrays in two dimensions, it is convenient to use the plane-wave decomposition of the free-space Green's function to write Eq.~(\ref{eq:G_coupling}) as (setting $\rr_\mm=0$)
\begin{align}
G(\rr) = -\frac{3\pi i\gamma_0}{2k_0^3} \int \frac{{\rm d}^2 Q_\parallel}{(2\pi)^2} \frac{k_0^2 - |\bar{\bf Q}\cdot\hat{\dpp}|^2}{ \sqrt{k_0^2 - {\bf Q}_\parallel^2} } e^{i\bar{\bf Q}\cdot\rr},
\label{eq.Greens_tensor_decomposition}
\end{align}
with $\bar{\bf Q} = [{\bf Q}_\parallel, \sqrt{k_0^2 - {\bf Q}_\parallel^2}{\rm sgn}(z)]$. To arrive at this expression, we use the Weyl representation of a spherical wave~\cite{Mandel_1995}
\begin{align}
\frac{e^{ik_0r}}{r} = \frac{i}{2\pi}\int {\rm d}^2Q_\parallel \frac{1}{\sqrt{k_0^2 - Q_\parallel^2}} e^{i\bar{\vec{Q}}\cdot \vec{r}},
\end{align}
with $\bar{\vec{Q}} = [\vec{Q}_\parallel,\sqrt{k_0^2 - Q_\parallel^2} {\rm sgn}(z)]$, as well as the identity~\cite{AsenjoGarcia2017PRX} 
\begin{align}
\bar{\bar{G}}(\vec{r},\omega_0) = \left(k_0^2 \mathbb{1} + \mathbf{\nabla}\otimes\mathbf{\nabla}\right)\frac{e^{ik_0r}}{4\pi k_0^2r}.
\end{align}
Inserting these into \cref{eq:G_coupling} yields \cref{eq.Greens_tensor_decomposition}. Taking the derivative of the latter along a given spatial direction yields
\begin{align}
G^{\alpha}_{\nn{\bf 0}} &= -\frac{3\pi i\gamma_0}{2k_0^4} \int \frac{{\rm d}^2Q_\parallel}{(2\pi)^2} \frac{k_0^2-(\bar{\bf Q}\cdot\hat{\dpp})}{\sqrt{k_0^2-{\bf Q}^2_\parallel}}\partial_\alpha e^{i\bar{\bf Q}\cdot\rr} \nonumber \\
&= \frac{3\pi \gamma_0}{2k_0^4} \int \frac{{\rm d}^2Q_\parallel}{(2\pi)^2} \frac{k_0^2-(\bar{\bf Q}\cdot\hat{\dpp})}{\sqrt{k_0^2-{\bf Q}^2_\parallel}}Q_{\parallel,\alpha} e^{i{\bf Q}_\parallel\cdot\rr}.
\end{align}
We now use the identity 
\begin{equation}\label{eq:Latice_Delta}
    \sum_{\nn} e^{i({\bf Q}_\parallel - \pp)\cdot\rr_\nn} = \Big(\frac{2\pi}{d}\Big)^2 \sum_{{\bf g}\in{\rm RL}} \delta^{(2)}({\bf Q}_\parallel - \pp - {\bf g}),
\end{equation}
where RL is the set of all reciprocal lattice vectors, to obtain the following expression for the spin-phonon coupling rate to first order in the Lamb-Dicke parameter:
\begin{align}\label{eq:g_pq_alpha_free_space}
g^\alpha_{\qq,\pp} = \frac{3\pi\gamma_0}{2k_0^4}\sum_{{\bf g}\in{\rm RL}}\Bigg[ &\frac{k_0^2 - (\tilde{\qq}\cdot\hat{\dpp})^2}{\sqrt{k_0^2 - (\qq + {\bf g})^2}}(q_\alpha + g_\alpha)  \nonumber \\
- &\frac{k_0^2 - (\tilde{\pp}\cdot\hat{\dpp})^2}{\sqrt{k_0^2 - (\pp + {\bf g})^2}}(p_\alpha + g_\alpha) \Bigg].
\end{align}
Here, we let $\tilde{\qq} = [\qq + {\bf g}, \sqrt{k_0^2 - (\qq + {\bf g})^2}{\rm sgn}(z)|_{z = 0}]$, with the convention that ${\rm sgn}(z)^2 = 1$ and ${\rm sgn}(z) = 0$ in the plane of the array ($z = 0$). Equation~\eqref{eq:g_pq_alpha_free_space} is used to efficiently calculate the vertex function.

\subsection{Correction to zero-point motion}

Simple expressions for the zero-point motion corrections to the dispersion relation, Eq.~\eqref{eq:Correction_fast_motion_General}, can only be obtained in special cases. In the following, we focus on two such cases: the case of isotropic trap frequencies, which we assume in Sec.~\ref{sec:transport}, and the case where two trap frequencies are equal, which we use in Sec.~\ref{sec:input-output}.

For isotropic trap frequencies, such that $\eta_\alpha = \eta$ for all spatial directions $\alpha = x,y,z$, the second-order Lamb-Dicke correction to the dispersion relation takes the simple form
\begin{equation}\label{eq:H2_Approx_appendix}
    \Hop_2 = -\eta^2\sum_{\pp} \left[\varepsilon_\pp^{(0)} + \frac{i\gamma_0}{2}\right] \sdop_\pp \sop_\pp.
\end{equation}
Together with the zeroth-order result, $\varepsilon_\pp^{(0)}$, this shows that in the isotropic trapping limit, \cref{eq:epsilon_infinity_general} reads
\begin{equation}\label{eq:Isotropic_Fast_Motion_Dispersion}
    \varepsilon^{\infty}_\pp = (1 - \eta^2)\varepsilon_\pp^{(0)} - \frac{i\eta^2\gamma_0}{2}.
\end{equation}
We emphasize that this result holds for one-, two-, and three-dimensional atomic arrays in three-dimensional vacuum space.
Equation~\eqref{eq:H2_Approx_appendix} is obtained by noting that away from the origin ($\rr=0$), the Green's function satisfies $\nabla(\nabla\cdot \bar{\bar{G}})=0$, as can be immediately verified using \cref{eq:green}.
From \cref{eq:Helmholtz_Eq}, we then have exactly~\footnote{The original derivation of \cref{eq:H2_Approx_appendix} presented in Ref.~\cite{Guimond2019} considered only the far-field contribution to the Green's function in \cref{eq:green}. We find, however, that this is an unnecessary assumption.}
\begin{equation}
\begin{split}
    \sum_\alpha G^{\alpha\alpha}_{\nn\mm} &= \sum_\alpha k_0^{-2}\partial^2_{\alpha} G(\rr)|_{\rr = \rr_\nn-\rr_\mm}\\
    &= G(\rr)|_{\rr = \rr_\nn-\rr_\mm}= -G_{\nn\mm}.
\end{split}
\end{equation}
Substituting this result into \cref{eq.app_H2_general_expression} finally yields \cref{eq:H2_Approx_appendix}.
Equation~\eqref{eq:Isotropic_Fast_Motion_Dispersion} corresponds to the modified dispersion relation for a finite Lamb-Dicke parameter but in the limit of fast motion, $\nu\gg \gamma_0$, such that the contribution of the spin-phonon scattering in \cref{eq:H1_crystal_momenta} can be neglected.

Let us now consider a 2D array of atoms where the trap frequencies for motion in the array plane are equal ($\eta_x=\eta_y=\eta_\parallel \neq \eta_z$). To obtain concrete expressions, we focus here on the case of circularly polarized dipoles (in the plane of the array), which corresponds to the choice made in Sec.~\ref{sec:input-output}. To calculate this in the easiest possible manner, we first rewrite \cref{eq:Correction_fast_motion_General} as 
\begin{equation}\label{eq:Delta_Epsilon_p_mirror}
\begin{split}
\Delta \varepsilon_\pp &=  -\eta_\parallel^2 \varepsilon_\pp^{(0)} - \frac{i\eta_\parallel^2\gamma_0}{2}\\
&+ (\eta_z^2 - \eta_\parallel^2)\spare{ \sum_{\nn}  G^{zz}_{\nn{\bf 0}} e^{-i\pp\cdot\rr_\nn} - G^{zz}_\mathbf{00}}.
\end{split}
\end{equation}
There is, thus, a correction for $\eta_z \neq \eta_\parallel$, which we must compute independently. We will consider the specific case of circularly polarized atoms, $\hat{\dpp} =(\mathbf{e}_x + i\mathbf{e}_y)/\sqrt{2}$, corresponding to the situation analyzed in Sec.~\ref{sec:input-output}. A similar procedure can be applied to other polarizations. We start by computing the first term in the square brackets. Using the plane-wave decomposition in \cref{eq.Greens_tensor_decomposition}, we obtain 
\begin{equation}\label{eq:divergent_summation_Gzz}
\begin{split}
\sum_{\nn} G^{zz}_{\nn{\bf 0}} e^{-i\pp\cdot\rr_\nn} 
=  \frac{3\pi i\gamma_0}{2k_0^5d^2}\sum_{{\bf g}\in{\rm RL}}&  \sqrt{k_0^2 - (\pp + {\bf g})^2} \\
\times& \spare{k_0^2 - \frac{(\pp + {\bf g})^2}{2}}.
\end{split}
\end{equation}
This contribution is divergent, because its real part becomes arbitrarily large as $|{\bf g}|$ increases. We note, however, that \cref{eq:Delta_Epsilon_p_mirror} is finite, because the divergence in \cref{eq:divergent_summation_Gzz} is canceled by the second term, $G_\mathbf{00}^{zz}$.
Computing the contribution in \cref{eq:Delta_Epsilon_p_mirror} as the difference of two diverging terms is, however, not practically feasible. We thus employ a regularization procedure inspired by Ref.~\cite{Perczel2017}.
We introduce into \cref{eq.Greens_tensor_decomposition} a Gaussian cutoff $e^{-(\pp + {\bf g})^2/\Lambda^2}$ at a momentum range $\Lambda$. 
With this choice, \cref{eq:Delta_epsilon_p_reg} reads
\begin{equation}\label{eq:Delta_epsilon_p_reg}
\begin{split}
\!\!&\Delta \varepsilon_\pp
= -\eta_\parallel^2 \varepsilon_\pp^{(0)} - \frac{i\eta_\parallel^2\gamma_0}{2} + (\eta_z^2-\eta_\parallel^2)  \\
&\times \lim_{\Lambda \to \infty} \Bigg\{ \frac{3\pi i\gamma_0}{2k_0^5d^2}\sum_{{\bf g}\in {\rm RL}}\sqrt{k_0^2 - (\pp + {\bf g})^2}\\
&\times\left(k_0^2 - \frac{[\pp + {\bf g}]^2}{2}\right)e^{-(\pp + {\bf g})^2/\Lambda^2} - G^{zz}_\mathbf{00}(\Lambda)\Bigg\}.\!\!
\end{split}
\end{equation}
The last term in the curly brackets represents the regularization of the term $G^{zz}_\mathbf{00}$ upon the introduction of the cutoff. It is defined as
\begin{align} \label{eq.G_zz_Lambda}
&G^{zz}_{{\bf 00}}(\Lambda) =\! \frac{3\pi i \gamma_0}{2k_0^5} \!\int\!\! \frac{{\rm d}^2 Q_\parallel}{(2\pi)^2} \sqrt{k_0^2\! -\! {\bf Q}_\parallel^2}\pare{k_0^2\! -\! |\bar{\bf Q}\cdot\hat{\dpp}|^2} e^{-{\bf Q}_\parallel^2/\Lambda^2} \nonumber \\
&=\! \frac{3 i \gamma_0}{4}\int_0^\infty dx \, x\sqrt{1-x^2}(1-x^2/2)e^{-k_0^2x^2/\Lambda^2} \nonumber \\
&=\! \frac{3 i \gamma_0}{8} e^{-k_0^2/\Lambda^2} \Big[\int_0^1 du \, u^2(1+u^2)e^{+k_0^2u^2/\Lambda^2} \nonumber \\
&\phantom{\frac{3\pi i \gamma_0}{4} e^{-k_0^2/\Lambda^2} } + i\int_0^\infty dv\, v^2(1-v^2) e^{-k_0^2 v^2/\Lambda^2}\Big].
\end{align}
In the second line, we define $x = Q_\parallel / k_0$. In the third line, we let $u^2 = 1 - x^2$ on the interval $x \in [0,1]$, whereas we let $v^2 = x^2 - 1$ on the interval $x\in[1,\infty]$. 
Only the first integral in the square brackets contributes to the imaginary part of $\Delta \varepsilon_\pp$ and, in the limit $\Lambda \to +\infty$, yields the contribution
\begin{align}\label{eq:Im_G00_zz}
{\rm Im}[G^{zz}_{{\bf 00}}(\Lambda)] &\simeq \frac{\gamma_0}{5},
\end{align}
which is independent of the cutoff. This is expected, as the imaginary part of \cref{eq.Greens_tensor_decomposition} at the origin, as well as its first and second spatial derivatives, are always finite.
Substituting \cref{eq:Im_G00_zz} into \cref{eq:Delta_epsilon_p_reg}, the shift to the decay rate reads (taking $\Lambda \to +\infty$):
\begin{align}\label{eq:Im_Delta_epsilon_reg}
&-2{\rm Im}(\Delta \varepsilon_\pp)
=\eta_\parallel^2 (\gamma_0 \!-\! \Gamma_\pp^{(0)}) \!+\! (\eta_z^2\!-\!\eta_\parallel^2 ) \Bigg\{ \frac{2\gamma_0}{5} \nonumber \\
&-\!\frac{3\pi \gamma_0}{k_0^5d^2}\!\!\!\!\sum_{\substack{{\bf g}\in {\rm RL} \\ |\pp + {\bf g}| < k_0}} \!\!\!\!\sqrt{k_0^2 \!-\! (\pp \!+\! {\bf g})^2} \left[k_0^2 - \frac{(\pp + {\bf g})^2}{2}\right] \Bigg\}.
\end{align}
Note that the only contribution to the second term in the curly brackets comes from those reciprocal lattice vectors satisfying $|\pp+\mathbf{g}|<k_0$.
Finally, the contribution to the real part of $\Delta\varepsilon_\pp$ comes from the second integral in \cref{eq.G_zz_Lambda} and gives
\begin{align}\label{eq:Re_Gzz_reg}
\!\!{\rm Re}[G^{zz}_{{\bf 00}}(\Lambda)] &= -\frac{3 \sqrt{\pi} \gamma_0}{64} e^{-k_0^2/\Lambda^2} \frac{\Lambda^3}{k_0^3}\bigg(2 \!-\! 3\frac{\Lambda^2}{k_0^2}\bigg).
\end{align}
Substituting \cref{eq:Re_Gzz_reg} into \cref{eq:Delta_epsilon_p_reg}, we obtain an expression that is convergent in the limit $\Lambda\to\infty$:
\begin{align}\label{eq:Re_Delta_epsilon_reg}
{\rm Re}(\Delta \varepsilon_\pp)
&=-\eta_\parallel^2 J_\pp^{(0)} \!-\! (\eta_z^2\!-\!\eta_\parallel^2 ) \nonumber \\
\times&\Bigg\{\!\frac{3\pi \gamma_0}{2k_0^5d^2}\!\!\!\!\sum_{\substack{{\bf g}\in {\rm RL} \\ |\pp + {\bf g}| > k_0}} \!\!\!\!\sqrt{(\pp \!+\! {\bf g})^2 \!-\!k_0^2} \left[k_0^2 - \frac{(\pp + {\bf g})^2}{2}\right] \nonumber\\
&\;\;+{\rm Re}[G^{zz}_{{\bf 00}}(\Lambda)] \Bigg\}.
\end{align}

In Sec.~\ref{sec:input-output}, we use Eqs.~\eqref{eq:Im_Delta_epsilon_reg}--\eqref{eq:Re_Delta_epsilon_reg} to evaluate the zero-point motion correction in \cref{eq:Delta_Epsilon_p_mirror}.

\section{Calculation of self-energy and quasiparticle weight} \label{app:calculation_self_energy}

In this appendix, we provide additional details for the computation of the quasiparticle properties of the polaron-polariton excitation discussed in \ref{sec:subwavelength_arrays}. In particular, we describe in detail how we compute the self-energy, $\Sigma_\pp$, and the quasiparticle weight, $Z_\pp$. It is necessary to distinguish the cases of arrays in one and two dimensions.

\subsection{1D arrays}

{\bf Self-energy.} For a 1D free-space array, the self-energy is obtained from~\cref{eq:second_order_energy_D_dimension} by taking $D=1$. It reads
\begin{equation} \label{eq:Self_Energy_1D}
\begin{split}
\Sigma_p^{1\text{D}} \equiv 
- \eta^2 d \int_{-\pi/d}^{\pi/d} \frac{\text{d}q}{2\pi}\frac{(g^x_{q,p})^2}{\varepsilon_p^{(0)} - \varepsilon_q^{(0)} - \nu + i0^+},
\end{split}
\end{equation}
where we assumed the array to be aligned along the $x$-axis.
To correctly evaluate \cref{eq:Self_Energy_1D}, it is essential to handle the divergences in the integrand carefully. These are of two types.

First, poles arise from a divergence in $g^x_{q,p}$ and $\varepsilon_q^{(0)}$ for $q=k_0$ in the case of atoms polarized perpendicular to the array direction.
To evaluate this contribution, we expand the integrand in \cref{eq:Self_Energy_1D} around $q=k_0$ and obtain
\be\label{eq:Expansion_Integrand_k0}
    \frac{(g^x_{q,p})^2}{\varepsilon_p^{(0)} - \varepsilon_q^{(0)} - \nu} \simeq -\frac{[G^x_\perp(q)]^2}{\varepsilon_q^{(0)}}.
\ee
Here, we kept only the leading terms in both the numerator and denominator and used \cref{eq:H1_vertex} to approximate $g^x_{q,p}\simeq G^x_\perp(q) \equiv \sum_n G^x_{n0} e^{i q d n}$, where $G^x_\perp(q)$ is given in \cref{eq:G_x_q_perp}.
The dispersion relation of a polariton in a 1D array of atoms polarized perpendicular to the array axis reads~\cite{AsenjoGarcia2017PRX} 
\be\label{eq:epsilon_0_q}
    \varepsilon_q^{(0)} = \frac{3\gamma_0}{4}\sum_{s=\pm}\spare{\frac{\log(1-\lambda_s)}{k_0 d} - i \frac{\Li_2(\lambda_s)}{(k_0 d)^2} + \frac{\Li_3(\lambda_s)}{(k_0 d)^3}},
\ee
where $\lambda_\pm \equiv \exp[i d(k_0\pm q)]$ and $\Li_n(x)$ is defined after \cref{eq:G_x_q_parallel}.
Substituting Eqs.~\eqref{eq:epsilon_0_q} and \eqref{eq:G_x_q_perp} into \cref{eq:Expansion_Integrand_k0}, we obtain a logarithmic divergence for $q\rightarrow k_0$ at leading order, arising from the first term in the square brackets in Eqs.~\eqref{eq:epsilon_0_q} and \eqref{eq:G_x_q_perp}. 
This pole is thus integrable and does not contribute to \cref{eq:Self_Energy_1D}.

Second, poles also arise from the divergence of the denominator in \cref{eq:Self_Energy_1D}. For $\nu\neq0$, this occurs only for $|p|,|q|>k_0$ such that $\Im[\varepsilon_p^{(0)}]=0= \Im[\varepsilon_q^{(0)}]$, and it corresponds to the resonant scattering between two subradiant polaritons via the creation of a phonon.
Given the parity symmetry of the dispersion relation, for any given value $|p|>k_0$, the resonant wave vectors always come in pairs $q=\pm k_r$, corresponding to a pole in \cref{eq:Self_Energy_1D}. 
Using the result 
\be\label{eq:Identity_SP}
\frac{1}{x\pm i0^+} = \mp i\pi \delta(x)+\text{P}\pare{\frac{1}{x}},
\ee
we rewrite \cref{eq:Self_Energy_1D} as
\be\label{eq:Self_Energy_1D_Explicit}
\begin{split}
    \Sigma_p^{1\text{D}} = \Sigma_p^\text{res} + \Sigma_p^\text{off-res},
\end{split}
\ee
where we defined the resonant contribution 
\be\label{eq:Self_Energy_1D_res}
    \Sigma_p^\text{res} = \sum_{k_r\in \mathcal{R}}\frac{i\eta^2}{2}\frac{(g^x_{p,k_r})^2}{|\varepsilon'_{k_r}|},
\ee
with $\varepsilon'_{k_r}\equiv (\text{d}\varepsilon_q^{(0)}/\text{d}q)_{q=k_r}$, and $\mathcal{R}$ represents the set of wave vectors satisfying the resonance condition.
The off-resonant contribution in Eq.~\eqref{eq:Self_Energy_1D_Explicit} reads
\be\label{eq:Self_Energy_off-res}
    \Sigma_p^\text{off-res} =
    - \eta^2 d \, \mathrm{P}\int_{-\pi/d}^{\pi/d}\frac{\text{d}q}{2\pi} \frac{(g^x_{q,p})^2}{\varepsilon_p^{(0)} - \varepsilon_q^{(0)} - \nu},
\ee
where $\mathrm{P}$ indicates the Cauchy principal value of the integral.

The numerical evaluation of \cref{eq:Self_Energy_1D_res} and \cref{eq:Self_Energy_off-res} requires particular care in the vicinity of the light cone and the resonance points. Special attention must be devoted to the evaluation of the resonant momenta $\pm k_r$ for the case of perpendicularly polarized atoms. In this case, when the resonant quasimomentum falls close to the light line (where the dispersion relation diverges), greater accuracy is required.

{\bf Quasiparticle weight.} The quasiparticle weight $Z_p$ is obtained from \cref{eq:residue} by computing the numerical derivative of $\Sigma_p(\w)$. 

\subsection{2D arrays} \label{app:pole_contribution_2D}

The contribution to the second-order energy shift from the self-energy for a 2D array is obtained from \cref{eq:second_order_energy_D_dimension} by taking $D=2$:
\be\label{eq:Self_Energy_2D_App}
\Sigma_\pp(\varepsilon^{(0)}_\pp)= - \eta^2 d^2 \int_{\rm BZ}\frac{{\rm d}^2q}{(2\pi)^2} \frac{\sum_\alpha(g^\alpha_{\qq,\pp})^2}{\varepsilon^{(0)}_\pp - \varepsilon^{(0)}_\qq - \nu + i0^+}.
\ee
To numerically compute Eq.~\eqref{eq:Self_Energy_2D_App}, it is beneficial to write the integral in polar coordinates because a pole can appear as a function of $q = |\qq|$ for a fixed polar angle $\varphi$. Similar to the 1D array, we obtain resonant contributions,
\be\label{eq:2D_resonant_self_energy}
\Sigma_\pp^\text{res} = \frac{i \eta^2 d^2 }{2}\int_0^{2\pi} \frac{{\rm d}\varphi}{2\pi}
\sum_{\qq \in \mathcal{R}(\varphi)} \frac{q 
\, \sum_\alpha (g^\alpha_{\qq,\pp})^2}{|\partial_q \varepsilon^{(0)}_\qq|},
\ee
and off-resonant contributions
\be \label{eq:2D_off_resonant_self_energy}
\!\!\Sigma_\pp^\text{off-res} = -\eta^2 d^2 \! \int_0^{2\pi}\!\frac{\text{d}\varphi}{2\pi} \, {\rm P}\!\int_0^{q_{\max}}\!\frac{\text{d}q}{2\pi} \frac{q\, \sum_\alpha (g^\alpha_{\qq,\pp})^2}{\varepsilon^{(0)}_\pp \!-\! \varepsilon^{(0)}_\qq \!-\! \nu}. 
\ee
Here, $q_{\max}d = \min( \pi / |\cos \varphi|, \pi / |\sin \varphi|)$ is the maximum crystal momentum within the first Brillouin zone. Moreover, $\mathcal{R}(\varphi)$ denotes the set of resonances $\qq(\varphi)$ at a fixed angle $\varphi$ such that $\varepsilon^{(0)}_\pp = \varepsilon^{(0)}_\qq + \nu$. Strictly speaking, Eq.~\eqref{eq:2D_resonant_self_energy} is only correct for resonances outside the light cone, where $\varepsilon^{(0)}_\pp$ and $ \varepsilon^{(0)}_\qq$ are real. Indeed, a complex analysis argument shows that when $\varepsilon^{(0)}_\qq$ is complex at the resonance, the correct formula is
\be\label{eq:2D_resonant_self_energy_2}
\Sigma_\pp^\text{res} = \frac{i \eta^2 d^2 }{2}\int_0^{2\pi} \frac{{\rm d}\varphi}{2\pi}
\sum_{\qq \in \mathcal{R}(\varphi)} \!\! \frac{q 
\, \sum_\alpha (g^\alpha_{\qq,\pp})^2}{\partial_q \varepsilon^{(0)}_\qq} {\rm sgn}[\Re(\partial_q \varepsilon^{(0)}_\qq)].
\ee
This reduces to Eq.~\eqref{eq:2D_resonant_self_energy} when $\partial_q \varepsilon^{(0)}_\qq$ is real. As we mainly focus on the corrections to the subradiant states, Eq.~\eqref{eq:2D_resonant_self_energy} generally suffices for our purposes.

The positions of the resonances are found numerically. The principal value integral as well as the pole contribution are then evaluated for enough polar angles $\varphi$ to resolve its angular dependency. For perpendicularly polarized dipoles, the angular dependence is mild and around $100$ angles are adequate to obtain precise results. For polarizations in the plane of the array, the position of the resonances strongly depends on the polar angle, and $10^3$--$10^4$ sampled angles are used for the computations.

Moreover, special care has to be taken in the evaluation of the integral in \cref{eq:2D_off_resonant_self_energy}, because there are poles in both the interaction vertex $g^\alpha_{\qq,\pp}$ and the energy $\varepsilon_\qq^{(0)}$ at $|\qq| = k_0$. These poles arise from the divergence of the Green's function at the light cone. We note that this divergence is always present in 2D, whereas in 1D it only exists for perpendicular polarization. However, this divergence is integrable and does not contribute to the self-energy.

\section{Quasiparticle properties of polaron-polaritons in 1D arrays} \label{app:Polaron-Polariton_1D}

\begin{figure*}
    \centering
    \includegraphics[width=2\columnwidth]{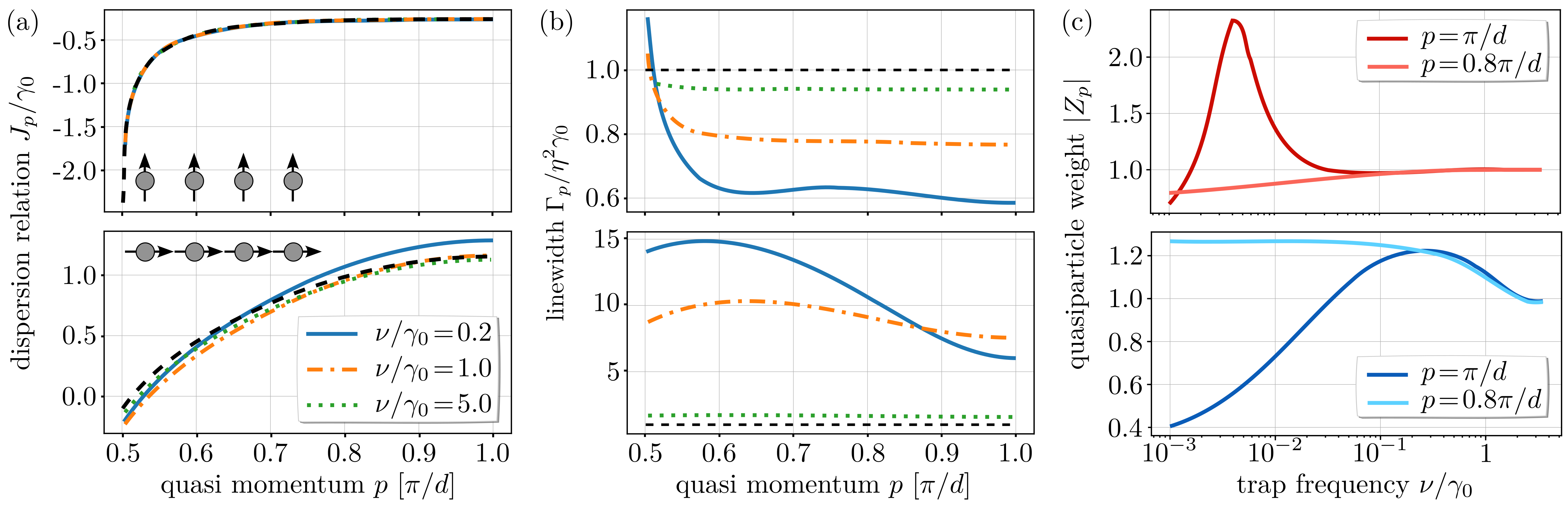}
    \caption{{\bf Quasiparticle properties of polaron-polaritons in 1D arrays.} Top and bottom panels represent the cases of atoms polarized perpendicular and parallel to the array axis, respectively. (a) Dispersion relation in the subradiant sector for pinned atoms (dashed black line) and $J_p ={\rm Re}(E_p)$ for finite trap frequencies $\nu/\gamma_0 = 0.2$ (solid blue line), $1.0$ (orange dot-dashed line), and $5.0$ (green dotted line). (b) Decay rate $\Gamma_p = -2\text{Im}(E_p)$ in the subradiant sector $p>k_0$ of a polaron-polariton for different trap frequencies as specified in the legend in panel (a). The black dashed line represents the decay rate in the fast-motion limit, $\eta^2\gamma_0$. (c) Absolute value of the quasiparticle weight, $|Z_p|$, as a function of the trap frequency for a polaron-polariton with quasimomentum $p=\pi/d$ and $p=0.8\pi/d$. Other parameters: $d=\lambda_0/4$ and $\eta=0.2$.}
    \label{fig:Polaron_1D}
\end{figure*}

In this appendix, we analyze the quasiparticle properties of polaron–polaritons in 1D subwavelength arrays, complementing the discussion of 2D arrays in Sec.~\ref{sec:subwavelength_arrays}. For a chain oriented along the $x$-axis, the resonant and off-resonant contributions in \cref{eq:Sigma_two_terms} take the explicit form:
\bea
    \Sigma_p^\text{res} &=&\frac{i\eta^2}{2} \sum_{k \in \mathcal{R}} \frac{(g^x_{p,k})^2}{|\varepsilon'_{k}|},\label{eq:main_Self_Energy_1D_res}\\
    \Sigma_p^\text{off-res} &=&
    - \eta^2 d \, \mathrm{P}\int_{-\pi/d}^{\pi/d}\frac{\text{d}q}{2\pi} \frac{(g^x_{q,p})^2}{\varepsilon_p^{(0)} - \varepsilon_q^{(0)} - \nu}.\label{eq:main_Self_Energy_1D_offres}
\eea
Here, $\mathcal{R}$ is the set of wave vectors satisfying the resonance condition $\varepsilon^{(0)}_k + \nu = \varepsilon^{(0)}_{p}$, and $1 / |\varepsilon'_{k}| = (\text{d}\varepsilon_q^{(0)}/\text{d}q)^{-1}_{q=k}$ is proportional to the density of states at resonance. For parallel dipole polarization, resonances are absent if $\nu > J^{(0)}_{p} - J^{(0)}_{k_0}$. In this case, $\Sigma_p^\text{res} = 0$ and the integrand in Eq.~(\ref{eq:main_Self_Energy_1D_offres}) has no divergences. In contrast, for perpendicular polarization, resonances always exist because the dispersion diverges near the light cone.

We evaluate Eqs.~\eqref{eq:main_Self_Energy_1D_res} and \eqref{eq:main_Self_Energy_1D_offres} numerically (see Appendix~\ref{app:calculation_self_energy}) and substitute the results into \cref{eq:second_order_energy_general} to obtain the modified eigenenergies $E_p$. In Figures~\ref{fig:Polaron_1D}(a--b), we show the resulting dispersion relation $J_p\equiv \Re(E_p)$ and linewidth $\Gamma_p\equiv -2\Im[E_p]$ as a function of the quasimomentum $p$ for several values of $\nu/\gamma_0$ and for perpendicular and parallel polarization. 

The dispersion relation $J_p$ remains remarkably robust even as the trap frequency decreases [Fig.~\ref{fig:Polaron_1D}(a)], due to the same mechanism as for 2D arrays: the mismatch in energy or decay between a polariton with momentum $p$ and a polariton-phonon pair with momenta $q$ and $p-q$, respectively, is much larger than the coupling for most modes, leading to suppressed scattering.

The decay rate of subradiant states [Fig.~\ref{fig:Polaron_1D}(b)] is modified by changes in the radiative decay, Eqs.~\eqref{eq:Isotropic_Fast_Motion_Dispersion} and \eqref{eq:main_Self_Energy_1D_offres}, and the appearance of additional non-radiative decay channels, which can significantly change the decay rate, Eq.~\eqref{eq:main_Self_Energy_1D_res}. The radiative decay rate depends on the interference between two processes: (1) the decay of $\ket{p, 0}$ under creation of a phonon by the recoil and (2) the decay from the state $\ket{q, p-q}$ without creating an additional phonon through the recoil.
In order to understand the interference between these two contributions, it is illustrative to evaluate the anti-Hermitian part of \cref{eq:H_expansion} on the Chevy ansatz \cref{eq:Chevy}, which yields:
\be \label{eq:app_decay_1D}
\begin{split}
-\frac{\im \Gamma_p^\infty}{2} |B_p|^2 - \frac{2 \im}{\sqrt{N}} \sum_q \Re(g_{q, p}^x) \Im\left(C_{q, p}^x\right) \\
- \sum_q \frac{\im \Gamma_q^\infty}{2} |C_{q, p}^x|^2.
\end{split}
\ee
The first and third terms describe decay associated with populating the polariton modes with momenta $p$ or $q$, respectively. The second term captures the interference between the phonon-producing processes mentioned above, and its sign determines whether the interference suppresses or enhances decay. From \cref{eq.polaron_state}: 
\be \label{eq:app_C_pq}
C_{q, p}^x = \frac{1}{\sqrt{N}} \frac{\eta_x g^x_{p,q}}{\varepsilon^{(0)}_p \!-\! \varepsilon^{(0)}_q \!-\! \nu_x}.
\ee
In general, the sign of Eq.~\eqref{eq:app_C_pq} depends on both $g_{p,q}^x$ and the denominator. 
In the fast-motion limit $\nu \gg \gamma_0$, this reduces to $C_{q, p}^x \approx - \eta_x g^x_{p,q} / (\sqrt{N} \nu_x)$, so that the sign of the interference is set entirely by the polarization-dependent couplings $g^x_{q, p}$. In this case, whether $\Gamma_p$ is larger or smaller than $\eta^2\gamma_0$ depends entirely on the sign of the dipole force between nearby atoms. In the subradiant sector, the dipole moments of neighboring atoms are approximately anti-aligned, so that when atoms are polarized parallel (perpendicular) to the array axis, the dipole force between nearby atoms is repulsive (attractive).

Non-radiative decay arises from resonant scattering via Eq.~\eqref{eq:main_Self_Energy_1D_res}. For parallel polarization, this channel is absent for large trap frequencies but dominates for intermediate ($\nu \sim \gamma_0$) or small ($\nu \ll \gamma_0$) trap frequencies.
For perpendicular polarization, a resonance always exists due to the diverging dispersion relation at the light cone. However, for large trap frequencies, the corresponding resonant points lie near $\pm k_0$, where the density of states vanishes, $1/|\varepsilon'_{k_r}|\approx 0$. 
We highlight that the combined decay (radiative and non-radiative) in Fig.~\ref{fig:Polaron_1D}(b) (top panel) is smaller than the decay in the fast-motion limit, $\eta^2 \gamma_0$, for all momenta $|p| \gtrsim 1.05~k_0$, despite the presence of the additional non-radiative decay channel.

Finally, Fig.~\ref{fig:Polaron_1D}(c) shows the quasiparticle weight $Z_p$ as a function of $\nu/\gamma_0$ for $p=0.8\pi/d$ and $p=\pi/d$. (Details on the underlying calculation are given in Appendix~\ref{app:calculation_self_energy}.) We find that $Z_p$ remains close to $1$ for $p=0.8 \pi / d$ throughout the entire range of trap frequencies, confirming the validity of the Chevy ansatz, \cref{eq:Chevy}. For $p=\pi / d$, $Z_p$ is close to unity for large and intermediate trap frequencies but deviates from $1$ significantly for small trap frequencies, indicating that our approach breaks down in this specific regime. 
For both polarizations, the dispersion relation has a maximum at the edge of the Brillouin zone \footnote{For perpendicularly polarized atoms, an additional local extremum of the dispersion relation appears for $k_0<|p|<\pi/d$ when $d/\lambda_0\lesssim 0.24$, leading to an additional divergence in the density of states. We do not consider this case here.}. 
In the limit $\nu/\gamma_0\rightarrow 0$, a polariton with $p=\pi / d$ can scatter resonantly into modes with $q \rightarrow \pm \pi / d$. This process is enhanced by the diverging density of states but simultaneously suppressed by vanishing coupling matrix elements. The coupling scales as $g_{q, \frac{\pi}{d}}^x \sim \im \gamma_0 (q - \frac{\pi}{d}) d$ for $q \rightarrow \frac{\pi}{d}$, and since the crystal momenta $\pi /d$ and $-\pi / d$ are physically equivalent, the coupling vanishes analogously for $q \rightarrow -\pi / d$. 
Expanding the dispersion $J_p$ around $p=\pm\pi / d$, we find that resonances occur for 
\be
k_\mathrm{r} \approx \frac{\pi}{d} \pm \sqrt{\frac{2 \nu}{|(\partial_p^2 J_p)_{p=\pm\pi/d}|}}
\ee
and the resonant contribution scales as $\Sigma^\mathrm{res}_p \sim \sqrt{\nu}$ for $\nu \rightarrow 0$.
Since $\partial_\omega \Sigma(\omega)|_{\omega=\varepsilon_p^{(0)}} = - \partial_{\nu} E_p$, the quasiparticle weight given in \cref{eq:residue} characterizes how sensitively the self-energy varies with $\nu$. The scaling of the resonant contribution with $\sqrt{\nu}$ leads to $Z_p \sim \sqrt{\nu}$ for $\nu \rightarrow 0$.
For perpendicular polarization, we observe that in addition to $Z_p \rightarrow 0$ for $\nu \rightarrow 0$, the quasiparticle weight exceeds $1$ significantly for small values of $\nu$. Due to the shape of the dispersion $J_p$ for perpendicular polarization, phonon scattering is weak for large and intermediate trap frequencies and rapidly surges once the density of states at resonance becomes large. This produces the pronounced peak in the upper panel of Fig.~\ref{fig:Polaron_1D}(c), which collapses once the coupling $g^x_{k_\mathrm{r}, p}$ diminishes for $k_\mathrm{r} \rightarrow \pm \pi / d$. 
In contrast, for $p=0.8\pi/d$, the quasiparticle weight remains close to one for all values of $\nu/\gamma_0$. In this case, the resonant contribution remains finite also in the limit $\nu \rightarrow 0$, since only the coupling to the resonance at $k_{\mathrm{r}} \approx p$ vanishes, while the coupling to the resonance at $k_\mathrm{r} \approx -p$ and the density of states at resonance remain finite. 

\section{Transport in free-space arrays}\label{app:Simulation_Transport}

In this appendix, we discuss additional details behind the results on transport presented in Sec.~\ref{sec:transport}.

\subsection{Details on the numerical simulations} \label{app:details_simulation}
\begin{figure}[t]
    \centering
    \includegraphics[width=\columnwidth]{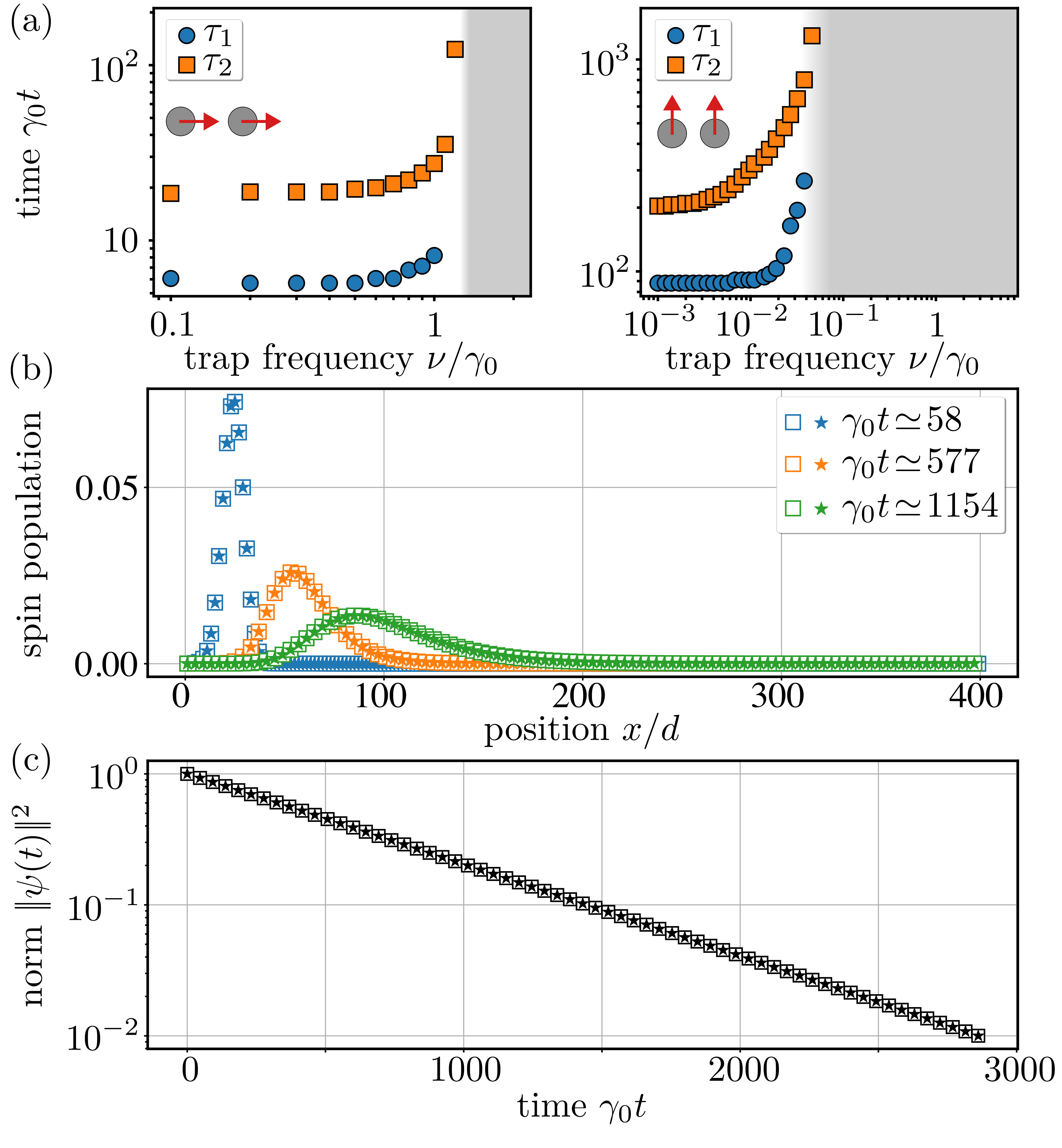}
    \caption{{\bf Details on the numerical simulations.}
    (a) Plot of the timescales $\tau_1$ and $\tau_2$ (as defined in the main text) for arrays with parallel (left panel) and perpendicular (right panel) polarization. (b,c) Comparison of the results obtained using the approximation in \cref{eq:H2_approx_simul} (square markers) and \cref{eq:H2_approx_polaron} (star markers) for the trap frequency $\nu/\gamma_0=0.2$. (b) Plot of the normalized spin population in \cref{eq:Spin_Distribution} at three different points in time as specified by the legend. (c) Time evolution of the norm $||\ket{\psi(t)}||^2$. Initial wave packet position $x/d=20$. Other parameters are chosen as in the caption of Fig.~\ref{fig:Transport_1D_perpendicular}.}
    \label{fig:Phonon_Hopping}
\end{figure}
Transport of an excitation in 1D and 2D arrays is computed by numerically integrating the Schrödinger equation with the non-Hermitian Hamiltonian 
\be \label{eq:H_expansion_simulations}
\Hop \simeq \Hop_0 + \eta\sum_\alpha \Hop^{\alpha}_1 + \frac{\eta^2}{2}\sum_{\alpha,\beta}  \Hop^{\alpha\beta}_{2},
\ee
where the terms $\Hop_0$, $\Hop^\alpha_1$, and $\Hop^{\alpha\beta}_{2}$ are given respectively by the real-space expressions in Eqs.~\eqref{eq:H0_diagonal}, \eqref{eq:H1_crystal_momenta}, and \eqref{eq:H2_appendix}. To simulate large system sizes, we reduce the dimensionality of the Hilbert space by projecting \cref{eq:H_expansion_simulations} onto the manifold containing at most one spin and one phonon excitation. 

While the spin population cannot grow as Eq.~\eqref{eq:H_expansion_simulations} commutes with $\hat{n}_\text{spin}=\sum_j\spl_j\smi_j$, phonons are instead created or destroyed by atomic recoil and dipole-forces. Thus, truncating at one phonon sets a critical time $t_c$, after which the simulations are no longer reliable due to a large phonon population. We define $t_c = \min(\tau_1, \tau_2)$, where $\tau_1$ is the time at which the phonon population reaches $0.1$, and $\tau_2$ indicates the time at which the ratio between the population in the one-phonon sector and the zero-phonon sector reaches $0.5$. The critical time $t_c$ depends on the trap frequency $\nu/\gamma_0$ as shown in Fig.~\ref{fig:Phonon_Hopping}(a), since the resonant phonon-scattering rate generically increases for smaller trap frequencies. Note that the absence of markers above a certain trap frequency (gray shaded region) corresponds to $\tau_1$ and $\tau_2$ exceeding the time of the simulations. Accordingly, for those values of $\nu/\gamma_0$, the phonon population never grew above the threshold value defined above during the simulated time-scale $\gamma_0t$.

For small values of $\nu/ \gamma_0\ll 1$, the simulation time exceeds $t_c$.
At time scales larger than $t_c$, the dynamics in the one phonon sector obtained from the numerical simulations cannot be trusted, since scattering processes into the two-phonon sector are discarded by the truncation of the Hilbert space, which may affect the dynamics in the one-phonon sector.
However, the dynamics of the zero-phonon sector is valid at much longer time scales than $t_c$: The rate at which excitations are scattered from the one-phonon sector to the zero-phonon sector is given by $\sim\eta^2\gamma_0$ scaled by the phonon excitation-density. This implies that errors in the dynamics of the one-phonon sector could only affect the dynamics in the zero-phonon sector at times much larger than $t_c$. Furthermore, the direct coupling between the zero-phonon sector and two-phonon sector in Eq.~\eqref{eq:H_expansion_simulations} would induce effects of order $O(\eta^4 \gamma_0)$, which lead to negligible corrections during the simulated time.

The projection on the single phonon sector removes all terms that create or destroy two phonons (i.e., terms of the form $\bdop_{\alpha,\nn}\bdop_{\beta,\mm}$), such that 
\be\label{eq:H2_approx_simul}
\begin{split}
    \Hop_2^{\alpha\beta} \simeq & 2\sum_{\nn\neq\mm} G^{\alpha\beta}_{\nn\mm}\spl_{\mm}\smi_{\nn} \Big[\delta_{\alpha\beta}+\bdop_{\alpha,\nn}\bop_{\alpha,\nn}+\bdop_{\alpha,\mm}\bop_{\alpha,\mm}\\
    & -\frac{1}{2}(\bdop_{\alpha,\nn}\bop_{\beta,\mm}+\bop_{\alpha,\mm}\bdop_{\beta,\nn}+\hc)\Big].
\end{split}
\ee
In computing the polariton-phonon properties in Sec.~\ref{app:Polaron-Polariton_1D}, we furthermore neglected terms proportional to the phonon number operator ($\bdop_{\alpha,\nn}\bop_{\alpha,\nn}$) and phonon hopping between sites or spatial directions ($\bdop_{\alpha,\nn}\bop_{\beta,\mm}$), as they contribute only to orders higher than $\eta^2$. That is, we approximated 
\be\label{eq:H2_approx_polaron}
    \sum_{\alpha\beta}\Hop_2^{\alpha\beta} \simeq -2\sum_{\nn\neq\mm} G_{\nn\mm}\spl_{\mm}\smi_{\nn}\,,
\ee
where we used Eq.~\eqref{eq:Helmholtz_Eq} to write $\sum_\alpha G_{\nn\mm}^{\alpha\alpha} = - G_{\nn\mm}$.

We compare the transport dynamics by numerically integrating the Schrödinger equation on the truncated Hilbert space using either \cref{eq:H2_approx_simul} or \cref{eq:H2_approx_polaron}. 
We simulate transport in an array of $N=400$ atoms polarized perpendicularly to the array axis with an interatomic distance $d=\lambda_0/4$.
We find no difference in the results when using \cref{eq:H2_approx_simul} or \cref{eq:H2_approx_polaron} [cf. empty squares and stars in Fig.~\ref{fig:Phonon_Hopping}(b,c)].

The results are presented for the case of $\nu=0.2\gamma_0$; however, we checked that the same conclusions hold for other trap frequencies in the regime where the phonon population is small, $\sum_{\nn,\alpha}\avg{\bdop_{\alpha,\nn}\bop_{\alpha,\nn}}\ll 1$.
The results presented in Sec.~\ref{sec:transport} and the remainder of this appendix are obtained by numerical integration in MATLAB, where the second-order Lamb-Dicke Hamiltonian is approximated by \cref{eq:H2_approx_polaron}.

\subsection{Transport in 1D arrays with parallel polarization}

\begin{figure}[!t]
    \centering
    \includegraphics[width=\columnwidth]{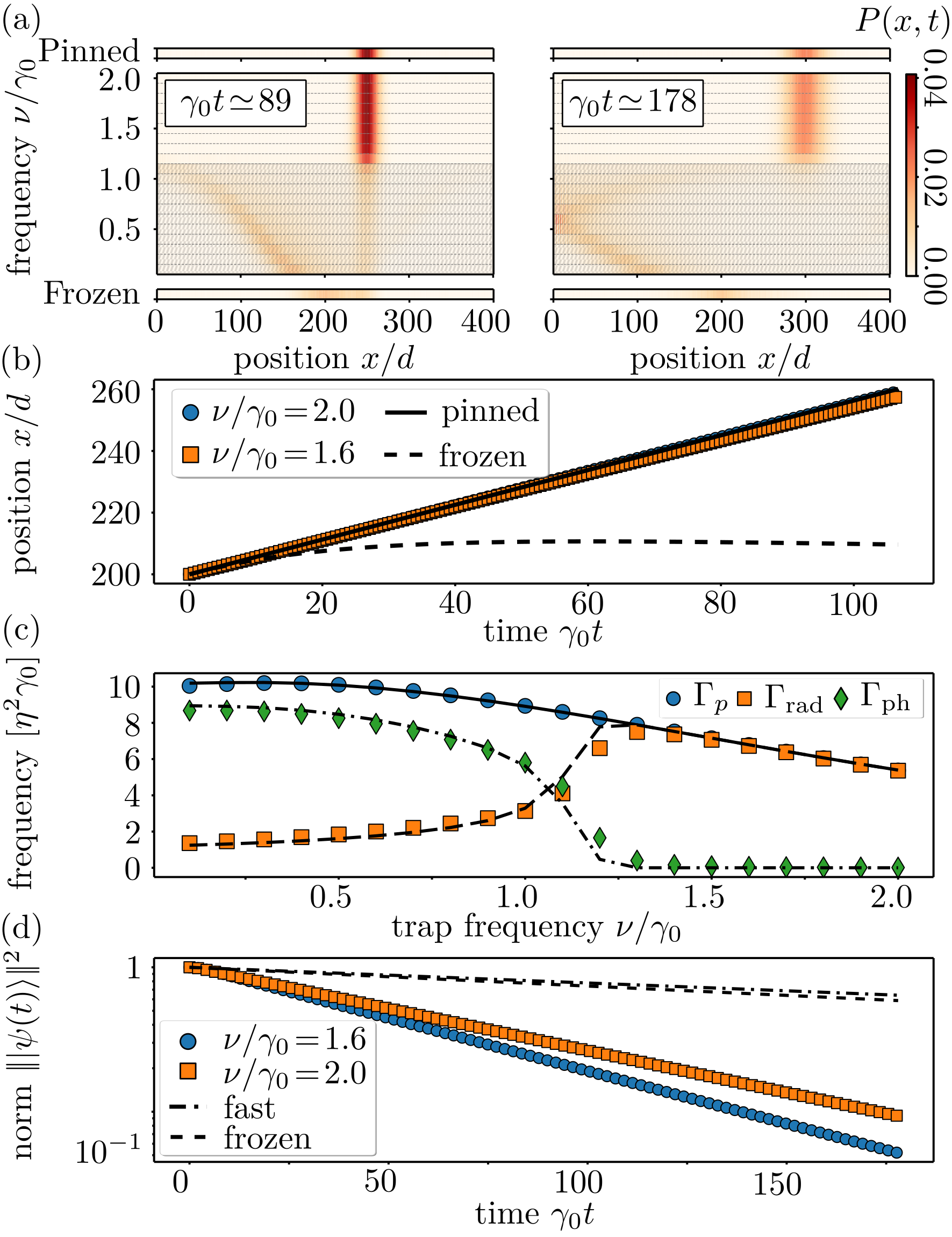}
    \caption{{\bf Transport for a 1D array with parallel polarization.} We consider an array of $N=400$ atoms with lattice spacing $d=\lambda_0/4$ and $\eta=0.05$. (a) Propagation of the normalized spin population along the array for different values of the trap frequency $\nu/\gamma_0$ at two instances of time as indicated in each panel. 
    The hatched region represents the regime where the numerical simulations cannot be trusted because the phonon population is significant. 
    (b) Evolution of the center $\bar{x}/d$ of the wave packet. The black dashed line corresponds to the prediction for pinned atoms. (c) Scattering rates as a function of trap frequency: depletion rate of the zero-phonon sector ($\Gamma_p$), radiative decay rate ($\Gamma_\text{rad}$), and rate of population of the one-phonon sector ($\Gamma_\text{ph}$). The values extracted from the numerical simulations (colored markers) are compared to values obtained from the polaron ansatz (Appendix~\ref{app:Polaron-Polariton_1D}): black solid line ($\Gamma_p$), dashed line ($\Gamma_\text{rad}$), and dot-dashed line ($\Gamma_\text{ph}$). (d) Survival probability of the polaron as a function of time for different trap frequencies (colored markers) and for the fast-motion (dash-dotted line) and frozen-motion approximations (dashed line). Other parameters: $k_s=0.8\pi/d$ and $\sigma_k = (\pi/d-k_s)/2$. For these values, the group velocity for a pinned atomic array is $v_g \simeq 0.5 d\gamma_0$ at $k=k_s$. The frozen-motion results are obtained assuming Gaussian disorder with standard deviation $\eta / k_0$ and averaging over 1000 repetitions.}
    \label{fig:Transport_1D_parallel}
\end{figure}

We now consider transport along a 1D array of parallel-polarized atoms. The results are shown in Fig.~\ref{fig:Transport_1D_parallel}.

Compared to the case of perpendicular polarization, excitation transport deviates from the pinned-atom limit at a much larger trap frequency [cf. Fig.~\ref{fig:Transport_1D_parallel}(a) and Fig.~\ref{fig:Transport_1D_perpendicular}(a)]. 
This behavior originates from the distinct features of the bare polariton dispersion relation $J_p^{(0)}$. For parallel-polarized atoms, $J_p^{(0)}$ has no divergence; thus, resonant scattering can only occur for a finite range of trap frequencies. For the lattice constant considered here, this occurs for $\nu\lesssim 1.2\gamma_0$. 
For trap frequencies where a resonance can occur, the density of states is always significant, leading to a strong contribution from resonant scattering [Fig.~\ref{fig:Transport_1D_parallel}]. In this regime, the large phonon scattering strongly modifies propagation properties [Fig.~\ref{fig:Transport_1D_parallel}(a)] and quickly leads to a breakdown of our perturbative ansatz [Fig.~\ref{fig:Phonon_Hopping}(a)].

Conversely, for $\nu\gtrsim 1.2\gamma_0$, resonant scattering is absent and transport is ballistic at a group velocity $v_g$ well-approximated by that of pinned atoms [see Fig.~\ref{fig:Transport_1D_parallel}(a,b)]. 
This is confirmed in Fig.~\ref{fig:Transport_1D_parallel}(c), where we see a large increase in the resonant phonon scattering $\Gamma_\text{ph}$ (green diamonds) as the frequency drops below $\nu\simeq 1.2\gamma_0$.
We also find good agreement between the different contributions to the polariton-phonon scattering rate extracted from the simulation and the predictions of polaron theory [see black lines and colored markers in Fig.~\ref{fig:Transport_1D_parallel}(c)].
This allows us to identify $\Gamma_\text{rad}=\Gamma_p-\Gamma_\text{res} = -2\Im[\varepsilon^{\infty}_p + \Sigma_p^\text{off-res}]$ and $\Gamma_\text{ph} = -2\Im[\Sigma_p^\text{res}]$ as the off-resonant and resonant contributions to polaron damping, respectively. 
This confirms the interpretation that the off-resonant scattering rate represents a direct radiative loss, while the resonant one describes an irreversible transfer of excitations to other modes that does not immediately correspond to radiative loss.

We analyze the radiative decay of the polaron-polariton excitation by plotting the norm $\norm{\ket{\psi(t)}}^2$ as a function of time in Fig.~\ref{fig:Transport_1D_parallel}(d) for the case of trap frequencies where propagation is valid at long time [Fig.~\ref{fig:Transport_1D_parallel}(a)]. This estimates the probability of finding an excitation in the array as a function of time. 

\subsection{Details on the fit of the decay rates}
\begin{figure}
    \centering
    \includegraphics[width=\columnwidth]{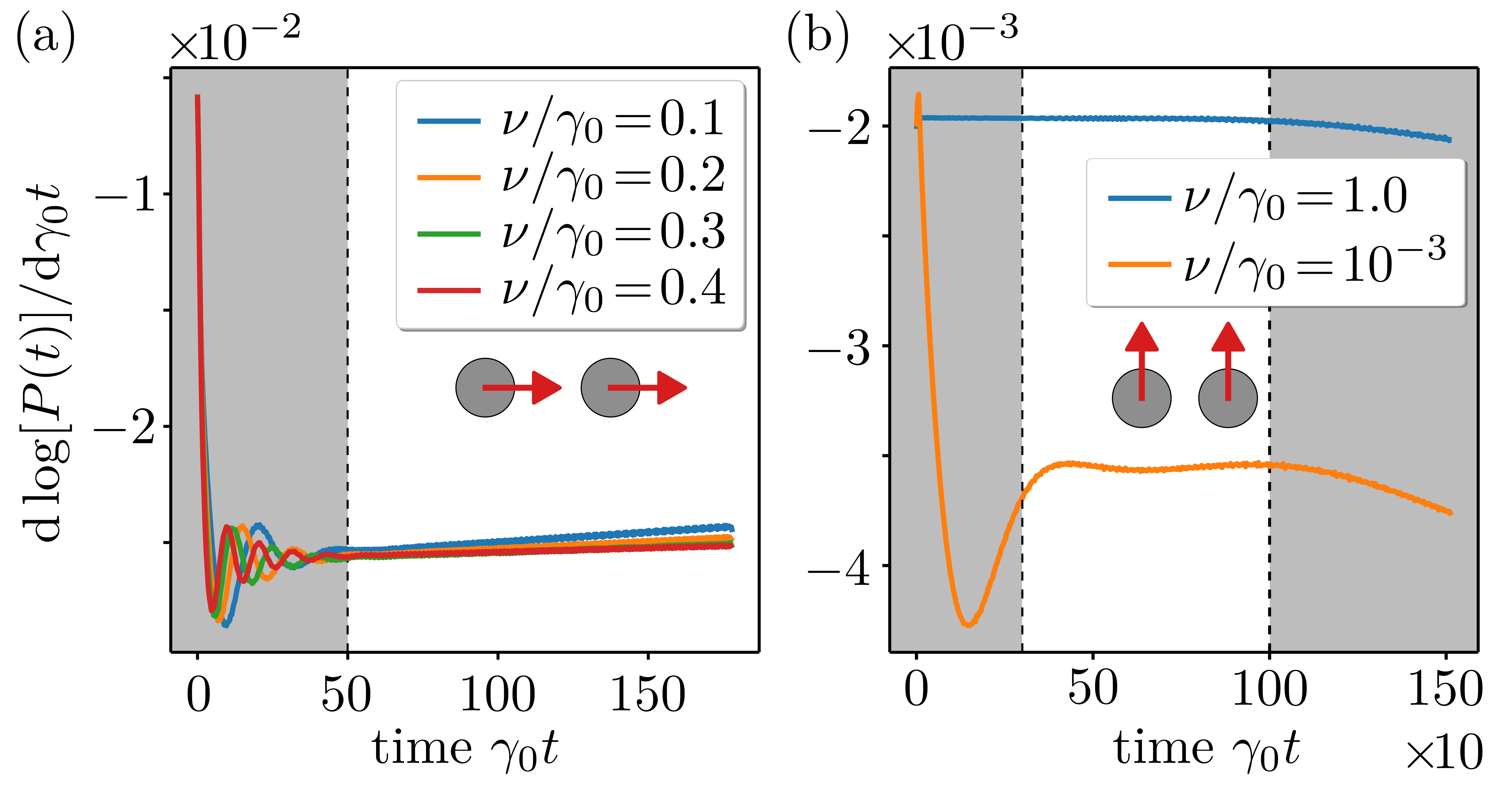}
    \caption{{\bf Timescales for the fit of decay rates.} Log-derivative of the total population in the zero-phonon sector for different trap frequencies (see legend) for (a) parallel and (b) perpendicular polarization, respectively. Other parameters as in Figs.~\ref{fig:Transport_1D_perpendicular} and \ref{fig:Transport_1D_parallel}.}
    \label{fig:Fig_Fit_time}
\end{figure}
We now describe the procedure for obtaining the rates $\Gamma_p$, $\Gamma_\text{rad}$, and $\Gamma_\text{ph}$ presented in Fig.~\ref{fig:Transport_1D_perpendicular}(b) and Fig.~\ref{fig:Transport_1D_parallel}(c). 
We extract the polaron-polariton decay $\Gamma_p$ and the radiative decay 
$\Gamma_{\text{rad}}$ by fitting an exponential to the population decay in the 
zero-phonon sector and the decay of the norm $\norm{\ket{\psi(t)}}^2$, respectively. 

The fit of the norm decay is limited to the maximum critical time $t_c$ defined above and plotted in Fig.~\ref{fig:Phonon_Hopping}(a), since otherwise inaccuracies in the dynamics of the one-phonon sector could affect the results.

In contrast, the time interval for the fit of the polaron-polariton decay rate can be extended to times later than $t_c$, since the dynamics in the zero-phonon sector is accurately captured also beyond $t_c$ [cf. Appendix~\ref{app:details_simulation}].
Since the polaron-polariton quasiparticle takes a finite time to form after the wave packet is prepared, 
it is necessary to exclude the polaron formation time from the fit. 
In Fig.~\ref{fig:Fig_Fit_time} we show the log-derivative of the population of the zero-phonon sector $P(t)$ for parallel and perpendicular polarization. The polaron formation time and the effects of the large phonon population can be observed as deviations from linearity in the log-derivative of $P(t)$.
Accordingly, in Figs.~\ref{fig:Fig_Fit_time}(a) and (b), we indicate the time intervals excluded from the fit in gray.
We observe that the formation time increases as the trap frequency decreases, suggesting it might be determined by the typical mechanical response time $1/\nu$. However, further study is necessary to determine the exact relation between the formation time and the atomic trap frequency $\nu$.

For 2D arrays, the phonon population remains below 0.1, and the ratio between the population in the one- and zero-phonon sector is well below 0.5 throughout the simulation for the values of $\nu / \gamma_0$ shown in Fig.~\ref{fig:Transport_2D}. We therefore extract $\Gamma_\text{rad}$ by fitting over the entire time interval. Consistent with the 1D case, the polaron formation time increases as $\nu/\gamma_0$ decreases. To isolate the effects of polaron formation from the fit of $\Gamma_p$, we restrict the fitting window to $\gamma_0 t > 250$. Visual inspection confirms that this threshold safely excludes the polaron formation time for all values of $\nu/\gamma_0$ presented in Fig.~\ref{fig:Transport_2D}(b).

\section{Input-output formalism including recoil}\label{app:input_output}

In this appendix, we derive in detail the input-output formalism for a 2D array of atoms driven at normal incidence in the presence of recoil effects. 
In the Lamb-Dicke regime, we approximate the total field of the system by expanding \cref{eq:input_output_formula_1} in powers of the atomic fluctuations to obtain \cref{eq:General_Input_Ouput_Formula}. The first term in \cref{eq:General_Input_Ouput_Formula} describes the input field (or driving field). We take this to be a classical field propagating in the direction $(\pp,p_z)$ with $p_z = \sqrt{(\omega_d/c)^2 - \pp^2} \simeq \sqrt{k_0^2 - \pp^2}$. Namely, we define
\be\label{eq:Def_Input_Field}
\!\!\!\! \EE_0^+(\rr,t) = \EE_0^+ e^{i\pp\cdot \rr_{xy} + ip_z z - i \omega_d t} = \EE_0^+(\rr_{xy}) e^{ip_z z - i \omega_d t}\!,\!\!\!
\ee
where $\rr_{xy} =(x,y)$. To compute the fields transmitted and reflected from the array, it is convenient to rewrite \cref{eq:General_Input_Ouput_Formula} in a mixed representation by taking the 2D Fourier transform in the $xy$-plane. We thus define
\be
    \hat{\EE}^+(\kk,z) \equiv \int_{\mathbb{R}^2}\text{d}\rr_{xy}\,e^{-i\kk\cdot\rr_{xy}}\hat{\EE}^+(\rr_{xy},z)\,.
\ee
We now proceed to take the Fourier transform of the right-hand side of \cref{eq:General_Input_Ouput_Formula}.
Using the properties of the Fourier transform, the calculations are straightforward for the input field and for the second term of \cref{eq:General_Input_Ouput_Formula} for $\alpha=x,y$. For the case of $\alpha=z$, we use the expression of the free-space Green's function~\cite{Novotny,Wild_dissertation}
\begin{align}\label{eq:Green_Function_Mixed_Rep}
    &\bar{\bar{G}}(\kk,z) = \frac{i}{2k_z}e^{ik_z|z|}\mathds{1} - \frac{i}{2k_zk_0^2}e^{ik_z|z|} \nonumber \\
    & \times \begin{bmatrix} k_x^2 & k_x k_y & k_x k_z {\rm sgn}(z) \\ k_x k_y & k_y^2 & k_yk_z {\rm sgn}(z) \\ k_x k_z {\rm sgn}(z) & k_y k_z {\rm sgn}(z) & k_z^2 + 2ik_z \delta(z) \end{bmatrix},
\end{align}
with $k_z = \sqrt{k_0^2 - \kk^2}$. Using \cref{eq:Green_Function_Mixed_Rep}, we have $k_0^{-1}\partial_z \bar{\bar{G}}(\kk,z) = \sgn z i(k_z/k_0)\bar{\bar{G}}(\kk,z)$ for $z \neq 0$. For the term in the second line of \cref{eq:General_Input_Ouput_Formula}, we utilize 
\begin{align}
\sum_\nn\!\int_{\mathbb{R}^2}\!\!\!\text{d}\rr_{xy}\,e^{-i\kk\cdot\rr_{xy}} \bar{\bar{G}}(\rr_{xy}\! -\! \rr_\nn,z) \dpp\, \smi_\nn \!=\! \sqrt{N} \bar{\bar{G}}(\kk,z)\dpp \, \sop_{\kk + \mathbf{g}},\!\!\! \nonumber
\end{align}
where we used $\sop_\kk = \sum_\nn e^{-i\kk\cdot\rr_\nn}\smi_\nn/\sqrt{N}$, and where $\mathbf{g}$ is a reciprocal lattice vector chosen such that $\kk + \mathbf{g}$ lies in the first Brillouin zone. We then obtain
\begin{align}\label{eq:General_Input_Ouput_Formula_k_space}
&\hat{\EE}^+(\kk,z) = \EE_0^+(\kk,z) \nonumber \\
& \!-\! \mu_0\w_0^2 \bar{\bar{G}}(\kk,z)\dpp\bigg[ \Big( 1 \!-\! \frac{\eta_\parallel^2}{2} \!-\! \frac{k_z^2}{k_0^2}\frac{\eta_z^2-\eta_\parallel^2}{2}\Big)\sop_{\kk + \mathbf{g}} \nonumber \\
&\!-\!\sum_{\qq,\alpha}\frac{i\eta_\alpha k_\alpha}{\sqrt{N}k_0}(1+\delta_{\alpha,3}({\rm sgn}(z) \!-\! 1)) (\bdop_{\alpha,\qq-\kk-\mathbf{g}}\!+\!\hc)\sop_\qq \bigg].
\end{align}
Assuming normal incidence, $\kk=\mathbf{0}$, we finally obtain \cref{eq:input_output_with_phonons_momentum_space}.

In the following, we will compute the output field for the steady state of the system $\rhop_\text{ss}$. Accordingly, we need to determine the steady-state values of $\avg{\sop_\mathbf{0}}_\text{ss}$, $\avg{\bop_{\alpha,-\qq}\sop_\qq}_\text{ss}$, and $\avg{\bdop_{\alpha,\qq}\sop_\qq}_\text{ss}$.

\subsection{Steady state of the array at normal incidence}

The dynamics of the expectation value $\braket{\hat{A}} = {\rm tr}[\hat{A}{\rho}]$ of a generic system operator follows from Eqs. \eqref{eq:ME} and \eqref{eq:Recycling_FreeSpace}, and reads 
\begin{align}
&\partial_t \braket{\hat{A}}\! = -i \big\langle\!\hat{A}(\Hop+ \Hop_d) - (\Hop+ \Hop_d)^\dagger\hat{A}\big\rangle \nonumber
\\
&+\gamma_0 \sum_{\nn,\mm}\int \!\!\text{d}\uu\, \mathcal{D}(\uu) \Big\langle e^{ik_0\uu\cdot \hat{\rr}_\mm}\spl_\mm\, \hat{A}\, e^{-ik_0  \uu \cdot \hat{\rr}_\nn} \smi_\nn \Big\rangle,
\label{eq:EoM_avg_A}
\end{align}
where we included the driving Hamiltonian in the rotating wave approximation 
\be\label{eq:H_drive_appendix}
    \Hop_d 
= -\sum_\nn \spare{\dpp^\dagger \EE^+_0(\rr_\perp,t)e^{i \eta_z (\bdop_z + \bop_z)}\spl_\nn+\hc}.
\ee
In the following, we consider the Lamb-Dicke regime defined in \cref{eq:LambDicke_condition} and expand \cref{eq:EoM_avg_A} to second order in the atomic center-of-mass fluctuations, as discussed in Sec.~\ref{sec:Model} and Appendix~\ref{app:master_equation}.
Assuming \emph{linear response}, we proceed to approximate $[\smi_\nn,\spl_\mm]\simeq\delta_{\nn\mm}$, and transform \cref{eq:EoM_avg_A} to momentum space following the procedure detailed in Sec.~\ref{sec:Model}. The driving Hamiltonian in \cref{eq:H_drive_appendix} then takes the form of \cref{eq:driving_Hamiltonian_normal_incidence}.
While working in momentum space is not strictly necessary, it is convenient as several of the approximations discussed below become more apparent.

We then proceed to derive the equations of motion for $\avg{\sop_\mathbf{0}}$ and $\avg{\bop_{\alpha,-\qq}\sop_\qq}$. For the former, we have
\begin{align}\label{eq:EoM_s0_general}
    \partial_t \!\braket{s_{{\bf 0}}} =\; & i \pare{\Delta-\varepsilon_{\mathbf{0}}^\infty}\!\braket{s_{{\bf 0}}}\! +\! i \Omega_{\mathbf{0}}\Big(1\!-\!\frac{\eta_z^2}{2}\Big)\!+\!\frac{\eta_z\Omega_{\mathbf{0}}}{\sqrt{N}}\braket{\bdop_{z,{\bf 0}}\!+\!\bop_{z,{\bf 0}}}\nonumber \\
    &+i\sum_{\qq,\alpha}\frac{\eta_\alpha g^\alpha_{\qq,{\bf 0}}}{\sqrt{N}} \Big\langle\!\!\pare{\bdop_{\alpha,\qq}+\bop_{\alpha,-\qq}}\!\sop_\qq\Big\rangle.
\end{align}
The equation for $\avg{\bop_{\alpha,-\qq}\sop_\qq}$ reads
\begin{align}\label{eq:EoM_bs_general}
    &\partial_t \braket{\bop_{\alpha,-\qq}\sop_\qq} \simeq i \pare{\Delta-\varepsilon_0^\infty-\nu_\alpha}\braket{\bop_{\alpha,-\qq}\sop_\qq} -i\frac{\eta_\alpha g^\alpha_{\qq,{\bf 0}}}{\sqrt{N}}\braket{s_{{\bf 0}}} \nonumber \\
    &+\frac{\eta_\alpha\Omega_{\mathbf{0}}}{\sqrt{N}}\bigg[\big\langle\pare{\bdop_{z,-\qq}+\bop_{z,\qq}}\bop_{\alpha,-\qq}\big\rangle +\delta_{\alpha z} (\braket{\sdop_\qq\sop_\qq}-1)\bigg]\nonumber \\
    & +\!i \delta_{\qq\mathbf{0}}\Omega_\mathbf{0}\pare{1\!-\! \frac{\eta_\parallel^2}{2}}\avg{\bop_{\alpha,\mathbf{0}}} \!-\! i \sum_{\beta, \kk} \!\frac{\eta_\beta g^\beta_{\bf \qq,\kk}}{\sqrt{N}}\braket{\bdop_{\beta,\kk-\qq}\bop_{\alpha,-\qq}\sop_\kk},
\end{align}
where we neglected a term proportional to $\avg{\sdop_{\kk+\qq}\sop_{\qq}\sop_\kk}$, which is irrelevant within the linear response regime.
The closed set of equations in Eq.~\eqref{eq.EOMS_mirror} is obtained from Eqs.~\eqref{eq:EoM_s0_general} and \eqref{eq:EoM_bs_general} by neglecting terms whose contribution is non-linear in the Rabi frequency $\Omega_{\mathbf{0}}$, higher than second order in the Lamb-Dicke parameter, or non-zero only in the presence of many phonons or spin excitations in the system.
For greater clarity, let us analyze how these conditions are reflected in Eqs.~\eqref{eq:EoM_s0_general} and \eqref{eq:EoM_bs_general}.

First, terms proportional to $\avg{\sdop_\qq\sop_\qq}$ with $\qq\neq 0$ are at least of order $O(\eta_\alpha)$, as they are generated either through driving a sideband or by phonon-assisted scattering of a spin-wave with $\kk=\mathbf{0}$. They can, therefore, be neglected, as they lead to corrections higher than second order in the Lamb-Dicke parameter for both Eq.~(\ref{eq:General_Input_Ouput_Formula_k_space}) and $\avg{\sop_\mathbf{0}}$. Terms proportional to $\avg{\sdop_\bold{0}\sop_\bold{0}}$ are also negligible, as they contribute only at $O(\Omega_\bold{0}^2)$ to $\avg{\bdop_{\alpha,\bold{0}}\sop_\bold{0}}$.

Second, the evolution of the phonon coherences $\avg{\bop_{\alpha,\mathbf{0}}}$ and $\avg{\bdop_{\alpha,\mathbf{0}}}$ is determined by the equation
\be\label{eq:eom_phonon-coherences}
\begin{split}
    \partial_t\avg{\bop_{\alpha,\mathbf{0}}} = & -i\nu_\alpha \avg{\bop_{\alpha, \mathbf{0}}} +\delta_{z\alpha}\frac{\eta_z\Omega_{\mathbf{0}}}{\sqrt{N}}\avg{\sdop_{0}-\sop_{0}}
\end{split}
\ee
and its complex conjugate, respectively. For atoms initially in their ground state, \cref{eq:eom_phonon-coherences} shows that the phonon coherences only contribute to \cref{eq:EoM_s0_general} and \eqref{eq:EoM_bs_general} via terms quadratic in $\eta_\alpha\Omega_{\mathbf{0}}$, and can thus be neglected.
For the same reason, we can also neglect $\braket{\bdop_{\beta,\qq}\bop_{\alpha,\qq}}$ and $\braket{\bop_{\beta,\qq}\bop_{\alpha,-\qq}}$.

The last term in \cref{eq:EoM_bs_general} originates from the scattering of a spin excitation by \emph{absorbing} a phonon. For atoms in their motional ground state, its contribution to \cref{eq:EoM_bs_general} is negligible to second order in the Lamb-Dicke parameter, as $\braket{\bdop_{\beta,\kk-\qq}\bop_{\alpha,-\qq}\sop_\kk}\sim O(\eta_\alpha^2)$. Note that, as discussed in Sec.~\ref{sec:Discussion} for the examples in Sec.~\ref{sec:input-output}, this contribution becomes relevant only after thousands of scattering events, which lead to significant phonon production.

Finally, the equation of motion for $\braket{\bdop_{\alpha,\qq}\sop_\qq}$ depends only on coherences of the form $\braket{\bdop \bop \sop}$ and $\braket{\sdop\sop\sop}$, and thus contributes negligibly to \cref{eq:EoM_s0_general}.
In this manner, Eqs.~\eqref{eq:EoM_s0_general} and \eqref{eq:EoM_bs_general} reduce to \cref{eq.EOMS_mirror}.
The steady-state solution is then obtained from \cref{eq.EOMS_mirror} by setting the left-hand side to zero.

\subsection{Frozen-motion limit} \label{app:frozen_motion_proof}

The general result for the steady state of the system in \cref{eq.steady_state}, together with the input-output equation \cref{eq:General_Input_Ouput_Formula_k_space}, allows us to determine the output field for any value of the atomic trap frequency. 
In the limit of $\nu / \gamma_0 \rightarrow 0$ and considering motion only perpendicular to the array ($\eta_\parallel=0$), we can show analytically that this result coincides with the output field predicted by the frozen-motion approximation to leading order in $\eta_z$. 

This agreement occurs because, in both approaches, an incoming photon couples equally strongly to all collective modes of the array with $\kk \neq 0$. 
In the quantum case, this appears as a constant coupling $\sim \eta_z\Omega$ to all sideband modes.
In the frozen-motion case, the coupling to $\kk \neq 0$ modes arises from the disorder in the lattice positions. 
Since atomic displacements are sampled independently (white noise), this results in momentum space in a constant coupling strength across the Brillouin zone. 

To prove this intuition, it is more convenient to work in the real-space representation. 
By taking the Fourier transform of \cref{eq.EOMS_mirror} and solving for the steady state, we obtain the solution
\be\label{eq:Steady_state_real_space}
\begin{split}
    \avg{\smi_\nn}_\text{ss} &= \pare{1-\frac{\eta_z^2}{2}}\Omega \sum_\mm K_{\nn\mm}^{-1}(0),\\
    \avg{\Rop_{z,\nn}\smi_{\jj}}_\text{ss} &= i \eta_z \Omega K_{\jj\nn}^{-1}(\nu_z),
\end{split}
\ee
where we defined the matrix 
\be\label{eq:kernel}
    K_{\nn\mm}(\nu_z) \equiv \pare{\nu_z+\Delta -i\frac{\gamma_0}{2}}\delta_{\nn\mm} + G_{\nn\mm} - \frac{\eta_z^2}{2}G^{zz}_{\nn\mm},
\ee
In Equation~\eqref{eq:Steady_state_real_space}, $K_{\nn\mm}^{-1}(0)$ indicates the element on the $\nn$-th row and $\mm$-th column of the inverse of \cref{eq:kernel}.

In the frozen-motion approximation, the atomic center-of-mass variables are randomly distributed parameters with zero mean $\avg{\!\avg{R_{z,\nn}}\!}=0$, where $\avg{\!\avg{\cdot}\!}$ indicates averaging over the atomic displacements.
Accordingly, solving for the steady state for evolution under Eqs.~\eqref{eq:H_expansion} and~\eqref{eq:driving_Lamb_Dicke_expansion} in the frozen-motion approximation yields the same result for $\avg{\smi_\nn}$ as \cref{eq:Steady_state_real_space}, while for the correlation between spin excitation and position we obtain
\be
    \avg{\!\avg{(R_{z,\nn}\avg{\smi_\jj}_\text{ss})}\!} = i \eta_z \Omega K^{-1}_{\jj\nn}(0)\,.
\ee
This expression coincides with the second line of \cref{eq:Steady_state_real_space} in the limit $\nu_z/\gamma_0\rightarrow 0$. Hence, the steady state and output field predicted by our polaron-polariton theory and the frozen-motion approximation agree for the case under consideration.

\subsection{Output field subject to the detection of a phonon} \label{app:output_field_with_phonon}

To study the effect of recoil on the scattering more closely, we derive the expression for the output field conditioned on the detection of a phonon at the site $\nn=\mathbf{0}$ of the array. For a field at position $\rr$, this reads
\begin{align}
\braket{\bop_{\alpha,\vec{n}={\bf 0}}\hat{\bf E}^+(\rr)} ={}& \mu_0\omega_0^2\Bigg[\sum_\nn \bar{\bar{G}}(\rr - \rr_\nn)\dpp \braket{\bop_{\alpha,\vec{n}={\bf 0}}\smi_\nn} \nonumber \\
& - \eta_\alpha \bar{\bar{G}}^\alpha(\rr) \braket{\smi_{\bf 0}}\Bigg].
\label{eq:input_output_with_phonon}
\end{align}
Equation~\eqref{eq:input_output_with_phonon} is derived from \cref{eq:General_Input_Ouput_Formula} to first order in the Lamb-Dicke parameter and neglecting expectation values of higher order in the phonon operators (as discussed above). We can rewrite the first term on the right-hand side in momentum space by using a 2D Fourier transform along the $xy$-plane:
\begin{align}\label{eq:second_term_approx}
\!\!\!\!\!\!&\sum_\nn \bar{\bar{G}}(\rr - \rr_\nn)\dpp \braket{\bop_{\alpha,\vec{n}={\bf 0}}\smi_\nn}  \nonumber \\
\!\!\!\!\!\!&= \frac{1}{N}\!\!\sum_{\pp,\kk\in\rm BZ}\!\int \!\!\frac{{\rm d}^2q}{(2\pi)^2} e^{i\qq\cdot\rr} \bar{\bar{G}}(\rr - \rr_\nn)\dpp \sum_\nn e^{i(\kk-\qq)\cdot\rr_\nn} \braket{\bop_{\alpha,\pp} \sop_{\kk}}\nonumber\\
\!\!\!\!\!\!&\simeq \int_{\rm  BZ} \!\frac{{\rm d}^2k}{(2\pi)^2}\!\braket{\bop_{\alpha,-\kk} \sop_{\kk}} \!\!\!\sum_{{\bf g}\in {\rm RL}} \! \! e^{i(\kk + {\bf g})\cdot\rr} \bar{\bar{G}}(\kk \!+\!{\bf g},z)\dpp.\!
\end{align}
In the last step, we first neglected all terms $\avg{\bop_{\alpha,\pp} \sop_{\kk}}$ with $\pp\neq -\kk$, as they only contribute to higher order in the Lamb-Dicke regime. We then used the identity in \cref{eq:Latice_Delta} and integrated in momentum space. Finally, we took the continuum limit by replacing $(N d^2)^{-1}\sum_{\kk\in \text{BZ}} \rightarrow \int_\text{BZ}d^2k/(2\pi)^2$.

For the second term in \cref{eq:input_output_with_phonon}, we use the approximation 
\begin{align}\label{eq:smi_0_approx}
\braket{\smi_{\bf 0}} = \frac{1}{\sqrt{N}} \sum_\kk \braket{\sop_\kk} \simeq \frac{1}{\sqrt{N}} \braket{\sop_{\bf 0}}\,,
\end{align}
since for a drive at normal incidence, all other modes contribute only as higher-order Lamb-Dicke corrections~\cite{Shahmoon2019}.

Let us now consider the field conditioned on the detection of a phonon along $\alpha = z$. Substituting the steady-state solution of Eq.~(\ref{eq.steady_state}) into Eqs.~\eqref{eq:second_term_approx} and \eqref{eq:smi_0_approx}, we obtain the following total field at $\rr$ with a $z$-polarized phonon:
\begin{align}
\braket{\bop_{z,\nn={\bf 0}}\hat{\EE}^+(\rr)} =&\, \frac{\mu_0\omega_0^2 \eta_z}{\sqrt{N}} \Big[ i \! \int\!\!\frac{{\rm d}^2q}{(2\pi)^2} e^{i\qq\cdot\rr} \bar{\bar{G}}(\qq,z)\hat{\dpp}\hat{\dpp}^\dagger \PP_{z,\qq}^{(1)}(\Delta)\nonumber \\
&- \frac{\partial_z \bar{\bar{G}}(\rr)}{k_0} \hat{\dpp}\hat{\dpp}^\dagger \PP^{(2)}(\Delta)\Big],
\end{align}
which is identical to \cref{eq:field_with_recoil}. Note that the integral in the first term is now extended to all momenta in the $xy$-plane, using the property $\varepsilon_{\kk+\mathbf{g}}= \varepsilon_\kk$.
Moreover, we used the definition $\Omega_\mathbf{0} = \dpp^\dagger \sum_\nn \EE_0^+(\rr_\nn,0) / \sqrt{N}$ and introduced the effective polarizations $\PP_{z,\qq}^{(1)}(\Delta)$ and $\PP^{(2)}(\Delta)$. The first is defined as
\be
\PP_{z,\qq}^{(1)}(\Delta) = -\frac{|\dpp|^2}{\Delta - \varepsilon^{\infty}_\qq - \nu_z}\frac{1}{\sqrt{N}}\sum_{\nn} \EE_0^+(\rr_\nn,0),
\ee
representing the response of the collective mode $\sdop_\qq \bdop_{z,-\qq}\ket{0}$ to the external driving field. The second is
\be\label{eq:P2_appendix}
\PP^{(2)}(\Delta) = -|\dpp|^2G_{\bf 0}^{\rm R}(\Delta)\frac{1}{\sqrt{N}} \sum_{\nn} \EE_0^+(\rr_\nn,0),
\ee
indicating that the excitation is stored at zero quasimomentum before a phonon is created as the \emph{photon} leaves the array. This describes a recoil event in the $z$-direction, captured by the derivative $\partial_z \bar{\bar{G}}$.

For a recoil occurring in the array plane ($\alpha = x,y$), a similar procedure yields 
\begin{align}
\braket{\bop_{\alpha,{\bf 0}}\hat{\EE}^+(\rr)} =& \, \frac{\mu_0 \omega_0^2 \eta_\alpha}{ \sqrt{N}}\Big[\int\!\!\frac{{\rm d}^2q}{(2\pi)^2} e^{i\qq\cdot\rr} \bar{\bar{G}}(\qq,z)\hat{\dpp}\hat{\dpp}^\dagger \PP_{\alpha,\qq}^{(1)}(\Delta)\nonumber \\
&- \frac{\partial_\alpha \bar{\bar{G}}(\rr)}{k_0} \hat{\dpp}\hat{\dpp}^\dagger \PP^{(2)}(\Delta)\Big].
\end{align}
In this case, the second term has the same interpretation as before, with $\PP^{(2)}$ still given by \cref{eq:P2_appendix}. The first term, instead, describes the excitation of a phonon via the spin-phonon Fr{\"o}hlich interaction $g^\alpha$, after the photon has been absorbed at quasimomentum $\qq = {\bf 0}$:
\be
\!\!\PP_{\alpha,\qq}^{(1)}(\Delta) = - \frac{G_{\bf 0}^{\rm R}(\Delta) g^{\alpha}_{\qq,{\bf 0}}|\dpp|^2}{\Delta - \varepsilon_\qq - \nu_\alpha}\frac{1}{\sqrt{N}}\sum_{\nn}\EE_0^+(\rr_\nn,0).\!\!
\ee
In this manner, we have derived the electric field structure subject to the presence of a phonon.

\section{Classical field of an oscillating electric dipole undergoing periodic motion}
\label{app:classical_dipole_field}

For comparison, here we compute the electric field of a \emph{classical} dipole undergoing periodic motion. We assume that the dipole, situated close to the origin ($\rr = {\bf 0}$), responds linearly to the field $\EE(\rr, \omega)$. This leads to a polarization field:
\begin{align}
\PP(\omega) = \alpha(\omega) \EE(\rr = {\bf 0}, \omega), 
\end{align}
with the standard electric polarizability $\alpha(\omega) = -|\dpp|^2[\omega - \omega_0 - i\gamma_0/2]^{-1}$ \cite{Wild_dissertation}. Furthermore, the dipole is assumed to undergo periodic motion along the $z$-direction in a harmonic potential (perpendicular to $\PP$) with $z(t) = l_{\rm osc} \cos(\nu_z t)$, where $l_{\rm osc} = 1/\sqrt{M\nu_z}$ is the oscillator length. The resulting electric field is given by the dipolar field:
\begin{align}
\EE(\rr,t) = \EE_{\rm dip}(\rr - z(t) \hat{z}, t), 
\end{align}
at a periodically varying distance $|\rr - z(t) \hat{z}|$. Here, $\EE_{\rm dip}(\rr, t)$ is the field produced at $\rr$ by a point dipole oscillating at $\rr = {\bf 0}$ with a polarization field $\PP(t) = \int d\omega \, e^{i\omega t} \PP(\omega)$. For this description to be accurate, we assume that $l_{\rm osc} \ll r$ and neglect the velocity and acceleration of the dipole. Indeed, this result follows directly by applying these approximations to the Liénard-Wiechert electric field obtained in Ref.~\cite{Power2015} for a moving electric dipole. To linear order, the corrected field is:
\begin{align}
\EE(\rr,t) &= \EE_{\rm dip}(\rr, t) - z(t)\partial_z \EE_{\rm dip}(\rr, t).
\end{align}

Since $z(t) = l_{\rm osc} \cos(\nu_z t)$, Fourier transforming yields:
\begin{equation}
\begin{aligned}
&\EE(\rr,\omega) - \EE_{\rm dip}(\rr, \omega)\\ 
&=- \frac{\eta_z}{\sqrt{2}k_0} \partial_z \left(\EE_{\rm dip}(\rr, \omega + \nu_z) + \EE_{\rm dip}(\rr, \omega - \nu_z)\right),
\end{aligned} \label{eq.E_osc_classical}
\end{equation}
using the Lamb-Dicke parameter $\eta_z = \sqrt{k_0^2/(2M\nu_z)}$. Note that the term proportional to $\partial_z\EE_{\rm dip}(\rr, \omega -\nu_z)$ is of \emph{exactly} the same form as the term proportional to $\partial_z \bar{\bar{G}}$ in Eq.~\eqref{eq:field_with_recoil}. Classically, as is evident from Eq.~\eqref{eq.E_osc_classical}, this correction to the field interferes with the bare field $\EE_{\rm dip}(\rr, \omega)$. This occurs because the mechanical driving is coherent. 

In the case considered in the main text, the system exists in a superposition of two terms: the bare field with no phonons and the modified field with one phonon. As a result, there is no interference between the bare and modified fields. Finally, the term proportional to $\partial_z\EE_{\rm dip}(\rr, \omega + \nu_z)$ corresponds to the absorption of a phonon. This process is not included in our quantum description, as we assume the array is initially in its motional ground state.

\bibliography{Bibliography}

\end{document}